\documentclass[sigconf]{acmart}
\usepackage{balance}       % to better equalize the last page
\usepackage{graphics}      % for EPS, load graphicx instead 
\usepackage{color}
\usepackage{booktabs}
\usepackage{textcomp}
\usepackage{subcaption}
\usepackage{enumerate}
\usepackage{makecell}
\usepackage{multicol}
\usepackage{multirow}
\usepackage{array}
\usepackage{fdsymbol}
\usepackage{enumitem}
\usepackage{amsmath}
\usepackage{arydshln}
\usepackage{stfloats}
\usepackage{graphicx}
\usepackage{amsthm}
\usepackage{listings}
\usepackage{caption} 
\usepackage{xspace}
\usepackage[linesnumbered]{algorithm2e}
\usepackage{csquotes}
\usepackage[bottom]{footmisc}

\newcommand*\textcircledmod[1]{\raisebox{.5pt}{\textcircled{\raisebox{-.9pt} {#1}}}}

\newcommand*{\eg}{\textit{e.g.},\xspace}
\newcommand*{\ie}{\textit{i.e.},\xspace}
\newcommand*{\vs}{\textit{vs.}\xspace}

\newcommand*{\etal}{\textit{et~al.}\xspace}

\newenvironment{s_itemize}{
\begin{itemize}[leftmargin=*]
  \setlength{\itemsep}{3pt}
  \setlength{\parskip}{0pt}
  \setlength{\parsep}{0pt}
}{\end{itemize}}

\newenvironment{s_enumerate}{
\begin{enumerate}[leftmargin=*]
  \setlength{\itemsep}{2pt}
  \setlength{\parskip}{0pt}
  \setlength{\parsep}{0pt}
}{\end{enumerate}}

\newcommand\pquote[2]{{``\textit{#2}'' (P#1)}}

\definecolor{DarkGreen}{HTML}{5DAC81}

\definecolor{XAIRGreen}{HTML}{1E6D30}
\definecolor{XAIRBlue}{HTML}{165DC2}
\definecolor{XAIRBrown}{HTML}{803F02}
\newcommand\colorwhen[1]{\textcolor{XAIRGreen}{#1}}
\newcommand\colorwhat[1]{\textcolor{XAIRBlue}{#1}}
\newcommand\colorhow[1]{\textcolor{XAIRBrown}{#1}}

\newcommand*{\gone}{\colorwhen{\textbf{\textit{G1}}}}
\newcommand*{\gtwo}{\colorwhen{\textbf{\textit{G2}}}}
\newcommand*{\gthree}{\colorwhen{\textbf{\textit{G2}}}}
\newcommand*{\gfour}{\colorwhat{\textbf{\textit{G3}}}}
\newcommand*{\gfive}{\colorwhat{\textbf{\textit{G4}}}}
\newcommand*{\gsix}{\colorwhat{\textbf{\textit{G5}}}}
\newcommand*{\gseven}{\colorhow{\textbf{\textit{G6}}}}
\newcommand*{\geight}{\colorhow{\textbf{\textit{G7}}}}
\newcommand*{\gnine}{\colorhow{\textbf{\textit{G8}}}}

\newcommand\review[1]{\textcolor{black}{#1}}

\setcitestyle{nocompress}

\AtBeginDocument{%
  \providecommand\BibTeX{{%
    \normalfont B\kern-0.5em{\scshape i\kern-0.25em b}\kern-0.8em\TeX}}}

\copyrightyear{2023}
\acmYear{2023}
\setcopyright{acmlicensed}
\acmConference[CHI '23]{Proceedings of the 2023
CHI Conference on Human Factors in Computing Systems}{April 23--28,
2023}{Hamburg, Germany}
\acmBooktitle{Proceedings of the 2023 CHI Conference on Human Factors in
Computing Systems (CHI '23), April 23--28, 2023, Hamburg, Germany}
\acmPrice{15.00}
\acmDOI{10.1145/3544548.3581500}
\acmISBN{978-1-4503-9421-5/23/04}

\begin{document}

\title{XAIR: A Framework of Explainable AI in Augmented Reality}


\author{Xuhai Xu}
\affiliation{%
  \institution{Meta Reality Labs \& UW}
  \city{} \state{} \country{}
}
\email{xuhaixu@uw.edu}

\author{Mengjie Yu}
\affiliation{%
  \institution{Meta Reality Labs}
  \city{} \state{} \country{}
}
\email{annaymj@meta.com}

\author{Tanya Jonker}
\affiliation{%
  \institution{Meta Reality Labs}
  \city{} \state{} \country{}
}
\email{tanya.jonker@meta.com}

\author{Kashyap Todi}
\affiliation{%
  \institution{Meta Reality Labs}
  \city{} \state{} \country{}
}
\email{kashyap.todi@gmail.com}

\author{Feiyu Lu}
\affiliation{%
  \institution{Meta Reality Labs \& VT}
  \city{} \state{} \country{}
}
\email{feiyulu@vt.edu}

\author{Xun Qian}
\affiliation{%
  \institution{Meta Reality Labs \& Purdue}
  \city{} \state{} \country{}
}
\email{qian85@purdue.edu}

\author{João Belo}
\affiliation{%
  \institution{Meta Reality Labs \& Aarhus Univ}
  \city{} \state{} \country{}
}
\email{joaobelo@cs.au.dk}

\author{Tianyi Wang}
\affiliation{%
  \institution{Meta Reality Labs}
  \city{} \state{} \country{}
}
\email{tianyiwang@meta.com}

\author{Michelle Li}
\affiliation{%
  \institution{Meta Reality Labs}
  \city{} \state{} \country{}
}
\email{michelleli@meta.com}

\author{Aran Mun}
\affiliation{%
  \institution{Meta Reality Labs}
  \city{} \state{} \country{}
}
\email{aranmun@meta.com}

\author{Te-Yen Wu}
\affiliation{%
  \institution{Meta Reality Labs \& Dartmouth}
  \city{} \state{} \country{}
}
\email{Te-yen.Wu.GR@dartmouth.edu}

\author{Junxiao Shen}
\affiliation{%
  \institution{Meta Reality Labs \& Cambridge}
  \city{} \state{} \country{}
}
\email{js2283@cam.ac.uk}

\author{Ting Zhang}
\affiliation{%
  \institution{Meta Reality Labs}
  \city{} \state{} \country{}
}
\email{tingzhang@meta.com}

\author{Narine Kokhlikyan}
\affiliation{%
  \institution{Meta Reality Labs}
  \city{} \state{} \country{}
}
\email{narine@meta.com}

\author{Fulton Wang}
\affiliation{%
  \institution{Meta Reality Labs}
  \city{} \state{} \country{}
}
\email{fultonwang@meta.com}

\author{Paul Sorenson}
\affiliation{%
  \institution{Meta Reality Labs}
  \city{} \state{} \country{}
}
\email{pfsorenson52@meta.com}

\author{Sophie Kahyun Kim}
\affiliation{%
  \institution{}
  \city{} \state{} \country{Meta Reality Labs}
}
\email{sophiekkim@meta.com}

\author{Hrvoje Benko}
\affiliation{%
  \institution{Meta Reality Labs}
  \city{} \state{} \country{}
}
\email{benko@meta.com}

% Mengjie Yu
% Tanya R. Jonker
% Kashyap Todi
% Feiyu Lu
% Xun Qian
% João Marcelo Evangelista Belo
% Tianyi Wang
% Michelle Li
% Aran Mun
% Te-Yen Wu
% Junxiao Shen
% Ting Zhang
% Narine Kokhlikyan
% Fulton Wang
% Paul Sorenson
% Sophie Kahyun Kim
% Hrvoje Benko

\renewcommand{\shortauthors}{Xu et al.}
% \renewcommand{\shorttitle}{XAIR}

\begin{abstract}

Explainable AI (XAI) has established itself as an important component of AI-driven interactive systems.
With Augmented Reality (AR) becoming more integrated in daily lives, the role of XAI also becomes essential in AR because end-users will frequently interact with intelligent services. 
However, it is unclear how to design effective XAI experiences for AR. 
We propose XAIR, a design framework that addresses \textit{when}, \textit{what}, and \textit{how} to provide explanations of AI output in AR.
The framework was based on a multi-disciplinary literature review of XAI and HCI research, a large-scale survey probing 500+ end-users’ preferences for AR-based explanations, and three workshops with 12 experts collecting their insights about XAI design in AR. 
XAIR's utility and effectiveness was verified via a study with 10 designers and another study with 12 end-users.
XAIR can provide guidelines for designers, inspiring them to identify new design opportunities and achieve effective XAI designs in AR.
\end{abstract}


% \ccsdesc[500]{Human-centered computing~Empirical studies in HCI}
% \ccsdesc[300]{Human-centered computing~Interaction techniques}


% \input{tex_fig_tab_alg/fig_overview}

\maketitle

\vspace{-0.1cm}
\section{Introduction}
\label{sec:introduction}

Breakthroughs in Artificial Intelligence (AI) and Machine Learning (ML) have considerably advanced the degree to which interactive systems can augment our lives~\cite{riedl2019human,li2020artificial}.
As black-box ML models are increasingly being employed, concerns about humans misusing AI and losing control have led to the need to make AI and ML algorithms easier for users to understand~\cite{mohseni_multidisciplinary_2021,barredo_arrieta_explainable_2020}.
This, in turn, has spurred rapidly growing interest into \textit{Explainable AI (XAI)} within academia~\cite{liao_questioning_2020, abdul_trends_2018, ehsan2021explainable} and industry~\cite{meta_ai, google_cloud_model_cards,arya2019one}, and by regulatory entities~\cite{gdpr_2019,gunning2017explainable,gunning2019darpa}.
\review{Earlier XAI research aims to help AI/ML developers on model debugging (\eg~\cite{Ribeiro2016,liu2017analyzing,zeiler2014visualizing,ribeiro2018anchors,kaur_interpreting_2020}) or assist domain experts such as clinicians by revealing more information such as causality and certainty (\eg~\cite{liao_personalized_2020,wang_designing_2019,eiband_bringing_2018,xie2020chexplain}).
Recently, there has been a growing amount of XAI research focusing on the non-expert end-users~\cite{jiang_who_2022,ehsan2021explainable,bhatt2020explainable}.}
Existing studies have found that XAI can help end-users resolve confusion and build trust~\cite{pu2006trust,dhanorkar_who_2021}.
Industrial practitioners have started to integrate XAI into everyday scenarios and improve user experiences, \eg by displaying the match rate of point-of-interest suggestions on map applications~\cite{google_map_match_rate_2018}.
% There is less XAI research on non-expert end-users, who represent a dominant body of audience potentially interested in XAI for intelligent interactive systems and have significantly different needs and goals~\cite{jiang_who_2022,ehsan2021explainable}.
% For instance, end-users may lose faith in the system and abandon products when they don't understand AI outcomes.
% XAI can help end-users resolve concerns and build trust~\cite{pu2006trust} when they encounter unexpected results~\cite{dhanorkar_who_2021} or seek additional information~\cite{burrell2016machine}.
% Industrial practitioners started to explore simple approaches to bring XAI to everyday living and improve user experience, \eg displaying the match rate of point-of-interest suggestions on map applications~\cite{google_map_match_rate_2018}, showing reasons for product recommendations on shopping websites~\cite{zhang_explainable_2020}. However, the study of XAI design for end-users is still at an early stage nowadays.

Alongside the surge of interest into XAI, \textit{Augmented Reality (AR)} is another technology making its way into everyday living~\cite{microsoft_hololens,google_glass}.
\review{Advances in more lightweight, powerful, and battery-efficient Head-Mounted Displays (HMDs) have brought us closer to the vision of pervasive AR~\cite{grubert2016towards}.}
As AI techniques are needed to enable context-aware, intelligent, everyday AR~\cite{cipresso2018past,rese2017augmented,abrash2021creating}, XAI will be essential because end-users will interact with outcomes of AI systems.
XAI could be used to make intelligent AR behavior interpretable, resolve confusion or surprise when encountering unexpected AI outcomes, promote privacy awareness, and build trust.
Therefore, we aim to answer the following research question: \textbf{How do we create effective XAI experiences for AR in everyday scenarios?}

% To answer our research question, a framework is needed as guidance for the design.
Researchers have developed several design spaces and frameworks to guide the design of XAI outside the context of AR~\cite{dhanorkar_who_2021,wang_designing_2019,mohseni_multidisciplinary_2021,barredo_arrieta_explainable_2020}.
% For example, Lim and Dey investigated end-users' preferences about different types of explanations and provided an XAI framework for intelligible context-aware systems~\cite{lim_assessing_2009}.
However, most previous work focused on identifying a taxonomy of explanation types or generation techniques. They did not consider everyday AR-specific factors such as
% Moreover, the majority of these frameworks are not targeted at non-expert end-users.
the rich sensory information that AR technologies have about users and contexts, and its always-on, adaptive nature.
These factors can not only support more personalized explanations, but also affect the design of an explanation interface. For example, one could render in-place explanations on related objects (\eg explaining a recipe recommendation by highlighting ingredients in the fridge).
In this paper, we provide a framework to guide the design of XAI in AR.
% To our knowledge, there is no previous work that aims to address the design of XAI in AR in a systematic way.

\begin{figure*}[t]
    \vspace{-0.3cm}
    \centering
    \includegraphics[width=1\textwidth]{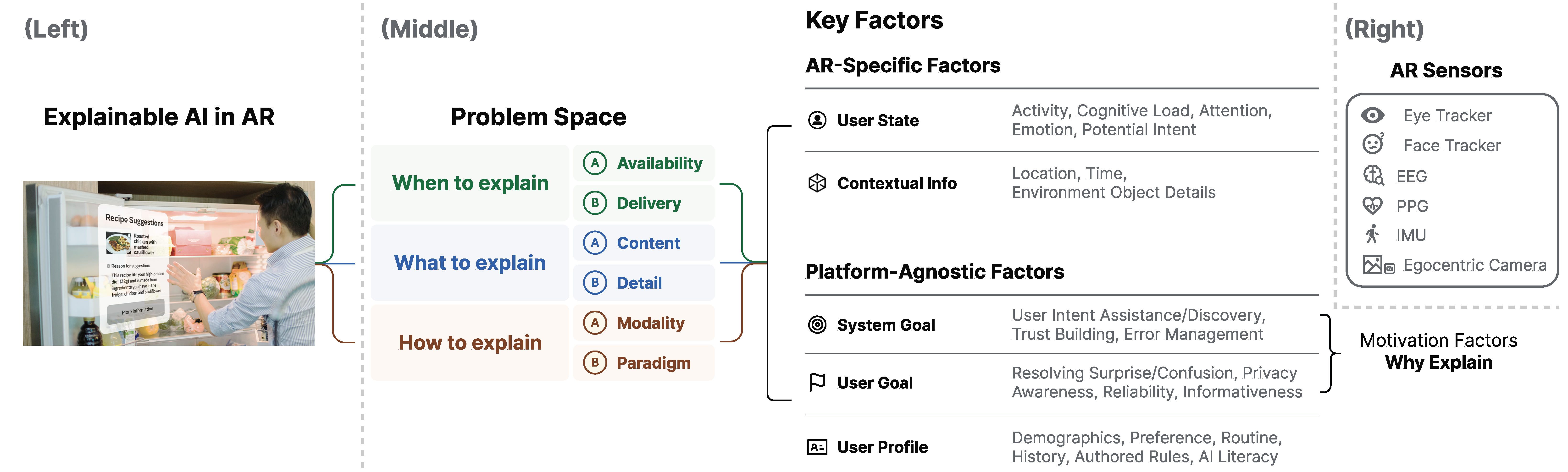}
    \vspace{-0.4cm}
    \caption{An Overview of XAIR Framework. (Left) An example of the AR interface with explanations. (Middle) The main structure of XAIR: the problem space and the key factors. (Right) Sensors that are integrated into AR.}
    \Description[An overview of the XAIR framework]{An overview of the XAIR framework.
Left) A man opening a fridge and reaching out to grab something. On his left, there is an AR interface showing a recommendation of a recipe (named "Roasted Chicken with mushed cauliflower", plus a figure of the recipe) and a paragraph of explanation "Reason for suggestion: This recipe fits your high-protein diet (32g) and is made from ingredients you have in the fridge: chicken and cauliflower".
Middle) The Problem Space of XAIR, showing three sections with detailed dimensions: when to explain (availability and delivery), what to explain (content and detail), and how to explain (modality and paradigm). It also contains a list of key factors, including user status, contextual info, system goal, user goal, and user profile.
Right) A list of AR sensors, including eye tracker, face tracker, EEG, PPG, IMU, and egocentric camera.}
    \vspace{-0.3cm}
    \label{fig:xair}
\end{figure*}

To answer the aforementioned research question, a design space analysis~\cite{qoc_1999} was used to break down the main research question into three sub-questions: 1) \textit{When to explain?}, 2) \textit{What to explain?}, and 3) \textit{How to explain?}
Previous research from the XAI and HCI communities has focused on one or two of these sub-questions (\eg when~\cite{pielot2017beyond,robbins2019misdirected}, what~\cite{liao_questioning_2020,barredo_arrieta_explainable_2020}). Although not within the context of AR, many of these findings can inform the design of XAI in AR.
\review{
Therefore, we first summarized related literature to identify the most important dimensions under each sub-question, as well as the factors that determine the answers to these questions, such as users' goals for having explanations (\ie why explain).
}
Then, we conducted two complementary studies to obtain insights from the perspectives of end-users and experts.
Specifically, we carried out a large-scale survey including over 500 end-users with different levels of knowledge of AI to collect user preferences about the timing (related to \textit{When}), content (related to \textit{What}), and modality (related to \textit{How}) of explanations in multiple AR scenarios.
In addition, we ran three workshops with twelve experts (\ie four experts per workshop) from different backgrounds, including algorithm developers, designers, UX professionals, and HCI researchers to iterate on the dimensions and generate guidelines to answer the \textit{When/What/How} questions.

Merging the insights obtained from these two studies, we developed the \textbf{XAIR} (e\underline{\textbf{X}}plainable \underline{\textbf{AI}} for Augmented \underline{\textbf{R}}eality) framework (Fig.~\ref{fig:xair}).
The framework can serve as a comprehensive reference that connects multiple disciplines across XAI and HCI. It also provides a set of guidelines to assist in the development of XAI designs in AR.
XAI researchers and designers can use the guidelines to enhance their design intuition and propose more effective and rigorous XAI designs for AR scenarios.

% We provide a few design examples to illustrate how to apply XAIR to various real-life AR scenarios.
We further conducted two user studies to evaluate XAIR.
To verify its utility to support designers, the first study focused on designers' perspectives. Ten designers were invited to use XAIR and design XAI experiences for two real-life AR scenarios.
To demonstrate its effectiveness in guiding the design of an actual AR system, a second study was conducted from the perspective of end-users. We implemented a real-time intelligent AR system based on the designers' proposals in the previous study using XAIR. The study measured the usability of the AR system with 12 end-users.
\review{
The results indicated that XAIR could provide meaningful and insightful support for designers to propose effective designs for XAI in AR, and that XAIR could lead to an easy-to-use AR system that was transparent and trustworthy.
}
% We end our paper by providing a series of additional design examples of various real-life AR scenarios using XAIR, followed by a series of discussions of our framework.

The contributions of this research are:
\begin{s_itemize}
\item \review{We summarized literature from multiple domains and identified the important dimensions for the when/what/how questions in the problem space when designing XAI in AR.}
\item Drawing the results from a large-scale survey with over 500 users and an iterative workshop study with 12 experts, we developed XAIR, the first framework for XAI design in AR scenarios. We also proposed a set of guidelines to support designers in their design thinking process.
% We applied XAIR to a series of everyday AR scenarios to illustrate its capability and practicality.
\item The results of design workshops with 10 designers indicated that XAIR could provide meaningful and insightful creativity support for designers. \review{The study with 12 end-users who used a real-time AR system showed that XAIR led to the design of AR systems that were transparent and trustworthy.}
\end{s_itemize}

\section{Background}
\label{sec:background}

In this section, we first introduce more background about XAI (Sec.~\ref{sub:background:xai}).
We then summarize existing XAI design frameworks and demonstrate the need for a new XAI framework that is specifically applicable to AR scenarios (Sec.~\ref{sub:background:xai_in_ar}).

\subsection{What is XAI?}
\label{sub:background:xai}
The notion of XAI can be traced back more than four decades~\cite{xu2019explainable}, where expert systems would explain output via a set of decision rules~\cite{scott1977explanation,swartout1985explaining}.
This concept has been brought back into focus by the success of black-box AI/ML models~\cite{confalonieri2021historical}.
The working definition of XAI used in this paper is: 
\textit{``given an audience, an explainable AI is one that produces details or reasons to make its functioning clear or easy to understand''}~\cite{barredo_arrieta_explainable_2020}.

With the increasing prevalence of advanced black-box models that make more critical predictions and decisions, the interpretability and transparency of AI systems has attracted increasing attention from various academic, industrial, and regulatory stakeholders~\cite{gregor1999explanations,preece2018stakeholders,gdpr_2019,gunning2019darpa}.
% For instance, the European Union General Data Protection Regulation (GDPR) commission established the legal right to obtain explanations~\cite{gdpr_2019}, and the Defense Advanced Research Projects Agency (DARPA) also formulated the XAI program to enable effective understanding and management of AI systems~\cite{gunning2019darpa}.
Addressing the broad vision of making AI more understandable for humans involves multidisciplinary research efforts.
ML researchers have developed algorithms that results in transparent models (\eg decision trees, Bayesian models~\cite{letham2015interpretable,caruana2015intelligible}) or used post-hoc explanation techniques (\eg feature importance, visual explanation, ~\cite{Lundberg2017,selvaraju2017grad,Shrikumar2017}) to generate explanations for users.
HCI researchers, on the other hand, have focused on improving user trust~\cite{pu2006trust,holliday2016user} and understanding~\cite{lim_assessing_2009,lim_why_2009} of machine generated explanations.
Psychology researchers have approached XAI from a more fundamental perspective and studied how people generate, communicate, and understand explanations~\cite{taylor2021artificial,yarkoni2017choosing}.
% We refer readers to several survey papers for more details: ML~\cite{barredo_arrieta_explainable_2020,holzinger_explainable_2018,zhang_explainable_2020,adadi_peeking_2018}, HCI~\cite{abdul_trends_2018,wang_designing_2019,amershi2014power,bellotti_intelligibility_2001}, and psychology~\cite{hoffman2018metrics,miller2019explanation,lepri2018fair}.

% , and can offer various benefits to different target audiences by providing transparency and interpretability.
% Depending on the target audience, XAI can have various benefits by providing more transparency and interpretability.
By providing more transparency and interpretability, XAI can offer different target audiences different benefits.
For instance, for algorithm developers and data scientists, XAI can provide more details for model debugging and improvement~\cite{lipton2018mythos} and increase production efficiency and robustness~\cite{samek2017explainable,yuan2019adversarial}.
For domain experts, XAI can reveal insights about causality~\cite{louizos2017causal}, transferability~\cite{tickle1998truth,chander2018working}, confidence~\cite{belle2017logic,pope2019explainability}, and also enhance the reliability of model output~\cite{dovsilovic2018explainable,Ribeiro2016,xu_leveraging_2019,xu_leveraging_2021}.
\review{
Early XAI research only focused on these two groups of users. Recently, there have been an 
% , \ie developers~\cite{Lundberg2017,Shrikumar2017,adadi_peeking_2018} and domain experts~\cite{liao_personalized_2020,wang_designing_2019,eiband_bringing_2018}.
increasing number of XAI studies that have focused on non-expert end-users who represent a large potential audience of XAI~\cite{jiang_who_2022,ehsan2021explainable}.
}
XAI has been found to improve reliance and build trust with non-experts~\cite{pu2006trust}, especially when users encounter unexpected AI outcomes~\cite{dhanorkar_who_2021}, have privacy concerns~\cite{edwards2017slave}, or seek additional information~\cite{doshi2017towards,burrell2016machine}.
Some companies have integrated XAI into products used by the general population~\cite{google_map_match_rate_2018,zhang_explainable_2020}, \eg visualizing the match rate of restaurant suggestions in a map application~\cite{google_map_match_rate_2018} or showing reasons for product recommendations on a  shopping website~\cite{zhang_explainable_2020}. However, these efforts are still at an early stage.

\subsection{Why do we need XAI in Everyday AR?}
\label{sub:background:xai_in_ar}
Since the first AR HMD was built in 1968~\cite{sutherland_head-mounted_1968}, researchers and engineers have been striving to integrate AR HMDs into everyday living.
Recent examples include simple head-mounted cameras and displays (\eg Google Glass Enterprise~\cite{google_glass} and Snap Spectacles~\cite{snap_spectacle}), as well as more advanced HMDs with 3D-depth sensing (\eg Microsoft Hololens~\cite{microsoft_hololens} and Magic Leap~\cite{magic_leap}).
% There is a growing number of examples of commercializing AR HMD for specific applications, such as assisting doctors with surgeries~\cite{vavra2017recent}, providing instructions for manufacturing workers~\cite{nee2012augmented}, or enabling convenient life logging for the general population~\cite{snap_spectacle}.
As hardware improves, it is foreseeable that AR will become an integral aspect of everyday living for general consumers and support a wide range of applications in the near future~\cite{cipresso2018past,rese2017augmented}.

\subsubsection{The Importance of AI and XAI in AR}
\label{subsub:background:xai_in_ar:importance}
The role of AI will be critical for AR devices if they are to provide intelligent services.
% For example, with sensors to track various signals (\eg motion, gaze, EEG), existing research has shown that AI can power AR devices to understand of user status in real-time, such as cognitive load~\cite{hess1964pupil,duchowski_index_2018} and attention~\cite{huang2018predicting,chong2018connecting,stappen2020x}.
% With a front-camera, existing research has shown the potential to infer users' daily activities~\cite{fathi2011understanding,singh2016first,schroder2017deep}, their context object details~\cite{redmon2016you,liu2020deep}, and semantics of the scene~\cite{grauman2022ego4d,pan2019content,bolanos2016toward,miech2020end}, \etc\
% This low-level intelligence would enable AR to provide an extensive variety of high-level smart services.
The integration of sensors enables AR systems to understand users' current states~\cite{huang2018predicting,stappen2020x,schroder2017deep} and their environment~\cite{liu2020deep,miech2020end} to provide a variety of intelligent functionalities.
For example, AR could infer user intent~\cite{admoni2016predicting} and provide contextual recommendations for daily activities (\eg recipe recommendations when a user opens the fridge during lunch)~\cite{adomavicius2011context,lam_a2w_2021,lee_all_2021}.
The rich interaction between the outcomes of AI and end-users requires effectively designed XAI that can support users in a variety of contexts, such as when users are confused or surprised while encountering an unexpected AI outcome, or when they want to make sure that an AI outcome is reliable and trustworthy~\cite{mohseni_multidisciplinary_2021,Amershi2019}.
% Like other AI systems outside AR, providing explanations of AI outcomes when needed can increase system transparency, resolve users' concerns, and build more trust~\cite{pu2006trust,holliday2016user}, thus improving user experience in AR.
Recent work has started to explore the application of XAI in AR~\cite{ahmed2022artificial}.
For instance, Wintersberger \etal found that showing traffic-related information in AR while driving can provide much needed explanation to users and improve user trust~\cite{wintersberger2018fostering}.
Danry \etal explored the use of an explainable AI assistant integrated within wearable glasses to enhance human rationality~\cite{danry2020wearable}.
Zimmermann \etal found that introducing XAI during an AR-based shopping process could improve user experiences~\cite{zimmermann2022enhancing}.
However, these studies proposed their own case-by-case XAI designs. In this research, we aggregated the major factors identified in the literature and studied the when/what/how questions systematically.

\subsubsection{The Need for A New XAI Framework for AR}
\label{subsub:background:xai_in_ar:need}
Researchers have proposed several XAI design spaces and frameworks for AI systems, \eg knowledge-based systems~\cite{graesser1992mechanisms}, decision support systems~\cite{amini2022discovering}, and recommendation systems~\cite{herlocker2000explaining}.
For instance, Wang~\etal proposed a conceptual framework for building user-centric XAI systems and put it into practice by implementing an explainable clinical diagnostic tool~\cite{wang_designing_2019}.
Eiband~\etal presented a stage-based participatory design process for designers to integrate transparency into systems~\cite{eiband_bringing_2018}. They evaluated the process using a commercial AI fitness coach.
Zhu~\etal proposed a co-creation design space between game designers using ML techniques and investigated the usability of XAI algorithms to support game designers~\cite{zhu_explainable_2018}.
Liao~\etal developed an algorithm-informed XAI question bank to support design practices for AI systems~\cite{liao_questioning_2020}.
\review{
Ehsan~\etal investigated how social transparency in AI systems supported sellers from technology companies and developed a conceptual framework to address what, who, why, when questions~\cite{ehsan2021expanding}.
Wolf proposed the concept of scenario-based XAI design and highlighted researchers' need to understand AI systems in specific scenarios such as when researchers are not uncertain or they want to explain data limitations~\cite{wolf_explainability_2019}.
These existing frameworks aim to guide XAI design for developers or domain experts for specific applications.
% Most of these existing frameworks aim to guide XAI design for developers or domain experts for specific applications, rather than non-expert end-users for everyday scenarios.
Focusing on non-expert end-users, Lim and Dey systematically investigated end-users' opinions and preferences about different types of explanations in multiple context-aware applications, and provided an XAI framework for intelligible context-aware systems~\cite{lim_assessing_2009}.
Moreover, recent industry practitioners have also made efforts towards a designing framework for end-user-facing explanations~\cite{ttc_labs}.
}

\begin{figure}[!t]
    \vspace{-0.2cm}
    \centering
    \includegraphics[width=1\columnwidth]{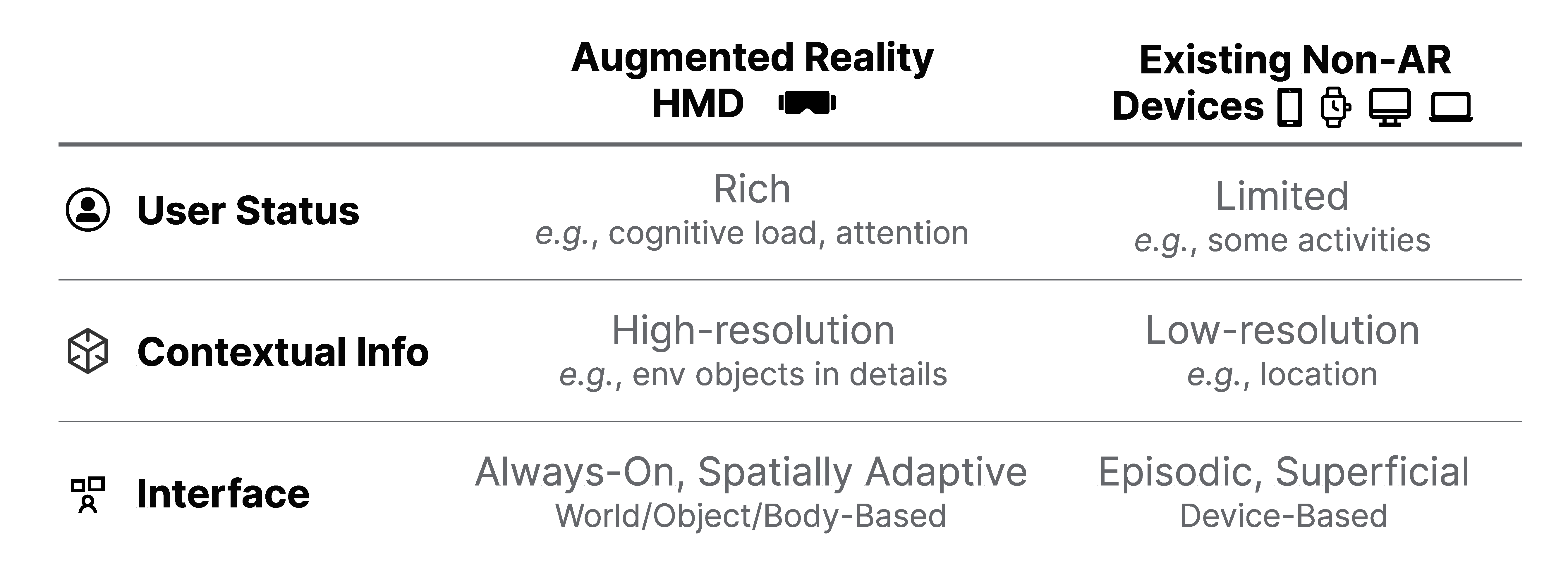}
    \caption{The Uniqueness of AR that Distinguishes XAI Design from Other Platforms.}
    \label{fig:uniqueness_ar}
    \Description[A table showing the difference between augmented reality devices and other non-AR devices.]{A table showing the difference between augmented reality devices (column 1) and other non-AR devices (column 2). There are three rows.
The first row is about user context. AR column says "Rich, e.g., congitive load, attention". Non-AR column says "Limited, e.g., some activities".
The second row is about contextual info. AR column says "High-resolution, e.g., env objects in details". Non-AR column says "Low-resolution, e.g., location".
The third row is about interface. AR column says "Always-On, Spatially Adaptive, World/Object/Body-Based". Non-AR column says "Episodic, Superficial, Device-Based".}
    \vspace{-0.2cm}
\end{figure}

Such XAI frameworks focused on the content design of XAI, which is mostly visualized on laptops or mobile phones, thus making them insufficient for the myriad of AR contexts.
There are several factors that distinguish AR from other platforms and necessitate the need for a new XAI design framework (see Fig.~\ref{fig:uniqueness_ar}).
First, AR has a much deeper real-time understanding of a user's current state via the sensors within an HMD~\cite{bonanni2005attention,samadiani2019review}.
Second, compared to other platforms, AR systems can develop a more fine-grained understanding of a user's context~\cite{liu2020deep,grauman2022ego4d,damen2022rescaling}.
This richer information not only provides new types of information that can be integrated into AR-based XAI explanations, but also influences the design of XAI as explanations need to be tailored to a user's state and context.
Third, from an interface perspective, the ability to be always-on and 3D-aware enables AR to present information at any time~\cite{zhu_bishare_2020,lu_glanceable_2021,azuma_survey_1997}, and spatially adapt explanations to the physical world~\cite{feiner1993windows,reitmayr2001mobile}.
% , such as displaying an interface on the wall (world-based) or over a certain object (object-based).
These factors influence the design of XAI in AR, as they need to be presented to users in an appropriate, efficient, and effective way.
Overall, these unique factors demonstrate how current frameworks are insufficient and there is a need for a new XAI framework specifically designed for AR scenarios.
% To the best of our knowledge, there is no prior work investigating this topic.

\section{XAIR Problem Space and Key Factors}
\label{sec:problem_space_factors}

Determining the way to create effective XAI experiences in AI is a complex challenge. Thus, it is important to first identify the problem space to bound the scope of our investigation.
We first summarize over 100 papers from the ML and HCI literature to identify the problem space and the main dimensions within each problem (Sec.~\ref{sub:problem_space_factors:problem_space}).
Then, we outline the key factors that determine the answers to the problems (Sec.~\ref{sub:problem_space_factors:factors}).

The problem space and key factors define the structure of XAIR (Fig.~\ref{fig:xair} middle).
In Sec.~\ref{sec:methodology}, we present two studies conducted to obtain insights from end-users and expert stakeholders about how to design XAI in AR.
Then, combining the structure and insights, we show how these factors are connected with the problem space, and provide design guidelines in Sec.~\ref{sec:framework}.

% We first identify the problem space of our research question about creating effective XAI experience in AR (Sec.~\ref{sub:problem_space_factors:problem_space}).
% We then spotlight the key factors that determine the answer to the research question (Sec.~\ref{sub:problem_space_factors:factors}).
% Fig.~\ref{fig:overview} presents a detailed overview of the problem space and the factors.
% After introducing the overall picture, the rest of the paper follows Fig.~\ref{fig:pipeline} to develop and evaluate XAIR.

% \input{tex_fig_tab_alg/fig_overview_details}

\subsection{Problem Space}
\label{sub:problem_space_factors:problem_space}

Following the design space analysis method~\cite{qoc_1999}, the research question was divided into three sub-questions: when to explain, what to explain, and how to explain~\cite{elliott2017living,nahum-shani_just--time_2018}.

\subsubsection{\colorwhen{When to Explain?}}
\label{subsub:problem_space_factors:problem_space:when}
The literature review revealed two aspects of ``when'' that were important to consider: the \textit{availability} of explanations (\ie whether to prepare explanations?), and the timing of the explanation's \textit{delivery} (\ie when to show explanations?).

\colorwhen{\textbf{\textit{Availability}}}.
% It is important to determine whether AI models should generate any explanation available for users to access.
Previous research has found that to maintain a positive user experience, supporting user agency and control is important during human-AI interaction~\cite{cai_impacts_2022,lee1992trust}.
Having explanations that are available and accessible is in line with the goal of supporting user agency.

\colorwhen{\textbf{\textit{Delivery}}}.
With the ability to show information at any time, AR can employ various timing strategies to present explanations. Thus, it is important to find the appropriate method to deliver explanations to users. Generally, there are two approaches, manual-trigger (\ie initiated by users) and auto-trigger (\ie initiated by the system)~\cite{cimolino_two_2022,yeh2022guide}.
On the one hand, researchers have found that explanations should not always be presented to users, because they can introduce unnecessary cognitive load and become overwhelming for non-expert end-users~\cite{chazette2019end,wagner2020regulating,bunt2012explanations,robbins2019misdirected,stumpf2016explanations}. This is especially important in AR, as users' cognitive capacity tends to be limited~\cite{buchner2022impact}.
Moreover, adopting manual triggers would enable users to choose to see explanations as needed, thus enabling them to exercise agency over their experience~\cite{roy_automation_2019,lu_exploring_2022}.
On the other hand, existing findings on just-in-time intelligent systems (\eg just-in-time recommendations~\cite{kapoor2015just,ma2020temporal} and just-in-time interventions~\cite{nahum-shani_just--time_2018,Sarker2014, xu_typeout_2022}) have suggested that automatically delivering explanations at the right time based on user intent and need (as detected via AR sensing that identifies a user's state and context) can provide a better user experience~\cite{bhattacharya2017intent,mehrotra2019jointly}.

% In Sec.~\ref{sec:methodology}, we will draw insights obtained from both end-users and expert stakeholders and propose our framework and guidelines to answer the two parts in Sec.~\ref{sec:framework}.

\subsubsection{\colorwhat{What to Explain?}}
\label{subsub:problem_space_factors:problem_space:what}
The literature review also found two important aspects of ``what'' to consider: First, the \textit{content} of the explanations (\ie what type of content to include?). Second, the level of \textit{detail} of the explanations (\ie how much detail should be explained?).

\colorwhat{\textbf{\textit{Content}}}.
Previous literature in XAI has identified several explanation content types~\cite{barredo_arrieta_explainable_2020,mohseni_multidisciplinary_2021}. 
% On include input/output, conceptual model (why/why not, how, what else, what if) and non-functional types (certainty, how-to).
\review{
The seven types are:
\begin{s_enumerate}
\item Input/Output. This type explains the details of input (\eg data sources, coverage, capabilities) or output (\eg additional details, options that the system could produce) of a model~\cite{lim_assessing_2009,lim_toolkit_2010}.
\item Why/Why-Not. This type explains the features in the input data~\cite{Ribeiro2016} or the model logic~\cite{ribeiro_anchors_2018} that have led or not led to a given AI outcome~\cite{myers2006answering} (also known as contrastive explanations). Showing feature importance is another commonly used technique to generate these explanations~\cite{schlegel2019towards,casalicchio2018visualizing}.
\item How. This type provides a holistic view to explain the overall logic of an algorithm or a model and illustrate how the AI model works. Typical techniques include model graphs~\cite{lakkaraju2016interpretable}, decision boundaries~\cite{lombrozo2009explanation}, or natural language explanations~\cite{berkovsky2017recommend}. 
\item Certainty. This type describes the confidence level of the model with its input (\eg for models whose input is not deterministic, explain how accurate the input of the model is) or output (\eg explain how accurate, or reliable the AI outcomes are)~\cite{google_map_match_rate_2018,schoonderwoerd2021human}. Scores based on softmax~\cite{bridle1989training} or calibration~\cite{platt1998sequential} are commonly used as the confidence/certainty score for ML models.
\item Example. This type presents similar input-output pairs from a model, \eg similar input that lead to the same output or similar output examples given the same input~\cite{cai2019human,keane2019case}. This is also known as the What-Else explanation. Example methods include influence functions~\cite{koh_understanding_2017} and Bayesian case modelling~\cite{kim2014bayesian}.
\item What-If. This type demonstrates how changing input or applying new input can affect model output~\cite{cai2019effects,lim_why_2009}.
\item How-To. In contrast to What-If, this type explains how to change input to achieve a target output~\cite{wang_designing_2019,liao_questioning_2020}, \eg how to change the output from X to Y? Common methods for What-If/How-To content include rule generation~\cite{guidotti2018local}, feature description~\cite{wachter2017counterfactual}, and input perturbation~\cite{zhang2018interpreting}.
\end{s_enumerate}
}
% Previous studies by Lim \etal offered suggestions about subsets of content types for context-aware systems~\cite{lim_toolkit_2010,lim_assessing_2009,lim2019these}.
Moreover, another aspect that is independent of the explanation content type is global \vs local explanations (explaining the general decision-making process \vs a single instance)~\cite{molnar2020interpretable}. In general, non-expert end-users were found to prefer local explanations ~\cite{lakkaraju2019faithful,dhanorkar_who_2021}.

\colorwhat{\textbf{\textit{Detail}}}.
Displaying every relevant explanation content type to an end-user can be overwhelming, especially with the limited cognitive capacity they have in AR~\cite{buchner2022impact,baumeister2017cognitive}. Explanations that extend a user's prior knowledge or fulfill their immediate needs should be prioritized~\cite{coppers2018intellingo}.
Moreover, previous research has suggested that presenting detailed and personalized explanations is useful for better understanding AI outcomes ~\cite{schneider2019personalized,kouki2019personalized,esteva2017dermatologist,jahanbakhsh_effects_2020,xu_understanding_2020}.

% We will verify whether these previous findings are transferable to AR scenarios in Sec.~\ref{sec:methodology}.
\review{
Our focus on \textit{content} and \textit{detail} is about choosing appropriate explanation content types and proper levels of detail, but not on picking which techniques to generate explanations.
From a technical perspective, there are interpretable models (\ie the model being transparent, such as linear regression or decision trees) and ad-hoc explainers (\ie generating explanations for complex, black-box models)~\cite{Lundberg2017, Ribeiro2016}. The latter can further be divided into model-specific and model-agnostic explanation methods~\cite{barredo_arrieta_explainable_2020}.
We refer readers to other surveys and toolkits for developing or selecting explanation generation algorithms \cite{arya2019one,liao2021human,dalex,h2oai,adadi_peeking_2018}.
}

\subsubsection{\colorhow{How to Explain?}}
\label{subsub:problem_space_factors:problem_space:how}
The last sub-question \colorhow{\textit{how}} focuses on the visual representation of the content in AR.
% Although Eiband \etal touched on the question of \textit{how to explain} in their participatory XAI design process outside of AR, they only involve a general ``iterative prototyping'' step~\cite{eiband_bringing_2018}.
% We summarize important dimensions that are needed to develop the framework and guidelines.
% The two important dimensions that are needed to develop the framework and guidelines are summarized below.
Two dimensions emerged from the literature review, \ie modality and paradigm.

\colorhow{\textbf{\textit{Modality}}}.
The multi-modal nature of AR enables it to support AI outcomes via various modalities (\eg visual, audio, or haptic)~\cite{chen2017multimodal,nizam2018review}.
Explanations are hard to convey using modalities with limited bandwidth (\eg haptic, olfactory, or even gustatory). Therefore, visual and audio are the two major modalities that should be employed for explanations.

\colorhow{\textbf{\textit{Paradigm}}}.
If explanations are presented using audio, the design space is relatively limited (\eg volume, speed). We refer readers to existing literature on audio design (\eg \cite{frauenberger2007survey, kern2009design}).
The design space of the visual paradigm for explanations, however, is much larger.
First, from a formatting perspective, explanation content can be presented in a textual format (\eg narrative, dialogue)~\cite{lakkaraju2016interpretable,myers2006answering}, graphical format (\eg icons, images, heatmaps)~\cite{zeiler2014visualizing,simonyan2013deep}, or a combination of both.
Second, from a pattern perspective, an explanation can be displayed either in an implicit way (\ie embedded in the environment, such as a boundary highlight of an object) or explicit way (\ie distinguished from the environment, such as a pop-up dialogue window)~\cite{lindlbauer_context-aware_2019,tatzgern_adaptive_2016,diverdi2004level}.
The pattern is closely related to the adaptiveness of the AR interface~\cite{dai2017scannet,wang_designing_2019}. With 3D sensing capabilities, the location of an explanation can be body-based (mostly explicit), object-based (implicit or explicit), or world-based (implicit or explicit) ~\cite{lu_exploring_2022,bonanni2005attention,laviola20173d,xu_hand_2018}. Prior AR research has explored adaptive interface locations~\cite{luo2022should,muller2016taxonomy}, \eg interfaces should be adaptive based on the semantic understanding of the ongoing interaction~\cite{cheng_semanticadapt_2021,qian_scalar_2022,rzayev2020effects} and ergonomic metrics~\cite{evangelista2021xrgonomics}.

% To answer \colorhow{\textit{how}} to explain, we will leverage end-users' preference and experts' knowledge collected from two user studies (see Sec.~\ref{sec:methodology}), and propose our framework to guide the choice of these options.

\subsection{Key Factors}
\label{sub:problem_space_factors:factors}
These three questions, and their dimensions, form the overall problem space of XAIR.
Another important aspect of XAIR is the factors that determine the answers to these questions.
We summarize these factors from two perspectives, one specific to AR platforms (Sec.~\ref{subsub:problem_space_factors:factors:ar_specific}), and the other agnostic to any platform (Sec.~\ref{subsub:problem_space_factors:factors:non_ar_specific}).
% The middle of Fig.~\ref{fig:overview} summarizes these factors.

\subsubsection{AR-Specific Factors}
\label{subsub:problem_space_factors:factors:ar_specific}
Fig.~\ref{fig:uniqueness_ar} summarizes the three main features that distinguish AR from other platforms: \textit{User State}, \textit{Contextual Information}, and \textit{Interface}.
As \textit{Interface} is an integral property of an AR platform, it remains invariant to external changes.
In contrast, the other two aspects are dynamic and would alter the design of XAI in AR.

\textbf{\textit{User State.}}
The sensors that could be integrated within future HMDs would empower an AR system to have a rich, instant understanding of user's state, such as activities (IMU~\cite{gjoreski2021head,windau2016walking}, camera~\cite{fathi2011understanding,singh2016first,schroder2017deep,liang_authtrack_2021}, microphone~\cite{xu_listen2cough_2021,xu_earbuddy_2020,wang_hearcough_2022,jin_earcommand_2022}), cognitive load (eye tracking~\cite{duchowski_index_2018,zagermann2018studying,joseph2020potential}, EEG~\cite{antonenko2010using,xu2018review}), attention (eye tracking~\cite{huang2018predicting,chong2018connecting,stappen2020x,xu_recognizing_2020}, IMU~\cite{leelasawassuk2015estimating}, EEG~\cite{vortmann2019eeg}), emotion (facial tracking~\cite{yong2019emotion, yan2022emoglass}, EEG~\cite{wan2021wearable,soleymani2015analysis}) and potential intent (the fusion of multiple sensors and low-level intelligence~\cite{tsai2018augmented,admoni2016predicting,kim2016understanding}).
Depending on a user's state, the design of explanations could be different. For example, as identified in previous research on ambient interfaces~\cite{pielot2017beyond,fogarty2005predicting}, when users engage in activities with a high cognitive load, explanations should not show up automatically to interrupt them
(related to \colorwhen{\textit{when}}).

\textbf{\textit{Contextual Information.}}
Compared to devices such as smartphones, AR HMDs have more context awareness. Other than having an awareness of location and time~\cite{dey_understanding_2001}, an egocentric camera and LiDAR, combined with other sensors (\eg Bluetooth, WiFi, RFID), can identify details about digital and non-digital objects in the environment~\cite{redmon2016you,liu2020deep,park2018high}, and have a better understanding of the semantics of a scene~\cite{grauman2022ego4d,pan2019content,bolanos2016toward,miech2020end}.
Such contextual information would also influence the design of XAI.
For instance, an explanation visualization about recipe recommendations that appears when users open the fridge may look differently from explanations about podcast suggestions that are shown while driving (related to \colorhow{\textit{how}}).

\subsubsection{Platform-Agnostic Factors}
\label{subsub:problem_space_factors:factors:non_ar_specific}
There are also other factors that are platform agnostic such as the motivation to present explanations (\ie \textit{why explain?}). We view this factor from two perspectives, one from the system side (\ie what are the \textit{system's goals} when presenting explanations?), and the other from the non-expert end-user side (\ie what are \textit{users' goals} when they want to see explanations?)~\cite{samek2019towards}.
The \textit{user profile} (\ie individual details) is another important factor related to personalized explanations~\cite{schneider2019personalized,kouki2019personalized}.

\textbf{\textit{System Goal.}}
Based on prior literature, we summarize four system goals that are desired when an AR system provides explanations for AI outcomes:
\begin{s_enumerate}
\item User Intent Discovery. When an AI model generates suggestions for a new topic, the system seeks to help users discover new intent~\cite{samek2019towards,gotz2009behavior,pu2006trust}.
For example, when a user is traveling in a city, the system recommends several attractions and local restaurants to visit.
Both the recommendation and explanations help the user explore new things that they were not aware of.
\item User Intent Assistance. When the target task has been already initiated by users, then the goal of generating AI outcomes and explanations assists users with existing intent~\cite{brown1998utility,bercher2014plan,das2021explainable}. For instance, when a user is making dinner, intelligent instructions and explanations would suggest alternative ingredients based on what a user has in their space.
\item Error Management. When a system has low confidence about input/output or makes a mistake, explanations can serve as error management and explain the process so that users can understand where an error comes from if it appears~\cite{xu2019explainable,adadi_peeking_2018}, how they might better collaborate with the system~\cite{das2021explainable}, or when to adjust their expectation of the system's intelligence~\cite{bertossi2020data, zhang2020effect}.
\item Trust Building. Various studies have found that explanations can help systems build user trust by offering transparency and increasing intelligibility~\cite{antifakos2005towards,mahbooba2021explainable,shin2021effects}. As a result, users’ trust in models leads them to rely on the system~\cite{berkovsky2017recommend,bussone2015role}.
\end{s_enumerate}
These four types of system goals are not exclusive. A system can seek to achieve multiple goals simultaneously.
Depending on the subset of system goals, the appropriate explanation timing and content types can differ~\cite{lim_assessing_2009,myers2006answering} (related to \colorwhen{\textit{when}} and \colorwhat{\textit{what}}).

\textbf{\textit{User Goal.}}
While a system has varying reasons to provide explanations, end-users also have varying reasons to have explanations. We summarize four types of user goals from literature.

\begin{s_enumerate}
\item Resolving Confusion/Surprise. Expectation mismatch is one of the main reasons to need explanations~\cite{dhanorkar_who_2021,ribera2019can,brennen2020people,langer_what_2021}. Users can become confused or surprised when AI outcomes are different from what users are expecting, and having explanations can help to resolve concerns~\cite{rai2020explainable,gervasio2018explanation}.
\item Privacy Awareness. As AI influences more aspects of daily living, concerns about invasion of one's privacy are also growing~\cite{manheim2019artificial}. Explanations could disclose which data is being used in a model's decision-making process to end-users ~\cite{eslami2015always,rader2018explanations,datta2014automated}. 
\review{
Researchers and designers are recommended to follow an existing privacy framework, such as contextual integrity~\cite{nissenbaum2009privacy}, to make privacy explanations more robust.}
\item Reliability. Ensuring the reliability of AI outcomes is essential for non-trivial decision-making processes so that users can rely on a trustworthy system~\cite{lepri2018fair,jiang_who_2022,Ribeiro2016}, \eg daily activity recommendations for personal health management or automatic emergency service contacting in safety-threatening incidents.
\item Informativeness. End-users can be curious about the reason or process behind an AI outcome~\cite{hoffman2018metrics,li2020survey}. Explanations can fulfill users' curiosity by providing more information~\cite{lage2019human,binns2018s,rader2018explanations}.
\end{s_enumerate}
Similar to the system goals, these user goals are not exclusive and users can have multiple goals at the same time. Different goals can require different explanation timings and content (\colorwhen{\textit{when}} and \colorwhat{\textit{what}}).
% However, it is worth noting that \textit{user goal} is a latent factor that can often be hard to detect by an AR system.

\textbf{\textit{User Profile.}}
This factor covers a range of individual details that influence the design of XAI.
For example, information such as demographics and user preferences is necessary to generate personalized explanations~\cite{schneider2019personalized,kouki2019personalized,gedikli2014should}.
End-users' familiarity with system outcomes is related to the need for explanations and \colorwhen{\textit{when}} to provide them~\cite{coppers2018intellingo}.
Users' digital literacy with AI also affects \colorwhat{\textit{what}} types of explanations are appropriate and would serve users' purposes~\cite{long_what_2020,ttc_labs,ehsan2021explainable}.
Moreover, users may have individual preferences about explanation visualizations, which may be closely related to \colorhow{\textit{how}}.
This factor takes these considerations into account.

It is worth noting that XAIR is proposed as a design framework.
In a context that AR can detect robustly, designers can use the framework to infer end-users' latent factors, such as \textit{User State} and \textit{User Goal}, based on their design expertise~\cite{eiband_bringing_2018}. 
For example, when users are driving (which can be easily detected by AR), designers can assess users' cognitive load to be high (\textit{User State}). For more complex factors such as \textit{User Goal}, designers can propose a set of potential goals in a given scenario and then refer to the framework to propose a set of designs.
% The automatic detection of these factors is not the focus of this paper.
As sensing and AI technology are maturing, the framework could be coupled with the automatic inference of these factors~\cite{tsai2018augmented,admoni2016predicting,kim2016understanding,gjoreski2021head,yan2022emoglass}.

% In the next section, we will leverage the insights acquired from end-users and expert stakeholders, and connect these key factors to answer the three XAI in AR design sub-questions.
\section{Methods}
\label{sec:methodology}
We conducted two studies after outlining the problem space, one from end-users' perspectives (Sec.~\ref{sub:methodology:end_user_surveys}), and the other from XAI/design/AR expert stakeholders' perspectives (Sec.~\ref{sub:methodology:expert_workshops}). The findings from the studies are complementary and provided insights that guided the development of the framework.

\subsection{Study 1: Large-Scale End-User Survey}
\label{sub:methodology:end_user_surveys}
In spite of the existing studies on XAI for end-users, it is unclear whether these findings hold for AR scenarios due to the unique features of AR systems.
Thus, we conducted a large-scale survey with end-users to collect their preferences on various aspects of XAI experiences for everyday AR.
% The survey results reveal the need of XAI in AR scenarios and are generally consistent with previous research outside XAI.

\subsubsection{Participants}
\label{subsub:methodology:end_user_surveys:participants}
\review{We recruited 506 participants from a third-party online user study platform (age 18 - 54, average 37$\pm$10), with a balanced gender distribution (Female 260, Male 241, Non-binary 5).}
Participants' digital literacy with AI varied, thus they were split into six groups: 1) unfamiliar with AI (12.2\%, 62), 2) heard of AI but never used AI-based products (23.5\%, 119), 3) used AI products occasionally a few times (23.1\%, 117), 4) used AI products on a regular basis (12.8\%, 65), 5) used AI products frequently (20.0\%, 101), and 6) worked on AI products (8.3\%, 42).
Participants were familiar with the concept of AR.
Among these groups, we further randomly sampled 20 participants (age 18 - 53, average 37$\pm$9, 11 Female, 9 Male) for a semi-structured interview to collect a more in-depth understanding about their preferences for XAI in AR.

\subsubsection{Design and Procedure}
\label{subsub:methodology:end_user_surveys:materials_design}
We prepared five sets of proof-of-concept descriptions and images with intelligent everyday AR services that represented five scenes in a typical weekday (\ie one set per scene). They included 1) music recommendations for the morning when users would be brushing their teeth, 2) podcast recommendations for when users would be driving to work, 3) music recommendations for when users would be working out, 4) recipe recommendations for when users would be making dinner, and 5) additional spice recommendations for when users would be making dinner.
In this study, we chose recommendations as the main AI service category, since it is arguably one of the most common AI applications in everyday AR~\cite{lam_a2w_2021,chatzopoulos2016readme} and users could easily contextualize these scenes in their mind.

% we pre-determined a set of explanations from different categories
For the AI outcome in each scene, participants were asked whether they wanted explanations (\ie yes, no, neutral). If their answer was yes, they would be directed to answer when they wanted it (\ie always/frequent, contextually dependent, rare/never), their preferred length of explanation (\ie concise \vs detailed) and the presenting modality (\eg visual, audio, neutral).
After viewing these scenes, they were asked to choose the explanation content types that they found useful. Participants were compensated \$5 USD for the task.
% In interviews, the experiment host would follow up with more questions based on participants' answers to collect detailed reasons behind their choices.

% including input/output, why/why-not, how, and certainty. The other three categories are not included as they may not compatible 

% \subsubsection{Procedure}
% \label{subsub:methodology:end_user_surveys:procedure}
% Participants were told to imagine themselves as end-users of AR devices. They then went through the five scenes and answered the questions.
% At the end of the survey, they were asked whether they would be interested in joining a follow-up interview study.
We randomly sampled 20 respondents who were willing to participate in a one-hour interview about the detailed reasons behind their survey responses.
These participants were compensated \$10 for the interview.
The interviews were video-recorded and manually transcribed.
\review{Two researchers collectively summarized and coded the data using a thematic analysis~\cite{braun2012thematic}. Specifically, they first met to establish an agreement on the themes and independently coded all the data. Then, they gathered to discuss and refine the coded data to resolve differences. Their inter-rater reliability ($\kappa$) was over 90\% after the refinement.
}

\subsubsection{Results}
\label{subsub:methodology:end_user_surveys:results}
The survey found that respondents had specific preferences for the timing, content, and modality of explanations.

\begin{figure*}[!b]
    \centering
    \vspace{-0.2cm}
    \begin{subfigure}[b]{.33\textwidth}
    \centering
    \includegraphics[width=1\columnwidth]{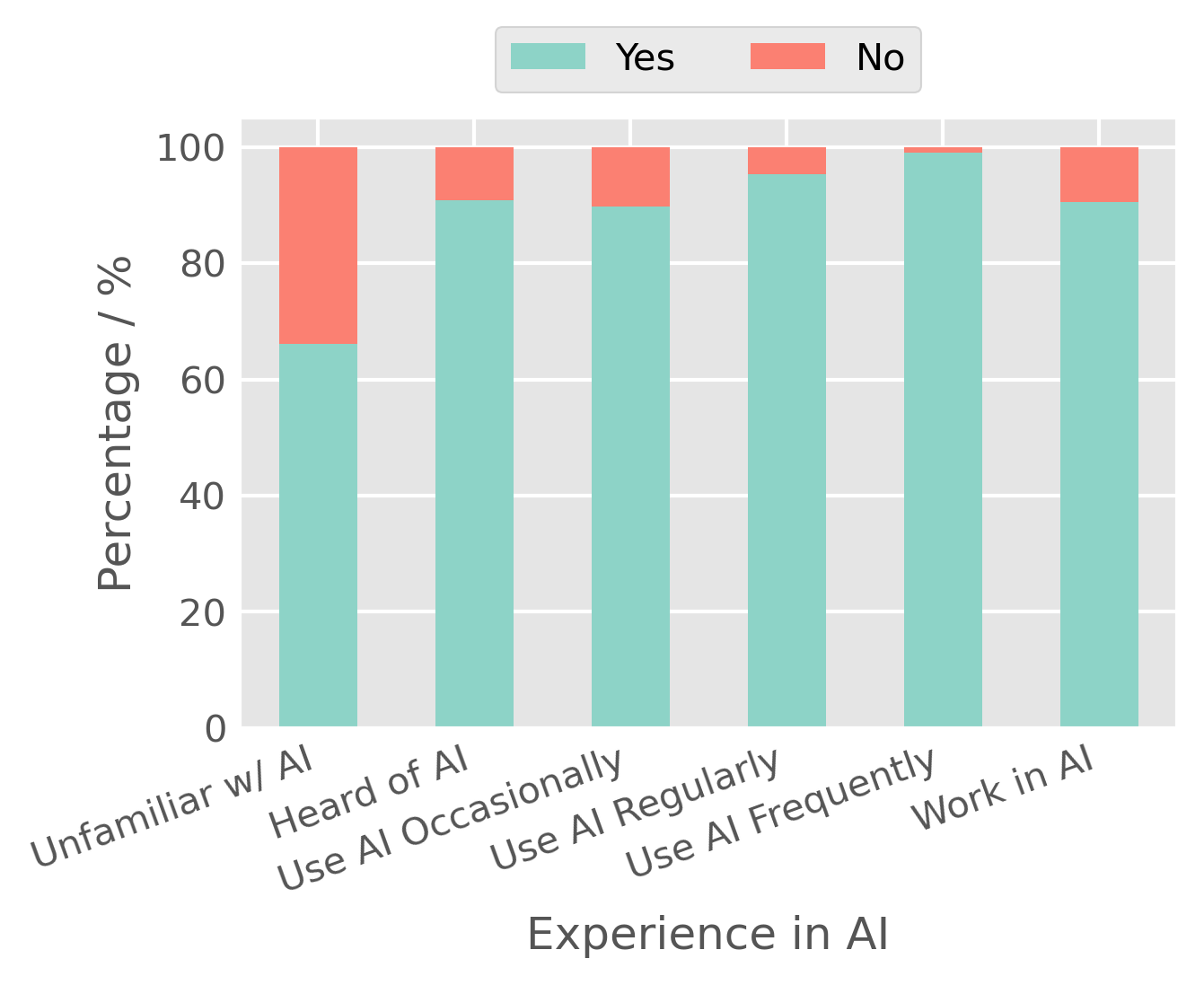}
    \caption{Need Explanations?}
    \label{subfig:survey_results:whether}
    \end{subfigure}
    \hfill
    \begin{subfigure}[b]{.34\textwidth}
    \centering
    \includegraphics[width=1\columnwidth]{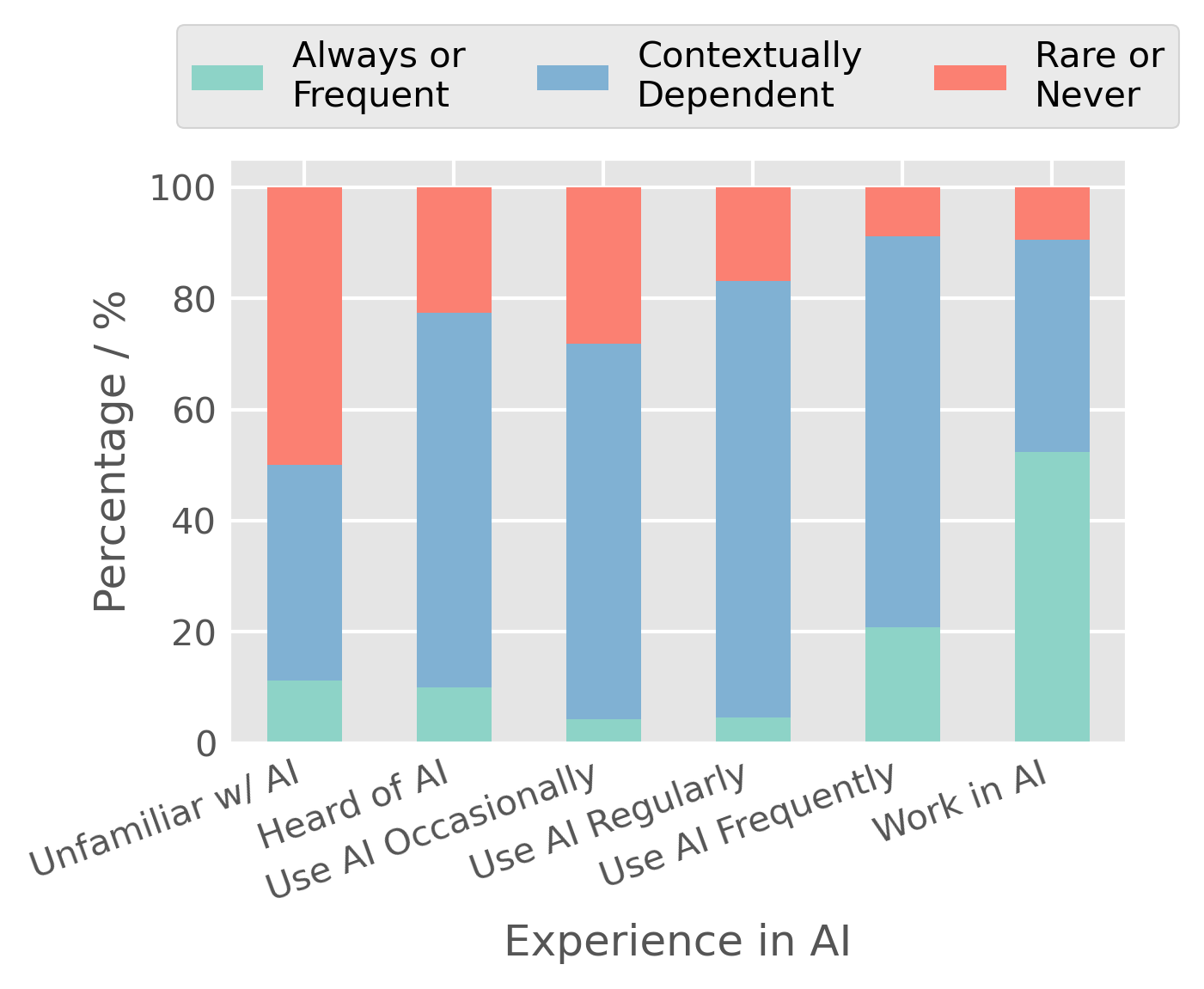}
    \caption{When to Have Explanations?}
    \label{subfig:survey_results:when}
    \end{subfigure}
    \hfill
    \begin{subfigure}[b]{.32\textwidth}
    \centering
    \includegraphics[width=1\columnwidth]{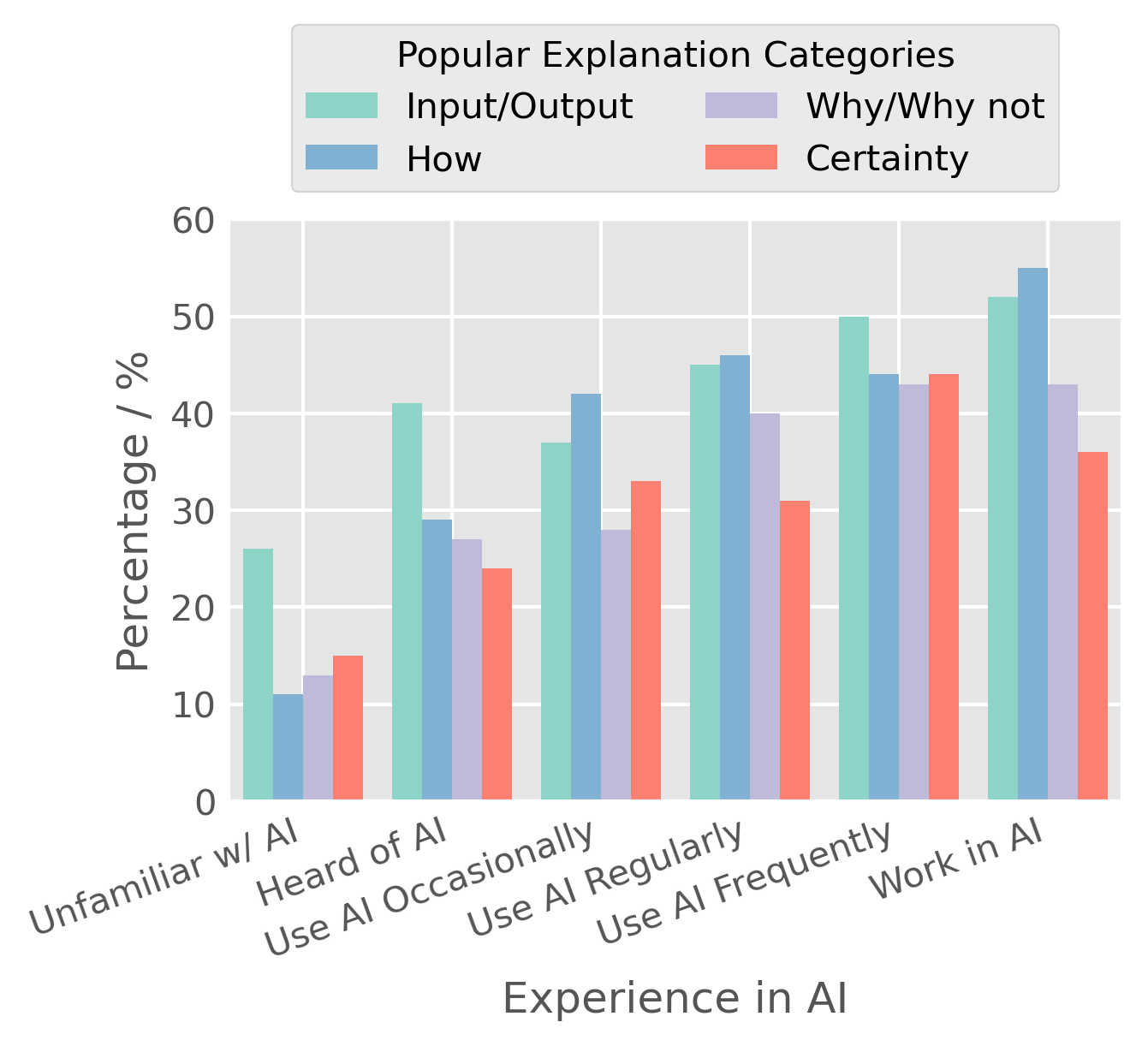}
    \caption{What Explanations Are Preferred?}
    \label{subfig:survey_results:what}
    \end{subfigure}
    \vspace{-0.4cm}
    \caption{Highlight of Survey Results with 506 End-Users about Their Needs and Preferences of XAI in everyday AR scenarios.}
    \label{fig:survey_results}
    \Description[Survey Results.]{(a) A stacked percentage bar plot with six bars showing whether users need XAI with their AI experience. The six X-axis stick labels are: unfamiliar with AI, never heard of AI, use AI occasionally, use AI regularly, user AI frequently, and work in AI. The Y-axis is about the percentage. The bar chart shows that around 37\% of the "unfamilar with AI" replied "No". All other types of users have less than 10\% replied "No".
(b) A stacked percentage bar plot with six bars showing when users need XAI with their AI experience. The six X-axis stick labels are the same. And it shows that over 60\% of the users prefer the explanations to be "contextually dependent". As users get more experience with AI, the percentage of choosing "always or frequently" increases, and the percentage of choosing "rare or never" decreases.
（c) A group bar plot with six groups showing when users need XAI with their AI experience. The six X-axis stick labels are the same. Each group has four categories: Input/Output, Why/Why-Not, How, and Certainty. As users get more experience with AI, the percentage of choosing these four categories increases (from around 15\% on average to 45\% on average).}
    \vspace{-0.3cm}
\end{figure*}

\review{\textbf{Finding 1: Most users wanted explanations of AI outputs in AR.}} (related to \colorwhen{\textit{when - availability}}).
A large proportion of respondents wanted explanations (89.7\%), motivating the need for XAI in everyday AR scenarios (see Fig.~\ref{subfig:survey_results:whether}).
Our findings were consistent with previous work on end-users' needs for XAI outside AR~\cite{ehsan2021explainable,ttc_labs}.
The results indicated that if respondents had at least heard of AI, they were more likely to express a need for XAI in AR compared to those who were not familiar with AI.
\review{
% An ANOVA with digital literacy as the main factor indicates statistical significance ($F_{5,500} = 10.8, p < 0.001$).
% Generally, the more familiar respondents were with AI, the more they felt there to be a need for explanations.
Our interviews found that respondents with little knowledge of AI didn't realize what explanations could be used for. 
Interestingly, around 10\% of respondents who worked on AI indicated that they didn't want explanations. Our interviews revealed the main reason being that some users were \textit{``familiar enough... with the algorithm''} (P2).
% For example, \pquote{4}{I can understand why... it's not really needed}. These participants already had knowledge of AI and thus didn't need additional explanations.
% This finding is consistent with previous work outside AR~\cite{ehsan2021explainable,ttc_labs}.
}
% 
% Moreover, our interview results also suggest that when participants were unsatisfied 

\textbf{Finding 2: The majority of users wanted explanations to be occasional and contextual, especially when they saw anomalies} (related to \colorwhen{\textit{when - delivery}}).
Although most respondents wanted explanations, only 13.8\% indicated that they needed explanations all the time.
% , mostly from participants who work on AI (see Fig.~\ref{subfig:survey_results:when}) -- an interesting polarization of XAI needs of users with high digital literacy when comparing \textbf{Finding 1} \& \textit{2}.
The majority of respondents (63.4\%) preferred for explanations to be presented contextually only when they have the need.
% For example, P7 didn't think there was a need for explanations in the morning music recommendation scene, \pquote{7}{I wouldn't be surprised if it knew what I wanted to listen to in the morning. I'm a creature of habit.}
The results of the interviews indicated that the need for explanations was mainly in cases where AI outcomes were new or anomalous to respondents. This finding is also in line with previous studies' findings outside AR~\cite{dhanorkar_who_2021,jiang_who_2022}.

\textbf{Finding 3: Users generally preferred specific types of explanations} (related to \colorwhat{\textit{what - content}}).
Four explanation content types stood out as useful: Input/Output (41.5\%), How (37.1\%), Why/Why-Not (31.6\%), and Certainty (30.6\%).
The first three types were highlighted in previous findings about context-aware systems~\cite{lim_assessing_2009,lim_toolkit_2010}, while the last type has been adopted by industrial practitioners~\cite{google_map_match_rate_2018,spotify_blend_taste}.
As shown in Fig.~\ref{subfig:survey_results:what}, respondents with more knowledge of AI would prefer having these explanation types more than those with less AI knowledge.
% (ANOVA $F_{5,2018} = 15.4, p < 0.001$),
% which aligns with \textbf{Finding 1}.

\textbf{Finding 4: Users found detailed and personalized explanations useful} (related to \colorwhat{\textit{what - detail}}).
Although showing more explanation content can introduce additional cognitive costs, 48.3\% of respondents reported that they would find detailed explanations with multiple content types to be useful.
% \pquote{1}{It gives me more context, substance for why I need to take this suggestion.}
Moreover, respondents indicated that explanations that included personal preferences would be more convincing, \eg \pquote{13}{more personable, more upbeat}.
These results suggest that there is a need to provide options to modulate the level of explanation detail (see Sec.~\ref{subsub:problem_space_factors:problem_space:what}) and the \textit{User Profile} factor in the framework).

\textbf{Finding 5: Users' preferences for modalities depended on the cognitive load in an AR scenario} (related to \colorhow{\textit{how - modality}}).
The five scenes introduced different levels of cognitive load, which led respondents' preferences for XAI modality to vary. We found that for scenes with complex visual stimuli such as driving, respondents tended to prefer audio explanations over visual ones by 40\%, as they were \pquote{8}{more easy and convenient}.
This suggests that it is necessary to take modality bandwidths into account when choosing \colorhow{\textit{how}} to present XAI in different AR scenarios~\cite{buchner2022impact}.

Overall, these findings motivated the need for XAI in AR (\textbf{Finding 1}).
% , and were aligned with previous studies on XAI outside of AR (\textbf{Finding 1, 2} \& \textit{3}).
% This verifies that previous findings are transferable to AR scenarios.
Moreover, these results (\textbf{Finding 2-5}) also provided guidance for design XAI for end-users in AR.

\subsection{Study 2: Iteration with Expert Workshops}
\label{sub:methodology:expert_workshops}
Based on the existing literature and the end-user survey results, we created an early draft of the framework. Since XAIR aims to support designers and researchers during their design process, we utilized our draft within three workshops with expert stakeholders to collect their insights and finalize the framework.

% \footnotetext{By expert stakeholders, we refer to experts related to our research problem (XAI, design, UX, HCI, AR), rather than domain experts as XAI users. We will use the word ``experts'' as this meaning for the rest of the paper.}

\subsubsection{Participants}
\label{subsub:methodology:expert_workshops:participants}
Twelve participants (7 Female, 5 Male, Age 35 $\pm$ 6) from a technology company volunteered to participate in the study. They came from four backgrounds, \ie 3 XAI algorithm developers, 3 designers, 3 UX professionals, and 3 HCI/AR researchers. Participants worked in their domains for at least five years. All participants were familiar with the concept of AI and AR. Participants were randomly assigned into three groups, with each group containing one expert from each domain.

\subsubsection{Design and Procedure}
\label{subsub:methodology:expert_workshops:materials_design}
We proposed a draft of the framework combining the summary of literature and the results of end-user study. It was an early version of XAIR that is introduced in Sec.~\ref{sec:framework} and can be found in Appendix~\ref{sec:appendix:earlier_frameworks}.
We also prepared a set of everyday AR scenarios similar to the ones used in the end-user survey (Sec~\ref{sub:methodology:end_user_surveys}) to provide more context and stimulate more insights from experts.
We utilized a Figma board to show images of the framework and experts could add in-place feedback to different areas of the framework.

We adopted an iterative process using three sequential workshops. \review{All workshops lasted about 90 minutes and were video-recorded. After each workshop, two researchers went through a similar coding and refining process as Sec.~\ref{subsub:methodology:end_user_surveys:materials_design}, to make sure the result achieved a inter-rater reliability ($\kappa$) over 90\%. We summarized experts' feedback, iterated on the framework, and presented the new version in the next workshop.}
% \subsubsection{Procedure}
% \label{subsub:methodology:expert_workshops:procedure}
% After participants signed the consent form, we conducted the workshop with a group of participants. 
% In each workshop, we introduced our framework and walked through an example intelligence everyday AR scenario to show the use case of our framework. We then asked participants to apply the framework on another scene to get a deeper understanding of the framework.
% Participants were encouraged to raise questions or give feedback at any time during the whole process.
% We repeated the process with the other two workshops, each time with an iterated version of the framework.
% met to establish an agreement on the themes. They independently coded all the data, and gathered together to refine the coded data to make sure the result achieved a inter-rater reliability $\kappa$ over 90\%.

% We then update the framework based on the final results.

\subsubsection{Results}
\label{subsub:methodology:expert_workshops:results}
Overall, experts found the framework to be \pquote{2, P6, P7}{useful} and that it would \pquote{11}{serve as a very good reference for design}.
Our framework converged as the workshops proceeded, with us receiving rich feedback during the first workshop, and participants in the last workshop only offering small suggestions. We briefly highlight the major comments that were made.

\textbf{Suggestion 1: Add Missing Pieces.}
Participants found a few factors missing in the early version of the framework.
% For example, they suggested that the \textit{System Goal} error management should also be considered for the automatic delivery (\colorwhen{\textit{when}}) of explanations if the model was uncertain.
% and that the 
\review{
For example, they pointed out that \textit{User Goal} and \textit{User Profile} needed to be considered for the \colorwhat{\textit{what}} part, and that the modality of AI output in AR needed to be taken into account for the \colorhow{\textit{how}} part.
}
% as it would indicate the need for different explanations.
They also provided suggestions on appropriate explanation content types with different system/user goals (\colorwhat{\textit{what - content}}).

\textbf{Suggestion 2: Remove Redundancy.}
Participants also found some parts unnecessarily complex. For example, four experts suggested removing the interface location from \colorhow{\textit{how}} part (\ie where to explain, mentioned in Sec.~\ref{subsub:problem_space_factors:problem_space:how}), because the location needed to be optimized with the whole interface including AI outcomes.

\textbf{Suggestion 3: Add Default Options.}
Participants provided advice for default options of different dimensions.
For instance, they recommended using the manual-trigger as the default delivery method (\colorwhen{\textit{when}}) due to users' limited cognitive capacity in AR.
% We worked with participants to summarize the advice into a set of guidelines.

\textbf{Suggestion 4: Connect across Sub-questions.}
Participants came to the consensus that the three sub-questions were interwoven.
For example, the choice of \colorwhat{\textit{what}} to explain would influence the design of \colorhow{\textit{how}} to explain, and the framework should capture and emphasize such connection.
% The interface design of the \colorhow{\textit{how}} part would also need to consider the manual-/auto-trigger mechanism for the \colorwhen{\textit{when}} part.

\textbf{Suggestion 5: Improve Visual Structure.}
Finally, participants also offered several suggestions about the visual simplification, clarification, and color choices. The figures in Appendix~\ref{sec:appendix:earlier_frameworks} show the evolution of the visual structure.

The results of the end-user study and expert workshops are complementary and guided the final version of the framework.
\section{XAIR Framework}
\label{sec:framework}
We introduced the structure of XAIR framework in Sec.~\ref{sec:problem_space_factors} (\ie problem space and key factors), and summarized insights from end-users and experts in Sec.~\ref{sec:methodology}.
\review{
Connecting the literature survey and studies' results (\textbf{Findings 1-5} in Sec.~\ref{subsub:methodology:end_user_surveys:results}, and \textbf{Suggestions 1-5} in Sec.~\ref{subsub:methodology:expert_workshops:results}), we introduce the details of XAIR, identify how the key factors determine the design choices for each dimension in the when/what/how questions, and present a set of guidelines.}

% Connecting between Sec.~\ref{sec:problem_space_factors} and Sec.~\ref{sec:methodology}, we develop XAIR, a framework for the design of XAI in AR. We further propose a series of guidelines in company with XAIR to answer the questions in the problem space.

\subsection{\colorwhen{When to Explain?}}
\label{sub:framework:when}
We first introduce the \textit{when} part and discuss how to make a choice for delivery options. Fig.~\ref{fig:overview_when} presents an overview.

\subsubsection{\colorwhen{\textcircledmod{A} \textbf{Availability}}}
\label{subsub:framework:when:availability}
\review{The end-user survey results suggested the need for explanations in AR for the majority end-users (\textbf{Finding 1})}. A system should always generate explanations with AI outcomes and make them accessible for users, so that they can have a better sense of agency whenever they need explanations~\cite{liu2021ai,zanzotto_viewpoint_2019,lee1992trust}.

\vspace{0.2cm}\noindent\colorwhen{\textbf{\textit{G1. Make explanations always accessible to provide user agency.}}}

\subsubsection{\colorwhen{\textcircledmod{B} \textbf{Delivery}}}
% \subsubsection{B - Delivery}
\label{subsub:framework:when:delivery}
Aligned with previous work~\cite{ibili2019effect,buchner2022impact,Pielot2017a}, experts also mentioned the risk of cognitive overload in AR \review{(\textbf{Suggestion 3})}. The default option should be to wait until users manually request explanations. An example could be a button with an information icon that enables users to click on it to see an explanation. 

% \colorwhen{\textbf{\textit{G2: By default, don’t trigger explanations automatically, wait until user’s request.}}}

However, there are cases where automatically presenting just-in-time explanations is beneficial~\cite{bhattacharya2017intent,mehrotra2019jointly}. \review{We summarize the three cases based on our two studies (\textbf{Finding 2} about the importance of contextual explanations, and \textbf{Suggestion 1} about the need of considering \textit{User Goal} and \textit{User Profile})}:

1) Cases when users have an expectation mismatch and become surprised/confused about AI outcomes~\cite{dhanorkar_who_2021,brennen2020people}, \review{\ie \textit{User Goal} as Resolving Surprise/Confusion (also reflected by \textit{User State}, which could be detected by AR HMDs using facial expressions and gaze patterns ~\cite{arguel2017inside,umemuro2003detection}).}
An example could be an intelligent reminder to bring umbrella when users are leaving home on a sunny morning (but it will rain in the afternoon). Automatic explanations of the weather forecast could help resolve users' confusion.

\begin{figure*}[!t]
    \centering
    \includegraphics[width=0.83\textwidth]{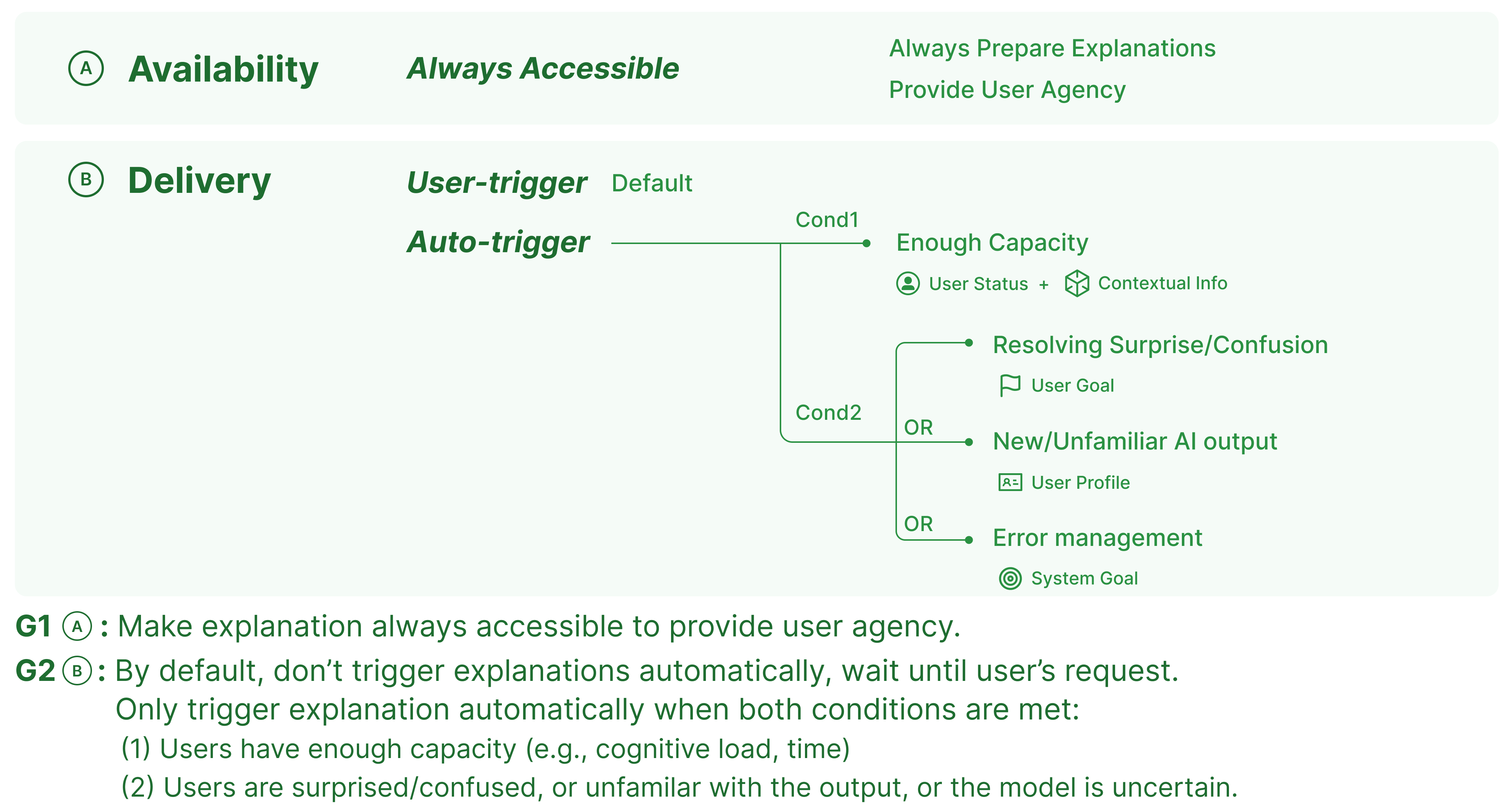}
    \caption{The "When" Part of XAIR. It contains two major dimensions: (A) Availability and (B) Delivery, highlighted in \textbf{bolded} texts.
    The design choice of dimensions are in \textbf{\textit{italic}} texts (same below for Fig.~\ref{fig:overview_what} and Fig.~\ref{fig:overview_how}).
    For example, \textbf{Delivery} can be either \textbf{\textit{User-trigger}} or \textbf{\textit{Auto-trigger}}.
    Each dimension has factors that should be considered for explanation designs.
    The guidelines G1 and G2 provide advice on these design choices.}
    \label{fig:overview_when}
    \Description[The "When" Part of XAIR.]{The "When" Part of XAIR. It contains two major dimensions: (A) Availability and (B) Delivery. Each dimension has its own design choice. Availability only has "Always Available". Delivery has the choices of "User-trigger" or "Auto-trigger". Each dimension has factors that should be considered for explanation designs. The guidelines G1 and G2 provide advice on these design choices.
G1: Make explanations always accessible to provide user agency.
G2: By default, don't trigger explanations automatically, wait until user's request.
Only trigger explanation automatically when both conditions are met:
(1) Users have enough capacity (e.g., cognitive load, urgency);
(2) Users are surprised/confused, or unfamiliar with the outcome, or the model is uncertain. }
\end{figure*}

2) Cases when users are unfamiliar with new AI outcomes (indicated via history information of \textit{User Profile}), \eg users receive a recommendation of a song that they have never heard before. Just-in-time explanations of the reason can help users to better understand the recommendation.

3) Cases when the model's input or output confidence is low and the model may make mistakes~\cite{bertossi2020data,kenny2021explaining}, \ie \textit{System Goal} as Error Management. \review{For instance, a system turning on a do-not-disturb mode when it detects a user working on a laptop in an office when the AR-based activity recognition confidence was low (\eg 80\%).} Explanations could be a gatekeeper if the detection was wrong and users could calibrate their expectations or adjust the system to improve the detection~\cite{das2021explainable,zhang2020effect}.

All of these cases have the prerequisite that users have enough capacity to consume explanations~\cite{schmidt_transparency_2020,Pielot2015}, \eg users' cognitive load is not high (\review{could be detected via gaze or EEG on wearable AR devices~\cite{xu2018review,zagermann2018studying}}), and users have enough time to do so (inferred based on context).
% We summarize these considerations in the third guideline.

\noindent\colorwhen{\textbf{\textit{G2. By default, don’t trigger explanations automatically, wait until users' request.
Only trigger explanations automatically when both conditions are met:\\
(1) Users have enough capacity (e.g., cognitive load, urgency);\\
(2) Users are surprised/confused, or unfamiliar with the outcome, or the model is uncertain.}}}

\subsection{\colorwhat{What to Explain?}}
\label{sub:framework:what}
In Sec.~\ref{subsub:problem_space_factors:problem_space:what}, we identified that \colorwhat{\textit{content}} and \colorwhat{\textit{detail}} were two dimensions of the \colorwhat{\textit{what}} part of the framework. We introduce how to choose among all explanation content types in Fig.~\ref{fig:overview_what}.

\subsubsection{\colorwhat{\textcircledmod{A} \textbf{Content}}}
\label{subsub:framework:what:content}
\begin{figure*}[!t]
    % \vspace{-0.4cm}
    \centering
    \includegraphics[width=1\textwidth]{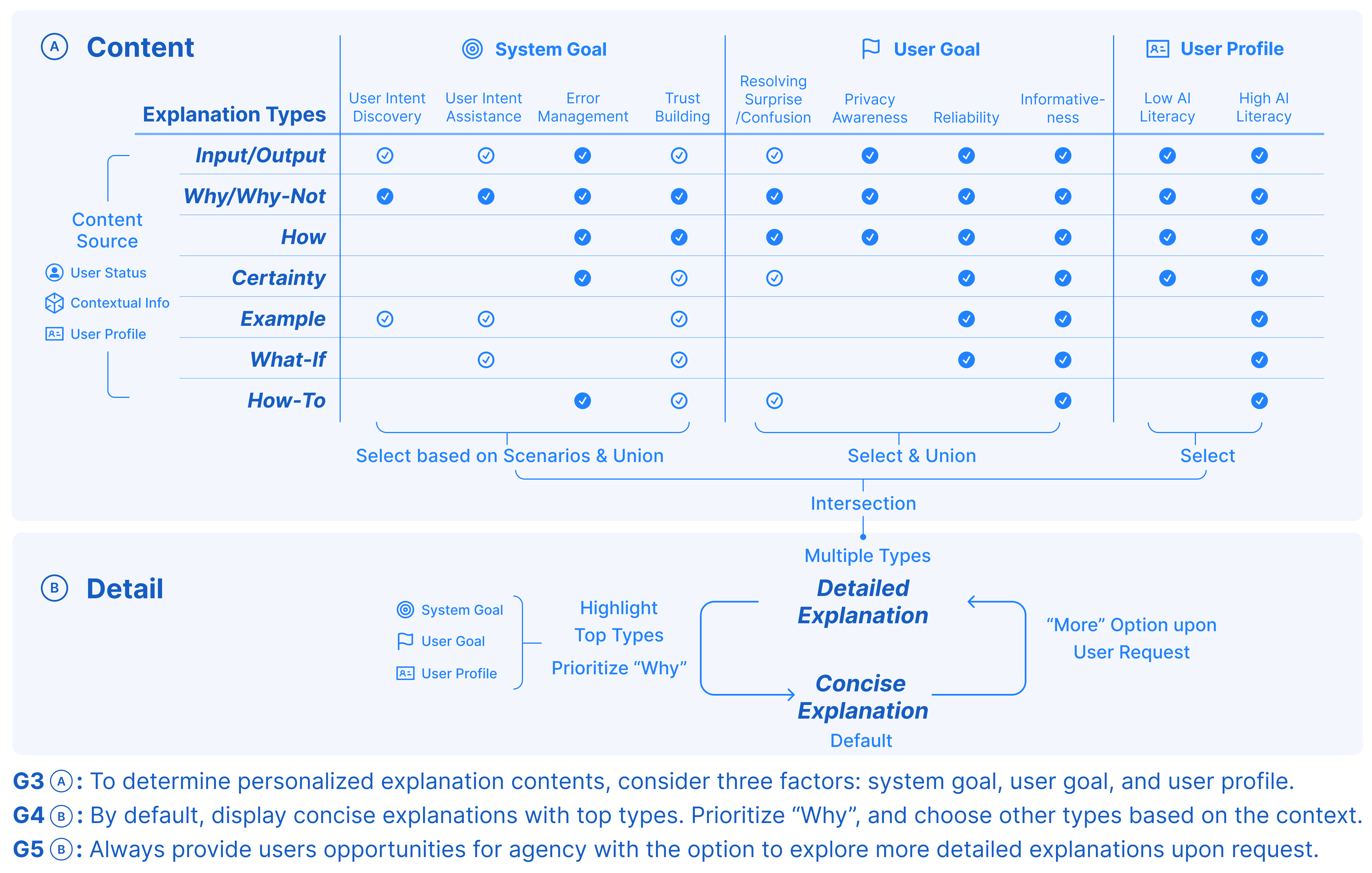}
    % \vspace{-0.5cm}
    \caption{The "What" Part of XAIR. (A) \textbf{Content} and (B) \textbf{Detail} are the two major dimensions. Combining the literature summary (shown as solid check marks) and findings from Study 1 \& 2 (shown as hollow check marks), G3-G5 provide guidelines on choosing appropriate explanation content types and length.}
    \label{fig:overview_what}
    \Description[The "What" Part of XAIR.]{The "What" Part of XAIR. (A) Content and (B) Detail are the two major dimensions. Content has the choice of seven categories, including Input/Output, Why/Why-Not, How, Certainty, Examples, What-If, and How-To. Detail has two choices: "Concise Explanations" and "Detailed Explanations".
Combining the literature summary and findings from Study 1 & 2, G3-G5 provide guidelines on choosing appropriate explanation categories and length.
G3. To determine personalized explanation content, consider three factors: system goal, user goal, and user profile.
G4. By default, display concise explanations by highlighting top categories. Prioritize Why, and choose other categories based on the context.
G5. Always provide users opportunities for agency with the option to explore more detailed explanations upon request.}
    % \vspace{-0.3cm}
\end{figure*}

In AR systems, the AI outcomes are based on factors such as \textit{User State} (\eg user activity), \textit{Contextual Information} (\eg the current environment), and \textit{User Profile} (\eg user preference). These factors also determine the content of different explanation content types.
\review{
To choose the right types, the framework lists three factors to consider and provides recommendations of personalized explanation content types based on the literature (shown as solid check marks in the top table in Fig.~\ref{fig:overview_what}), end-user survey, or expert advice (based on \textbf{Finding 3} and \textbf{Suggestion 1}, shown as hollow check marks).}

1) \textit{System Goal}. Different system goals need different explanations. For example, when a system recommends that users check out a new clothing store (User Intent Discovery), presenting {Examples} of similar stores that users are interested in and {Why} this store is attractive to users can be helpful. When a system wants to calibrate users' expectations about uncertain recipe recommendations (Error Management), showing {Examples} is less meaningful than presenting {How} and {Why} the system recommended this recipe, and {How To} change output if users want to. We leverage some literature on contextualized explanation content types to support our recommendations in the framework~\cite{lim_toolkit_2010,lim_assessing_2009,das2021explainable}.

2) \textit{User Goal}. Similarly, different user goals also require different explanations. For instance, {Certainty} explanations are helpful when users want to make sure an exercise recommendation fits their health plan (Reliability), while such explanations would be not useful when users want to be more aware of which data an AR system uses (Privacy Awareness). Most of these recommendations are supported by previous studies~\cite{lim_assessing_2009,barredo_arrieta_explainable_2020,mohseni_multidisciplinary_2021,rai2020explainable,langer_what_2021,wang_designing_2019}.
Regarding how to identify the user goals to choose explanations, designers can use their expertise to infer them in the context determined by AR systems. \review{In the future, it is also possible for AR systems to combine a range of sensor signals to detect/predict users' goals}~\cite{tsai2018augmented,admoni2016predicting,kim2016understanding}.

% As we mentioned earlier in Sec.~\ref{sec:problem_space_factors}, designers can infer these goals in a given context based on their expertise. Future AR systems may combine various sensor signals to infer users' goal~\cite{tsai2018augmented,admoni2016predicting,kim2016understanding}.

3) \textit{User Profile}, specifically user literacy with AI. For the majority of end-users who are unfamiliar with the AI techniques, we recommend only considering the four content types that users indicated that they would find useful: Input/Output, Why/Why-Not, How, and Certainty \review{(as shown in \textbf{Findings 3} and Fig.~\ref{subfig:survey_results:what}).} If users have high AI literacy, then all types could be considered~\cite{ttc_labs,ehsan2021explainable}.

Elements in \textit{System/User Goal} are not exclusive to each other. If there is more than one goal, these columns can be merged within each factor section to find the union (\ie content types checked in at least one column). Then, one can find the intersection among the three factors' content type sets (\ie overlapping types in all sets) to ensure that these explanations can fulfill all factors simultaneously. We show complete examples in later sections (Sec.~\ref{sec:applications} and Sec.~\ref{sec:evaluation}).
% These considerations lead to our next guideline.

\colorwhat{\textbf{\textit{G3. To determine personalized explanation content, consider three factors: system goal, user goal, and user profile.}}}

\subsubsection{\colorwhat{\textcircledmod{B} \textbf{Detail}}}
\label{subsub:framework:what:detail}
After selecting the appropriate content, default explanations need to be concise and can be further simplified by highlighting the most important types~\cite{buchner2022impact,baumeister2017cognitive}. General end-users are primarily interested in Why, \review{which is in line with experts' advice (\textbf{Suggestion 3} about default options) and previous literature~\cite{jiang_who_2022,lim_assessing_2009,lim_toolkit_2010}.}
Designers' can leverage their expertise to determine whether other types should be omitted or combined with Why in a specific context.

\colorwhat{\textbf{\textit{G4. By default, display concise explanations with top types. Prioritize Why, and choose other types based on the context.}}}

\review{As a large proportion of Study 1 participants indicated that detailed explanations could be useful (\textbf{Finding 4}), AR systems need to provide an easy portal (an interface widget such as a button) for end-users to explore more details. This can also provide user agency~\cite{lu_exploring_2022}.}

\colorwhat{\textbf{\textit{G5. Always provide users opportunities for agency with the option to explore more detailed explanations upon request.}}}

% Essentially it is a recommendation system with additional input from AR~\cite{da2020recommendation,manotumruksa2018contextual}
% Mix-initiative~\cite{horvitz_principles_1999}
% Personalization \& Customization~\cite{cui2020personalized,qian2013personalized}

% Focus of local explanations for end-users~\cite{lakkaraju2019faithful} Global for experienced \cite{dhanorkar_who_2021,long_what_2020}

% Different types of explanations~\cite{barredo_arrieta_explainable_2020,langer_what_2021,wang_designing_2019}

% Even after selecting an appropriate category subset, showing all of them can be overwhelming for end users, especially with limited cognitive capacity in AR~\cite{buchner2022impact,baumeister2017cognitive}. A summary with key explanation elements would be needed

% ``Findings include only showing explanations that extend prior knowledge~\cite{coppers2018intellingo}, and not to be too ``creepy'' by disclosing too much information~\cite{esteva2017dermatologist}.''~\cite{wang_designing_2019}

\subsection{\colorhow{How to Explain?}}
\label{sub:framework:how}
\begin{figure*}[!t]
    \centering
    \vspace{-0.2cm}
    \includegraphics[width=0.85\textwidth]{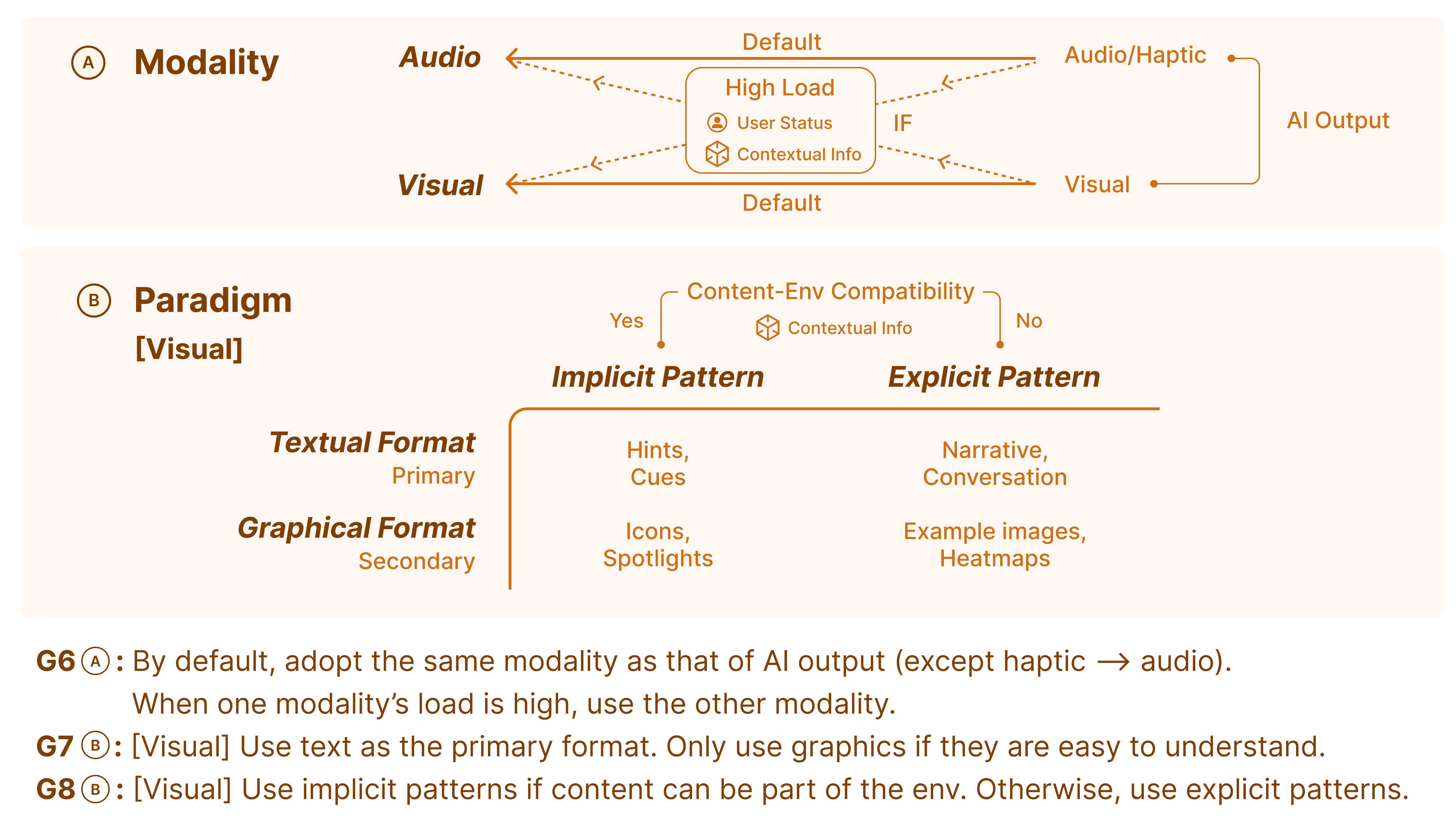}
    \caption{The "How" Part of XAIR. (A) \textbf{Modality} and (B) \textbf{Paradigm} are the major dimensions. Note that \textbf{Paradigm} is only for the visual modality, and it is further broken down into two perspectives: format and pattern. G6-G8 provides guidelines on making the proper design choices.}
    \Description[The "How" Part of XAIR.]{The "How" Part of XAIR. (A) Modality and (B) Paradigm are the major dimensions. Modality has the choice of "Audio" or "Visual". Paradigm is only for the visual modality, and it is further broken down into two perspectives: format and pattern. Format has the choices of "Textual Format" or "Graphical Format". Pattern has the choices of "Implicit Pattern" or "Explicit Pattern". G6-G8 provides guidelines for proper design choices.
G6. By default, adopt the same explanation modality as that of AI output (except haptic→audio). When one modality's load is high, use the other modality.
G7. [Visual] Use text as the primary format. Only use graphics if they are easy to understand.
G8. [Visual] Use implicit patterns if content can be part of the environment. Otherwise, use explicit patterns.}
    \label{fig:overview_how}
\end{figure*}

Finally, we introduce the \colorhow{\textit{how}} part and elaborate from the \colorhow{\textit{modality}} and \colorhow{\textit{paradigm}} perspective (see Fig.~\ref{fig:overview_how}).

\subsubsection{\colorhow{\textcircledmod{A} \textbf{Modality}}}
\label{subsub:framework:how:modality}
Considering channel bandwidth, the visual and audio modalities are the two most feasible modalities for AR.
Since explanations usually come during or after AI outcomes, to maintain consistency, the default modality of an explanation should be the same\review{(\textbf{Finding 5} and \textbf{Suggestion 3}).}
In cases when outcomes use a haptic modality (\eg vibration as a reminder), audio channels could be used as necessary (although this should be rare), since the choice of the haptic channel already conveys the need to be subtle.

However, there are also cases where one modality could be overloaded (based on \textit{User State} and \textit{Contextual Information}). For example, when users are driving and a navigation app suggests an alternate detour route, although the AI outcome is visual, the explanation should be audio to avoid visual overload.
When users are in a loud environment, a vibration-based AI outcome needs to use the visual modality for explanations.
\review{These scenarios can be easily detected by AR HMDs.}

\colorhow{\textbf{\textit{G6. By default, adopt the same explanation modality as that of the AI output (except for haptic$\rightarrow$audio). When one modality’s load is high, use another modality.}}}

Note that the modality choice also applies to the manual-trigger case when explanations are not automatically delivered (\gtwo), \eg a button icon for visual modality, a voice trigger for audio modality.

\subsubsection{\colorhow{\textcircledmod{B} \textbf{Paradigm}}}
\label{subsub:framework:how:paradigm}

Experts agreed that the audio design space does not belong within this framework. 
\review{For visual design, after removing the location from our framework (\textbf{Suggestion 2} of redundancy removal)}, we mainly focused on two aspects: \colorhow{format} and \colorhow{pattern}.
Depending on the content (\gfour), the explanation \colorhow{format} can be textual~\cite{lakkaraju2016interpretable,myers2006answering}, graphical~\cite{zeiler2014visualizing,simonyan2013deep}, or both.
Based on the consensus of experts in Study 2, text should be the primary format.
Experts suggested several reasons for this. Text takes up less space in a limited AR interface, and can introduce relatively less cognitive load. Moreover, the textual format is more universal and can cover all types.
Graphics can be used as the secondary format.
For default explanations (\gfive), in addition to displaying a short and concise textual paragraph, simple graphics such as icons can be used to provide additional information.
For detailed explanations (\gsix), more complex graphical formats (\eg example images or heatmaps) can be used as long as they are easy for end-users to understand.

\colorhow{\textbf{\textit{G7. [Visual] Use text as the primary format. Only use graphics if they are easy to understand.}}}

Independent of the \colorhow{format}, explanations can be presented in an implicit or explicit \colorhow{pattern}~\cite{lindlbauer_context-aware_2019,tatzgern_adaptive_2016}. Given the capability of depth sensing and 3D registration in AR, we recommend using the implicit pattern when the explanation content is compatible with the environment (\ie can be naturally embedded as a part of the environment).
For example, for book recommendations, a text cue or a small icon can float on the book to indicate the book's topic that users like (belonging to the Why explanation content type).
When explanations and the environment are not compatible, using an explicit pattern (\eg a dialogue window) can be the back-up option.
With regard to what explanation content is compatible with the environment, designers can leverage their expertise and intuition to propose appropriate embedding patterns for given a context. \review{Future AR systems may first understand the environment using object detection and context recognition algorithms, and then utilize techniques such as knowledge graphs (\ie networks of real-world entities and their relationships)~\cite{chen2020review} to assess the compatibility between the content and the environment.}

\colorhow{\textbf{\textit{G8. [Visual] Use implicit patterns if content can be embedded in the environment. Otherwise, use explicit patterns.}}}
\\

\noindent XAIR can not only serve as a summary of the study findings and the multidisciplinary literature across XAI and HCI, but also guide effective XAI design in AR.
In the next two sections, we provide examples of XAIR-supported applications (Sec.~\ref{sec:applications}), and evaluate XAIR from both designers' and end-users' perspectives (Sec.~\ref{sec:evaluation}).
% In Sec.~\ref{sec:evaluation}, we evaluate our framework from various perspectives.

\section{Applications}
\label{sec:applications}
To demonstrate how to leverage XAIR for XAI design, we present two examples that showcase potential workflows that use XAIR for everyday AR applications (Fig.~\ref{fig:applications}). More details can be found in Appendix~\ref{sub:appendix:details_scenarios:more_details_of_application}.
After determining the key factors for a given scenario, we used the framework (Fig.~\ref{fig:overview_when}-Fig.~\ref{fig:overview_how}) to make design choices based on the factors.

\begin{figure*}[!t]
    \centering
    % \vspace{-0.3cm}
    % \includegraphics[width=1\columnwidth]{figures/application_scenarios.png}
    % \includegraphics[width=0.8\columnwidth]{figures/application_scenarios_2.png}
    \begin{subfigure}[b]{0.49\textwidth}
    \includegraphics[width=1\columnwidth]{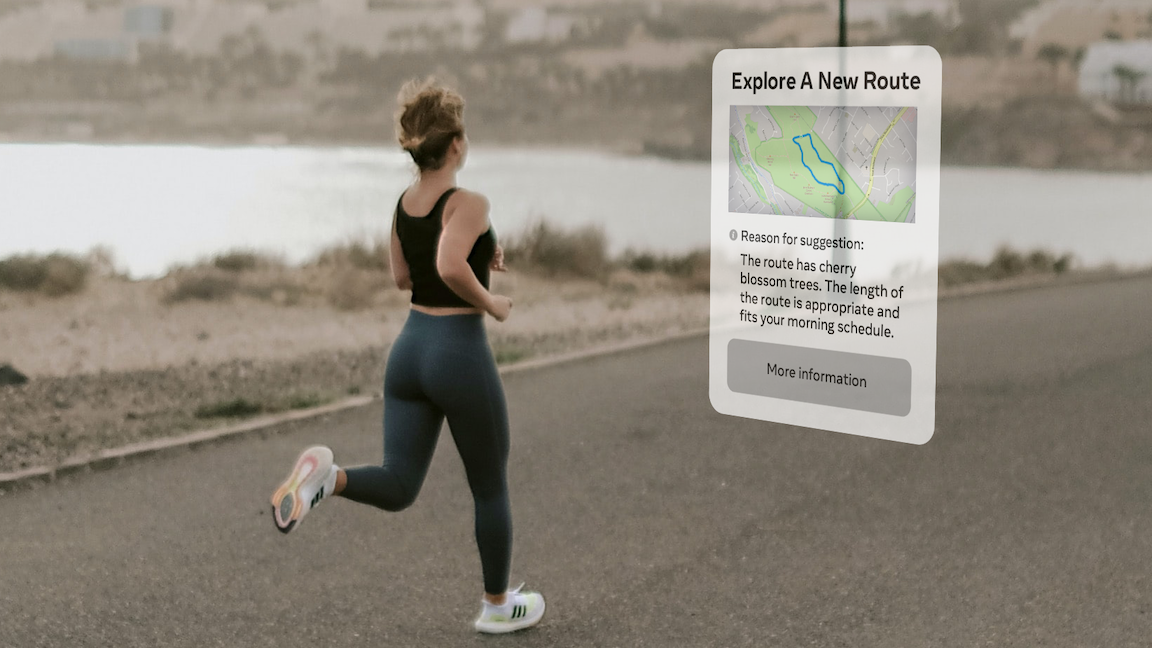}
    \caption{Scenario 1: Route Suggestion when Jogging}
    \label{subfig:applications:jogging}
    \end{subfigure}
    \hfill
    \begin{subfigure}[b]{0.49\textwidth}
    \centering
    \includegraphics[width=1\columnwidth]{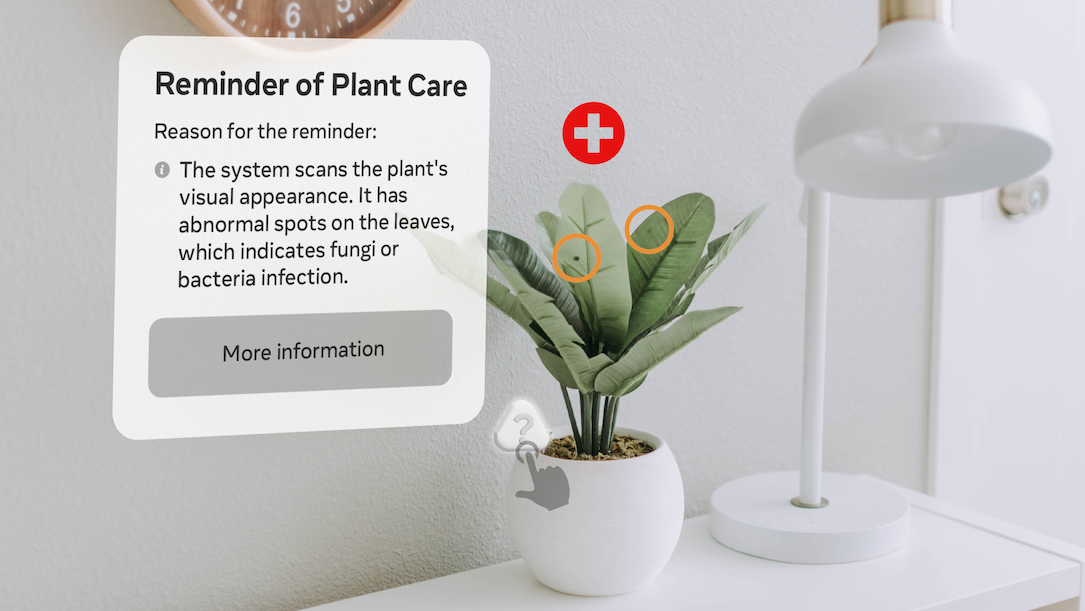}
    \caption{Scenario 2: Plant Fertilization Reminder.}
    \label{subfig:applications:plant}
    \end{subfigure}
    % \vspace{-0.3cm}
    \caption{Application of XAIR on Two Everyday AR Scenarios. In the second scenario, the hand icon indicates that explanations are manually triggered (the same below). Figures only present the default, concise explanations. Detailed explanations are described in the main text of Sec.~\ref{sec:applications}.}
    \label{fig:applications}
    \Description[Application of XAIR on Two Everyday AR Scenarios.]{The figure illustrates two everyday AR scenarios of how XAIR could be applied. The left subfigure shows a scenario described in section 6.1 in the paper. In the scenario, Nanci is jogging on a quiet trial. Her AR glasses display a map beside her view and recommend a detour. Nancy is curious to know the reason why the new route is recommended. The AR glasses prompt explanations with the user goal of resolving surprise, and the system goal of user intent discovery. The right subfigure shows a scenario described in section 6.2 in the paper. In the scenario, Sarah is recommended by her AR glasses a plant fertilization reminder after chatting with her neighbor on relevant topics. She is concerned about her privacy being invaded. In this case, the system displays an explanation with the user goal of privacy awareness and the system goal of trust building. Both scenarios demonstrate how XAIR could be used to support the design of explainable interfaces in AR glasses.}
\end{figure*}

\subsection{Scenario 1: Route Suggestion while Jogging}
\label{sub:applications:scenario_jogging}
\textit{Scene}. Nancy (AI expert, high AI literacy) is jogging in the morning on a quiet trail. Since it is the cherry-blossom season and Nancy loves cherries, her AR glasses display a map beside her and recommend a detour. Nancy is surprised since this route is different from her regular one, but she is happy to explore it. She is also curious to know the reason this new route was recommended.

\colorwhen{\textit{\textbf{When}}}.
\colorwhen{Delivery}. Nancy has enough cognitive capacity in this scenario. Her \textit{User Goal} is Resolving Surprise. Therefore, an explanation is automatically triggered because the two conditions are met (\gthree).

\colorwhat{\textit{\textbf{What}}}.
\colorwhat{Content}. Other than the \textit{User Goal}, the \textit{System Goal} is User Intent Discovery (exploring a new route to see cherry blossom). Considering Nancy's \textit{User Profile}, she is an expert in AI, so the appropriate explanation content types (\gfour) are Input/Output (\eg ``This route is recommended based on seasons, your routine, and preferences.'') and Why/Why-Not (\eg ``The route has cherry blossom trees that you can enjoy. The length of the route is appropriate and fits your morning schedule.'').
Examples for all seven explanation content types can be found in Appendix~\ref{sub:appendix:details_scenarios:more_details_of_application}.
\\
\colorwhat{Detail}. The AR interface shows the Why as default (\gfive), and can be expanded to show both types in detail (\gsix). Nancy can slow down and click the ``More'' button to see more detailed explanations while standing or walking.

\colorhow{\textit{\textbf{How}}}.
\colorhow{Modality}. The explanation is presented visually, the same as the recommendation (\gseven).
\\
\colorhow{Format}. The default explanation uses text, while the detailed explanation contains cherry-blossom pictures of the new route to help explain the Why (\geight).\\
\colorhow{Pattern}. The explanation is shown explicitly within the route recommendation window (\gnine).

\subsection{Scenario 2: Plant Fertilization Reminder}
\label{sub:applications:scenario_reminder}
\textit{Scene}. Sarah (general end-user, low AI literacy) was chatting with her neighbor about gardening. After she returned home and sat on the sofa, her AR glasses recommended instructions about plant fertilization by showing a care icon on the plant. Sarah is concerned about technology invading her privacy, and wants to know the reason behind the recommendation.

\colorwhen{\textit{\textbf{When}}}.
\colorwhen{\textit{Delivery}}. Although Sarah has enough cognitive capacity, none of the three cases in the second condition of \gthree\ are met (\ie she was familiar with the recommendation and not confused, and the model didn't make a mistake). Therefore, the explanation needs to be manually triggered (\gtwo).

\colorwhat{\textit{\textbf{What}}}.
\colorwhat{Content}. In this case, the \textit{System Goal} is Trust Building (clarifying the usage of data), and the \textit{User Goal} is Privacy Awareness. Sarah's \textit{User Profile} indicates that she is not an expert in AI. According to \gfour, the explanation content type list contains Input/Output, Why/Why-Not, and How.\\
\colorwhat{Detail}. Considering Sarah's concern, the default explanation merges Why and How: ``The system scans the plant's visual appearance. It has abnormal spots on the leaves, which indicate fungi or bacteria infection.'' (\gfive).
For the detailed explanation, the full content of the three types is presented in a drop-down list upon her request (\gsix).

\colorhow{\textit{\textbf{How}}}.
\colorhow{Modality}. Following \gseven, the visual modality is used for both the explanation and the manual trigger (a button beside the plant care icon).\\
\colorhow{Format}. Other than using text as the primary format, the abnormal spots on the leaves are also highlighted via circles to provide an in-situ explanation (\geight).\\
\colorhow{Pattern}. Since the highlighting of spots is compatible with the environment (shown on leaves), it adopts the implicit pattern (\gnine). The rest of the texts of the explanation uses the explicit pattern.

Our two examples demonstrate XAIR's ability to guide XAI design in AR in various scenarios. In Appendix~\ref{sub:appendix:details_scenarios:more_scenarios}, we provide additional everyday AR scenarios to further illustrate its practicality.
\section{Evaluation}
\label{sec:evaluation}

In addition to showing examples to illustrate the use case of XAIR, we also conducted two user studies to evaluate XAIR.
The first study was from the perspective of designers (as XAIR users) to evaluate XAIR's ability to assist designers during their design processes (Sec.~\ref{sub:evaluation:utility}).
The second study was from an end-user perspective and evaluated XAIR's effectiveness at achieving a user-friendly XAI experience in AR. We measured the usability of the real-time AR experiences that were developed based on the design examples proposed by designers (Sec.~\ref{sub:evaluation:effectiveness}).
% The results indicate positive utility and strong effectiveness of XAIR.

\subsection{Study 3: Design Workshops}
\label{sub:evaluation:utility}
We conducted one-on-one design workshops with designers to investigate whether the framework could support their design processes, inspire them to identify new design opportunities, and achieve effective designs.

\subsubsection{Participants}
\label{subsub:evaluation:utility:participants}
Future XAI and AR designers can come from various backgrounds, so we recruited 10 participants (4 Female, 6 Male, Age 32 $\pm$ 6) from a technology company as volunteers.
Three were XAI algorithm researchers, four were product designers, and three were HCI/AR researchers.
All participants were familiar with AI and AR, and none had participated in previous studies.

\subsubsection{Design and Procedure}
\label{subsub:evaluation:utility:materials}
We prepared two AR scenarios, both related to recipe recommendations while preparing meals.

\textbf{Case 1: Reliable Recipe Recommendation}.
Michael works in a sales company (general end-user, low digital literacy). He recently started a high-protein diet due to his workout routine. He opens the fridge and wants to make lunch. His AR glasses present a window on the fridge door and recommend an option that Michael usually has, but Michael wants to make sure that this option fits his recent diet changes.

\textbf{Case 2: Wrong Recipe Recommendation}.
Mary works in an AI company (high AI literacy) and has friends coming over for dinner, who are beef lovers. She opens the fridge and sees steak. However, her AR glasses mistakenly recognize steak as salmon with a medium level of confidence~\footnote{If the system has low-level confidence, the expected cost of making mistakes will be higher than the cost of asking for users' input, so the system should ask for users' confirmation about the ingredients they have on hand before presenting recommendations (\eg asking ``Is this salmon or steak?''). In this scenario, the confidence is at the medium level, thus the system provides recommendations, but  is still aware of the potential to make mistakes.}, and recommends a few recipes that use salmon. She is confused and wonders how she can correct the recommendations.

Since generating explanations is not the focus of the framework, we prepared examples for the seven explanation content types (Appendix~\ref{sub:appendix:details_scenarios:more_details_of_study}). Participants were free to use our examples, or propose their own (without the need to design how an algorithm could generate them).

% We adopted a process similar to the within-subject design: 
Participants first used their expertise and intuition to propose XAI designs for the two cases before being shown the framework. They spent 10 minutes on each case.
Participants were encouraged to think aloud and describe their design via text and simple sketches.
Then, after XAIR was introduced, they spent another 10 minutes following the three parts and eight guidelines and applied them to the two cases, resulting in another version of the design. The order of the two cases was counterbalanced.

To quantify the utility of XAIR, we employed the Creativity Support Index (CSI, 1-10 Likert scale)~\cite{cherry2014quantifying} and System Usability Scale (SUS)~\cite{bangor2008empirical}.
Since both scales were originally designed for tools or systems, the language was modified from ``tools'' and ``system'' to ``framework'' and ``guidelines''.
At the end of the workshop, we conducted a semi-structured interview that began with the question: ``Do you think the framework and guidelines are helpful? If so, in what aspects they are helpful?''
Each workshop lasted 90 minutes. Two researchers independently coded the qualitative data using thematic analysis and discussed it to reach an agreement.

% \subsubsection{Procedure}
% \label{subsub:evaluation:utility:procedure}
% We introduced the background and two scenarios. Participants first spent around 10 minutes on each task and proposed designs of the three sub-questions from scratch.
% After we presented our framework and guidelines, participants spent another 10 minutes on each task and proposed a new version.
% They completed the questionnaire and did the interview after finishing the designs.

% \input{tex_fig_tab_alg/fig_design_examples}

\subsubsection{Design Results}
\label{subsub:evaluation:utility:design_results}
After using XAIR, nine out of ten participants modified their designs and preferred the updated version. One participant (P7) liked the design as it was and thought that the framework \textit{``perfectly supported the design''}.
Consistency was found among the designs, which indicated that XAIR could effectively guide users through the design process.
For example, Tab.~\ref{tab:design_example_case1} presents two designers' designs (images are rendered based on their proposals) of the reliable recommendation case. Their designs of the \colorwhen{\textit{when}} part and most of the \colorwhat{\textit{what}} part were the same.
Tab.~\ref{tab:design_example_case2} presents another two designers' designs of the wrong recommendation case (Case 2). Similarly, we also found consistent design choices between the two examples.

\renewcommand{\arraystretch}{1.6}
\begin{table*}[t]
\centering
\resizebox{1\textwidth}{!}{
\begin{tabular}{c|c|m{7.3cm}|m{7.3cm}}
\hline \hline
\multicolumn{2}{c|}{\makecell{\text{ }}} & \includegraphics[width=7.3cm]{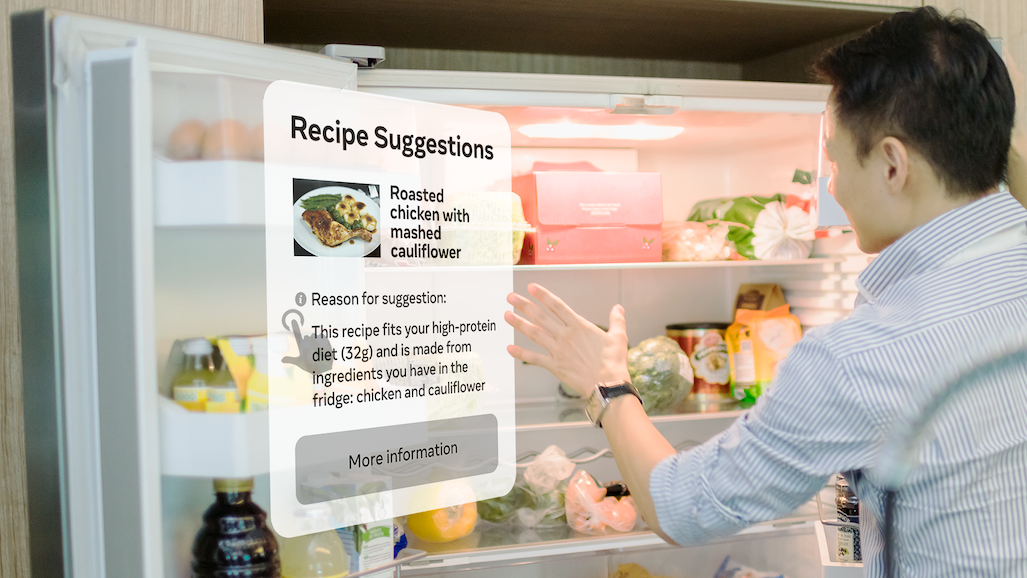} & \includegraphics[width=7.3cm]{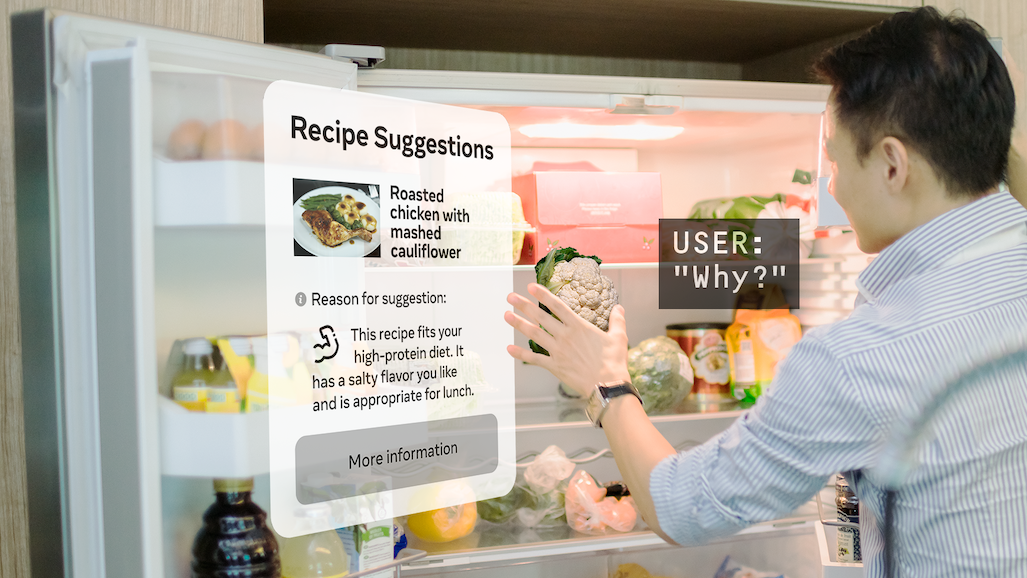}\\
\multicolumn{2}{c|}{\makecell{\textbf{Designer}}} & \makecell{\textbf{P2, Product Designer}} & \makecell{\textbf{P6, XAI Researcher}} \\
 \hline
\multirow{3}{*}{\makecell[c]{\\\textbf{Platform-}\\\textbf{Agnostic}\\\textbf{Key Factors}}}  & \textbf{System Goal} & \multicolumn{2}{c}{\makecell[c]{\textbf{User Intent Assistance} (to find a good recipe)}} \\
 & \textbf{User Goal} & \multicolumn{2}{c}{\makecell[c]{\textbf{Reliability} (to make sure the recipe fits the diet)}} \\
 & \textbf{User Profile} & \multicolumn{2}{c}{\makecell[c]{\textbf{User Preference}: High protein food; \textbf{History}: Know these recommended recipes;\\\textbf{AI Literacy}: General end-user, low}} \\ \hdashline
\multirow{2}{*}{\makecell[c]{\\\textbf{AR-Specific}\\\textbf{Key Factors}}} & \textbf{Contextual Info} & \multicolumn{2}{c}{\makecell[c]{\textbf{Location}: Kitchen; \textbf{Time}: Noon; \textbf{Environment}: Various ingredients in the fridge}} \\
 & \textbf{User State} & \makecell[l]{\textbf{Activity}: Opening the fridge to make lunch;\\\textbf{Cognitive Load}: Low} & \makecell[l]{\textbf{Activity}: Opening the fridge to make lunch,\\possibly holding something\\\textbf{Cognitive Load}: Low} \\ \hline
\multirow{2}{*}{\makecell[c]{\\\\\textbf{XAI Designs}\\\textbf{in AR}: \colorwhen{\textbf{\textit{When}}}}} & \colorwhen{Availability} (\gone) & Always available & Same as P2\\
 & \colorwhen{Delivery} (\gtwo) & Manual-trigger, because the second condition of auto-trigger was not met given the \textit{System Goal}, \textit{User Goal}, and \textit{User Profile}. & Same as P2 \\ \hdashline
\multirow{3}{*}{\makecell[c]{\\\\\textbf{XAI Designs}\\\textbf{in AR}: \colorwhat{\textbf{\textit{What}}}}} & \colorwhat{Content} (\gfour) & Input/Output \& Why/Why-Not based on Fig.~\ref{fig:overview_what}'s table & Same as P2\\
 & \colorwhat{Detail - Concise} (\gfive) & An explanation merging the Why and Input content types, as explaining \textit{``showing ingredients is also important''} & An explanation of the Why part as \textit{``it needs to be prioritized''}\\
 & \colorwhat{Detail - Detailed} (\gsix) & A list of the two explanation types in detail & Same as P2; Cherry flower pictures to support the Why explanation \\ \hdashline
\multirow{3}{*}{\makecell[c]{\\\\\textbf{XAI Designs}\\\textbf{in AR}: \colorhow{\textbf{\textit{How}}}}} & \colorhow{Modality} (\gseven) & Visual modality to ensure consistency with the recommendation interface & \makecell[l]{Visual modality for explanations;\\Audio/visual modality for manual trigger if\\the user is/isn't holding something}\\
 & \colorhow{Paradigm - Format} (\geight) & Textual format & Textual format as the primary format; Graphic format (protein icon) to support explanations\\
 & \colorhow{Paradigm - Pattern} (\gnine) & Explicit pattern, presenting texts in the same window as the recommendations & Same as P2\\
\hline \hline
\end{tabular}
}
\vspace{0.1cm}
\caption{Two Design Examples of Case 1: Reliable Recipe Recommendation. Participants' quotes were presented in italic font. Among key factors, P2 and P6 had different thoughts on \textit{User State}, which leads to different design choices of \colorhow{how - modality}. The comparison indicates both consistency and variance between two designers' examples.}
\label{tab:design_example_case1}
\Description{Two Design Examples of Case 1: Reliable Recipe Recommendation: Michael works in a sales company (general end-user, low digital literacy). He recently started a high-protein diet due to his workout routine. He opens the fridge and wants to make lunch. The AR glasses present a window on the fridge door, and recommend an option that Michael usually has, but Michael wants to make sure that this option fits his recent diet changes.
The examples are from P2 and P6. Among key factors, P2 and P6 had different thoughts on User State, which leads to different design choices of "how - modality". The comparison indicates both consistency and variance between the two designers' examples.
P2's design is as follows:
When. P2 thought that Michael's User Goal was to ensure the recipes went along with his diet (Reliability), and the System Goal was to help Michael find suitable recipes (User Intent Assistant). Moreover, Michael was pretty familiar with these recommendations. Therefore, the second condition of G2 was not met, and P2 agreed that the manual trigger was a better idea for delivery.
What. Using XAIR, he identified the three factors (i.e., the goals of the system and the user, plus Michael had a low AI literacy) and narrowed down the content to be Input/Output, and Why/Why-Not. For detail, besides prioritizing Why, P2 also proposed a short paragraph by merging the Input category as he thought explaining ingredients was also important. As for detailed explanations, he decided to show the list of the two explanation categories in detail.
How. P2's design is to adopt the visual modality, which was in line with the recommendation interface. P2 designed a simple button icon to support the manual trigger. The textual format was appropriate to explain the Why/Why-Not and Input/Output reasons. Since a textual paragraph was not easily compatible with the environment, P2 adopted the explicit pattern that presented texts in the same window as the recommendations.
P6's design is as follows:
When. Same as P2, P6 also examined the conditions and chose the manual-trigger option for delivery.
What. P6 selected the same categories as Input/Output and Why/Why-Not, and adopted the same detailed explanations design. However, for the default concise explanations in the detail dimension, P6 mainly focused on the Why part and proposed to add an icon to indicate the high-protein feature (also about the Why explanation).
How. P6 proposed the same visual modality and explicit pattern. Moreover, he also brought up an interesting case when Michael's hands were busy holding ingredients. In this case, P6 proposed to support an audio trigger as an alternative. As for the format, other than using texts as the primary format, P6 further proposed using a simple icon as a secondary format to support explanations.}
% \vspace{-0.7cm}
\end{table*}
\renewcommand{\arraystretch}{1.0}

Meanwhile, we also found variance across participants' designs.
For instance, in Case 1, P6 had a different consideration of \textit{User State} than P2, in which P6 brought up a case where the user could hold something in their hand. In this case, P6 adopted the audio modality for manual trigger (the rightmost column of Tab.~\ref{tab:design_example_case1}).
Moreover, as shown in the rightmost column of Tab.~\ref{tab:design_example_case2}, P9 proposed an interesting tweak that always highlighted ingredients (Input explanation type). Her reason was that it introduced \textit{``ultra-low cognitive cost''}, thus there was no need to check the second auto-trigger condition. \textit{``I don't think it is a violation of the guideline. Instead, I was inspired by the framework to consider this case.''}
This reveals that XAIR is flexible and can support the diverse creativity of users.

\renewcommand{\arraystretch}{1.6}
\begin{table*}[t]
\vspace{-0.1cm}
\centering
\resizebox{1\textwidth}{!}{
\begin{tabular}{c|c|m{7.3cm}|m{7.3cm}}
\hline \hline
\multicolumn{2}{c|}{\makecell{\text{ }}} & \includegraphics[width=7.3cm]{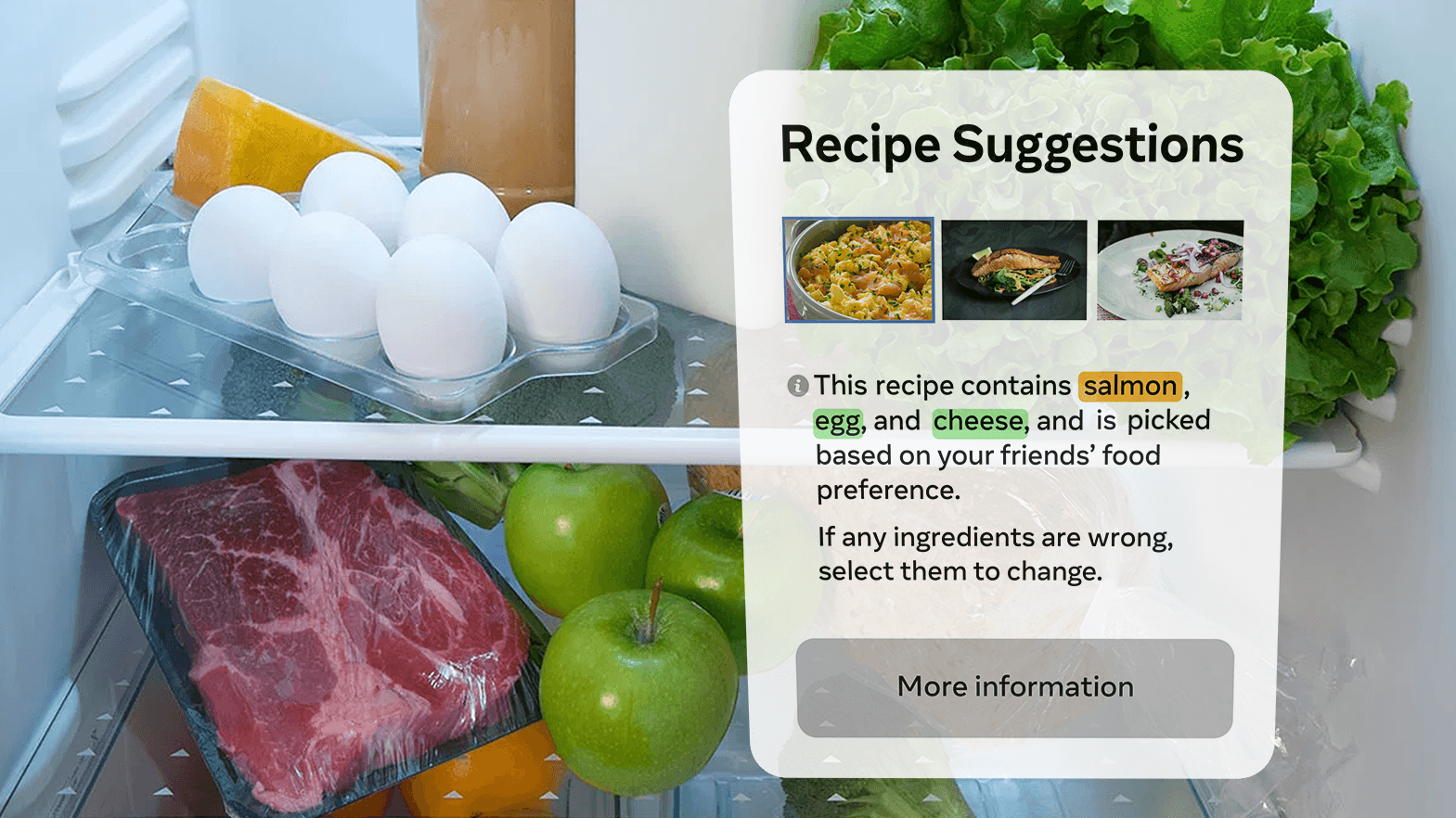} & \includegraphics[width=7.3cm]{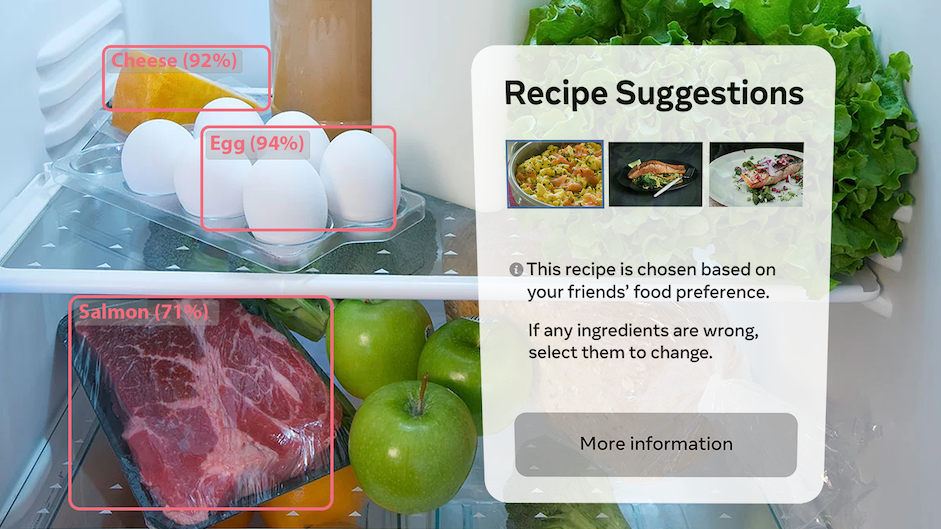}\\
\multicolumn{2}{c|}{\makecell{\textbf{Designer}}} & \makecell{\textbf{P5, HCI/AR Researcher}} & \makecell{\textbf{P9, Product Designer}} \\
 \hline
\multirow{3}{*}{\makecell[c]{\\\textbf{Platform-}\\\textbf{Agnostic}\\\textbf{Key Factors}}}  & \textbf{System Goal} & \multicolumn{2}{c}{\makecell[c]{\textbf{User Intent Assistance} (to find a good recipe for friends)\\\textbf{Error Management} (to calibrate the user's trust for mid-level recognition confidence)}} \\
 & \textbf{User Goal} & \multicolumn{2}{c}{\makecell[c]{\textbf{Resolve Confusion} (to understand why the recommendations are wrong)}} \\
 & \textbf{User Profile} & \multicolumn{2}{c}{\makecell[c]{\textbf{User Preference}: Meet-lovers (friends); \textbf{AI Literacy}: Expert, high}} \\ \hdashline
\multirow{2}{*}{\makecell[c]{\textbf{AR-Specific}\\\textbf{Key Factors}}} & \textbf{Contextual Info} & \multicolumn{2}{c}{\makecell[c]{\textbf{Location}: Kitchen; \textbf{Time}: Evening; \textbf{Environment}: Various ingredients in the fridge}} \\
 & \textbf{User State} & \multicolumn{2}{c}{\makecell[c]{\textbf{Activity}: Opening the fridge to make dinner; \textbf{Cognitive Load}: Low}} \\ \hline
\multirow{2}{*}{\makecell[c]{\\\\\textbf{XAI Designs}\\\textbf{in AR}: \colorwhen{\textbf{\textit{When}}}}} & \colorwhen{Availability} (\gone) & Always available & Same as P5\\
 & \colorwhen{Delivery} (\gtwo) & Auto-trigger, because both conditions were met given the \textit{System Goal} and \textit{User Goal} & Auto-trigger; Besides, a new tweak to always spotlight ingredients automatically, since it introduced \textit{``ultra-low cognitive cost''} \\ \hdashline
\multirow{3}{*}{\makecell[c]{\\\\\\\textbf{XAI Designs}\\\textbf{in AR}: \colorwhat{\textbf{\textit{What}}}}} & \colorwhat{Content} (\gfour) & Five Types: Input/Output, Why/Why-Not, How-To, Certainty, and How & Same as P5\\
 & \colorwhat{Detail - Concise} (\gfive) & An explanation merging Why, Input, Certainty (color-coding to show ingredient with a mid-level confidence), and How-To (selecting ingredients to change) & An explanation Why and How-To; Besides, Input explanations were shown by spotlighting ingredients, which can be selected and changed (How-To)\\
 & \colorwhat{Detail - Detailed} (\gsix) & A drop down menu of the five types & Same as P5\\ \hdashline
\multirow{3}{*}{\makecell[c]{\\\\\textbf{XAI Designs}\\\textbf{in AR}: \colorhow{\textbf{\textit{How}}}}} & \colorhow{Modality} (\gseven) & Visual modality & Same as P5\\
 & \colorhow{Paradigm - Format} (\geight) & Textual format & Textual format as the primary format; Graphic format (spotlighting boundaries) to denote ingredients\\
 & \colorhow{Paradigm - Pattern} (\gnine) & Explicit pattern, presenting texts in the same window as the recommendations & Explicit pattern for texts (same as P5); Implicit pattern for graphic spotlights\\
\hline \hline
\end{tabular}
}
\vspace{0.1cm}
\caption{Two Design Examples of Case 2: Wrong Recipe Recommendation.}
\label{tab:design_example_case2}
\Description{Two Design Examples of Case 2: Wrong Recipe Recommendation: Mary works in an AI company (high AI literacy) and has friends coming over for dinner, who are beef lovers. She opens the fridge and sees steak. However, the AR glasses mistakenly recognize steak as salmon with mid-level confidence, and recommend a few recipes using salmon. She is confused and wonders how she can correct them.
The examples are from P5 and P9. The comparison again indicates both consistency and variance between two designers' examples.
P5's design is as follows:
When. P5 speculated that the major changes between the two tasks include: the System Goal (User Intent Assistance for finding a good recipe for friends, and Error Management for mid-level confidence), the User Goal (Resolve Confusion), and the User Profile (Mary's high AI literacy, friends' food preference). Both conditions of G3 were fulfilled. Thus P5 chose to deliver explanations automatically.
What. According to the table in the what part of the framework, the three factors led to five categories, including Input/Output, Why/Why-Not, How-To, Certainty, and How. For default explanations, in addition to Why, P5 proposed to color-code the Input to emphasize the ingredient with the mid-level confidence (Certainty), and to add a simple selection-based way to allow Mary to change the salmon (How-To). For detailed explanations, P5 proposed using a drop-down menu to show the five categories.
How. P5 chose to use visual modality, textual format, and explicit pattern to present explanations.
P9's design is as follows:
When. P9 had a similar analysis of the System Goal and User Goal as P5. She further proposed an interesting tweak of delivery: always spotlighting ingredients by showing simple information around them (Input category, G4), since it introduced "ultra-low cognitive cost, thus didn't need to follow G2".
What. P9 proposed the consistent category list following G3, and thus suggested the same detailed explanations design as P5. She decided to present Why and How-To as the default explanations. Moreover, as mentioned in the when part, P9 also proposed to show names and recognition certainty on ingredients as "low-cost" explanations.
How. The three sub-questions in P9's design are closely related. Besides displaying main textual explanations explicitly with recommendations, P9's proposed to use simple graphics in an implicit pattern for low-cost explanations}
\vspace{-0.7cm}
% \vspace{-0.7cm}
\end{table*}
\renewcommand{\arraystretch}{1.0}

\begin{figure}[b]
    \centering
% \begin{minipage}[t]{.49\columnwidth}
    \centering
    \includegraphics[width=1\columnwidth]{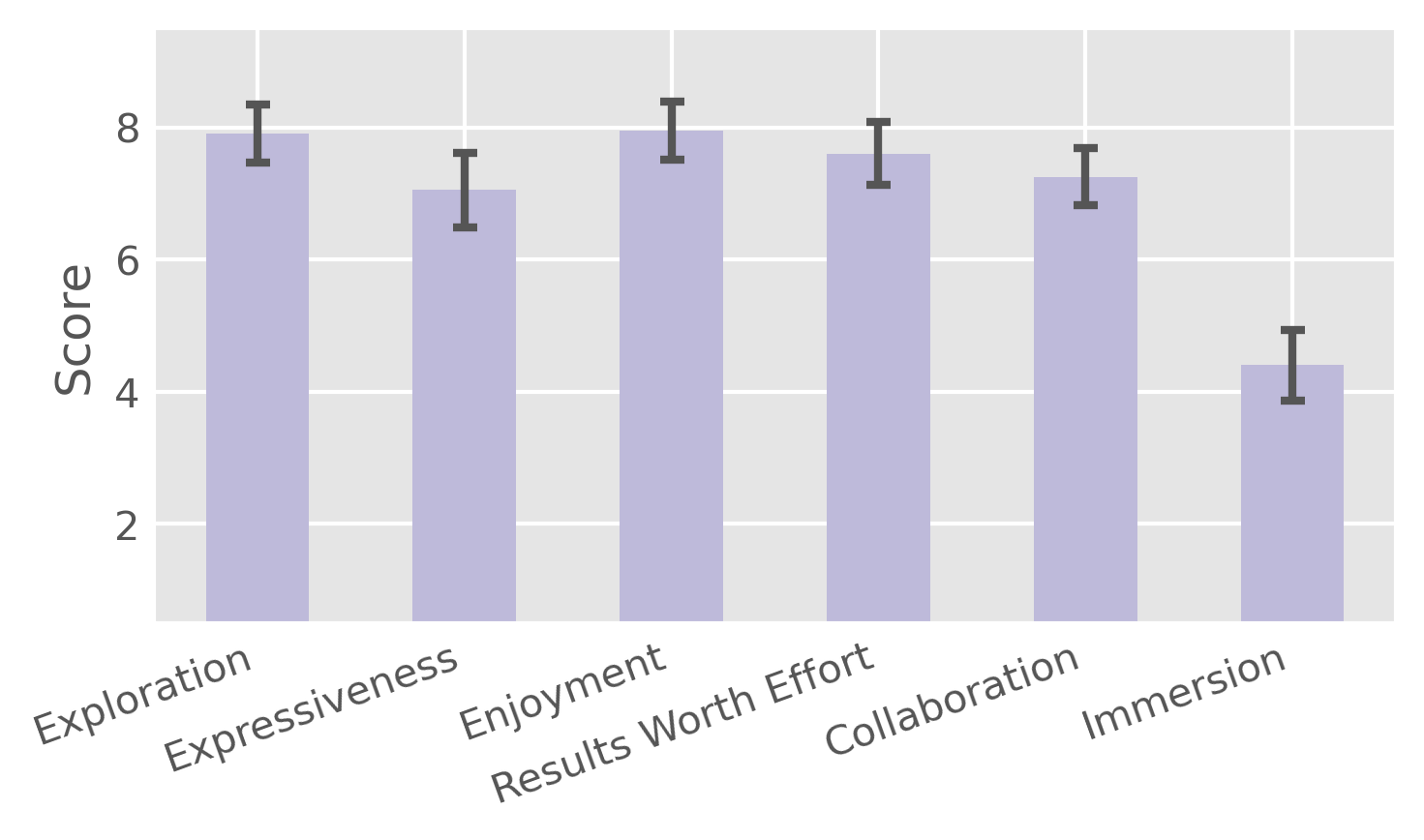}
    \vspace{-0.3cm}
    \label{subfig:design_examples_task1:example2}
    \caption{CSI Scores of Design Workshops in Study 3}
    \label{fig:workshop_utility_csi}
    \vspace{-0.3cm}
    \Description[CSI Scores of Design Workshops in Study 3]{The figure shows a bar chart of the mean ratings from experts on Creativity SUpport Index CSI scores in study 3. The Y-axis is the mean score on a 10-point Likert scale. The X-axis contains six categories, including Exploration, Expressiveness, Enjoyment, Results in Worth Effort, Collaboration, and Immersion. The Exploration category received a mean rating of 7.9, and the enjoyment category received a mean rating of 8.0, which indicates the good utility of XAIR framework in supporting creative designs by experts in the domain. The immersion category received a lower score with a mean of 4.4, which is due to the comprehensiveness of the framework.}
% \end{minipage}
% \hfill
% \begin{minipage}[t]{.49\columnwidth}
%     \centering
%     \includegraphics[width=0.8\columnwidth]{figures/enduser_study_xai.png}
%     \vspace{0pt}
%     \caption{End-user Evaluation of The AR System in Study 4. Users had positive experience in both tasks. Note that tasks were evaluated separately and not meant to compare against each other.}
%     \label{fig:enduser_usability_xai}
% \end{minipage}
\end{figure}

\subsubsection{Feedback Results}
\label{subsub:evaluation:utility:feedback_results}
Participants provided positive feedback about the framework. Eight participants explicitly commented that XAIR was \textit{``useful/helpful''}.
The results of the CSI scores (Fig.~\ref{fig:workshop_utility_csi}) and the SUS scores (74 $\pm$ 6 out of 100, indicating good usability) both illustrate the good utility of XAIR.
Four themes emerged in participants' feedback.

\textit{The Framework as a Useful and Comprehensive Reference.}
Consistent with the feedback from the experts in Study 2 (Sec.~\ref{sub:methodology:expert_workshops}), participants also found that the framework was a valuable handbook. For example, \pquote{4}{This framework is an excellent reference point for people getting started designing XAI experiences... to check if they have missed things} and \pquote{7}{I may not use it for every design decision, but I would refer to it when I want to make sure that I have considered everything.}
The comprehensiveness of XAIR thus helped participants perform a sanity check of their designs.

\textit{Design Opportunity Inspiration.}
Participants also leveraged XAIR to inspire new ideas.
P6's original design did not consider the case where users' hands could be busily holding ingredients. But the modality in the \colorhow{how} part inspired him, \ie
\textit{``The framework reminded me to realize potential alternatives. It inspired me to think about not just one design, but a set of designs.''}
Moreover, participants found that XAIR could help generate baseline designs. \pquote{8}{I could then further customize it for various scenarios.}
The high scores for exploration (7.9$\pm$0.4 out of 10) and expressiveness (7.1$\pm$0.6) on the CSI also support this observation.

\textit{Backing Up Design Intuitions.}
Some participants also found that the guidelines in XAIR could support their intuition.
For instance, P7 did not change her design after using XAIR, but was very excited to see the alignment, \eg \textit{``Sometimes I am not sure whether my design intuition is right. It feels great that the framework can support it.''}
This could be part of the reason for the positive enjoyment score on the CSI (8.0$\pm$0.4).

\textit{Time to Learn The Framework.}
Participants also commented that XAIR incorporates a lot of information and that they needed time to digest it, \eg \pquote{10}{I need to go back and forth between the visual diagrams} and \pquote{4}{the table \text{[in Fig.~\ref{fig:overview_what}]} is useful but also pretty complex}.
This may explain the relatively low immersion score (4.4$\pm$0.5) on the CSI. Moreover, six participants Agreed or Strongly Agreed in response to the question \textit{``Need to learn a lot...''} on the SUS.
On the one hand, this shows XAIR's comprehensiveness (covering multiple research domains), whereas on the other hand, this illuminates future directions to convert XAIR into a design tool.

% \textit{Potential of An Automatic Framework.}
% Interestingly, several participants asked about the potential of using XAIR as an automatic toolkit. 
% For example, P3 was thinking aloud when using XAIR in the study, \textit{``If this framework is described as an algorithm, the five key factors can be viewed as the input of the algorithm... and the output is the design of the three questions.''}
% We elaborate more on this topic in the discussion section.

\subsection{Study 4: Intelligent AR System Evaluation}
\label{sub:evaluation:effectiveness}
To demonstrate XAIR's effectiveness, we show that the designers' proposals using XAIR could achieve a positive XAI user experience in AR for end-users.
Based on the designs proposed in Study 3, we took one example from each case and implemented a real-time intelligent AR system. We then evaluated the system's usability.

\subsubsection{System Implementation}
We selected one reliable recipe recommendation example from the left of Tab.~\ref{tab:design_example_case1} and one wrong recipe recommendation example from the left of Tab.~\ref{tab:design_example_case2}. We then instantiated the examples by implementing a real-time system on a Microsoft Hololens V2.
The system had three major modules: a recognition module, a recommendation module, and an interface module.

For ingredient recognition, we trained a vision-based object detection model that was a variant of the Vision Transformer from CLIP~\cite{radford2021learning} on the LVIS~\cite{gupta2019lvis} and Objects365~\cite{shao2019objects365} datasets.
We then added ImageNet22k and performed weakly-supervised training with both box and image level annotations~\cite{zhou2022detecting}.
The top 50 ingredient-related classes from LVIS were retained, with an average F1 score of 81.1\%.
The model was run on Hololens' egocentric camera stream at 5 FPS to recognize ingredients.
The model was used in Case 1, while in Case 2, misrecognition (\ie recognizing steak as salmon) was manually inserted to create the designed experience.

For recipe recommendation, the Spoonacular Food API~\cite{spoonacular} was used to obtain potential recipes given a set of ingredients. We then implemented an algorithm to rank the recipes based on user preference and recommend the top recipes (\eg if a user prefers food that is fast to prepare, the recipes are sorted based on the cooking time). For the explanations, we developed a template-based explanation generation technique~\cite{zhang_explainable_2020} to cover different 
types.

Finally, the interface followed the designs in Tab.~\ref{tab:design_example_case1} and Tab.~\ref{tab:design_example_case2}.
Clicking on one recipe's image would show the detailed instructions. An icon button under each recipe could be triggered to present short default explanations, followed by another button to display detailed explanations as a list of content types.

\begin{figure}
    \vspace{-0.1cm}
    \centering
    \begin{subfigure}[b]{1\columnwidth}
    \centering
    \includegraphics[width=0.7\columnwidth]{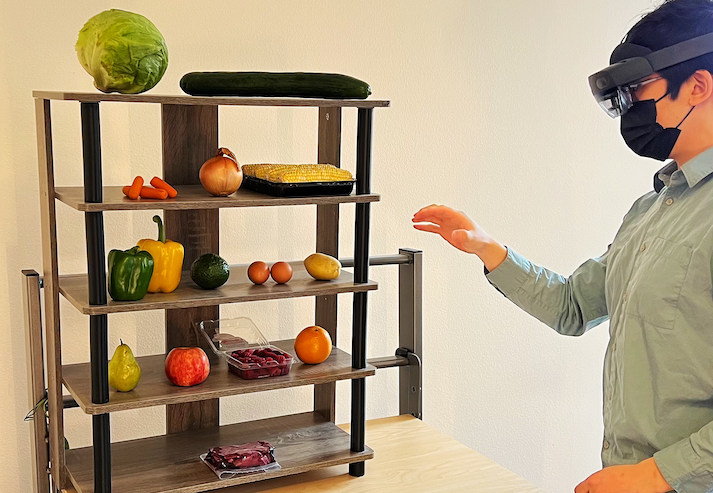}
    % \vspace{0.3cm}
    \caption{Evaluation Setup}
    \label{subfig:study_end_user_eval:setup}
    \end{subfigure}
    \begin{subfigure}[b]{1\columnwidth}
    \centering
    \includegraphics[width=1\columnwidth]{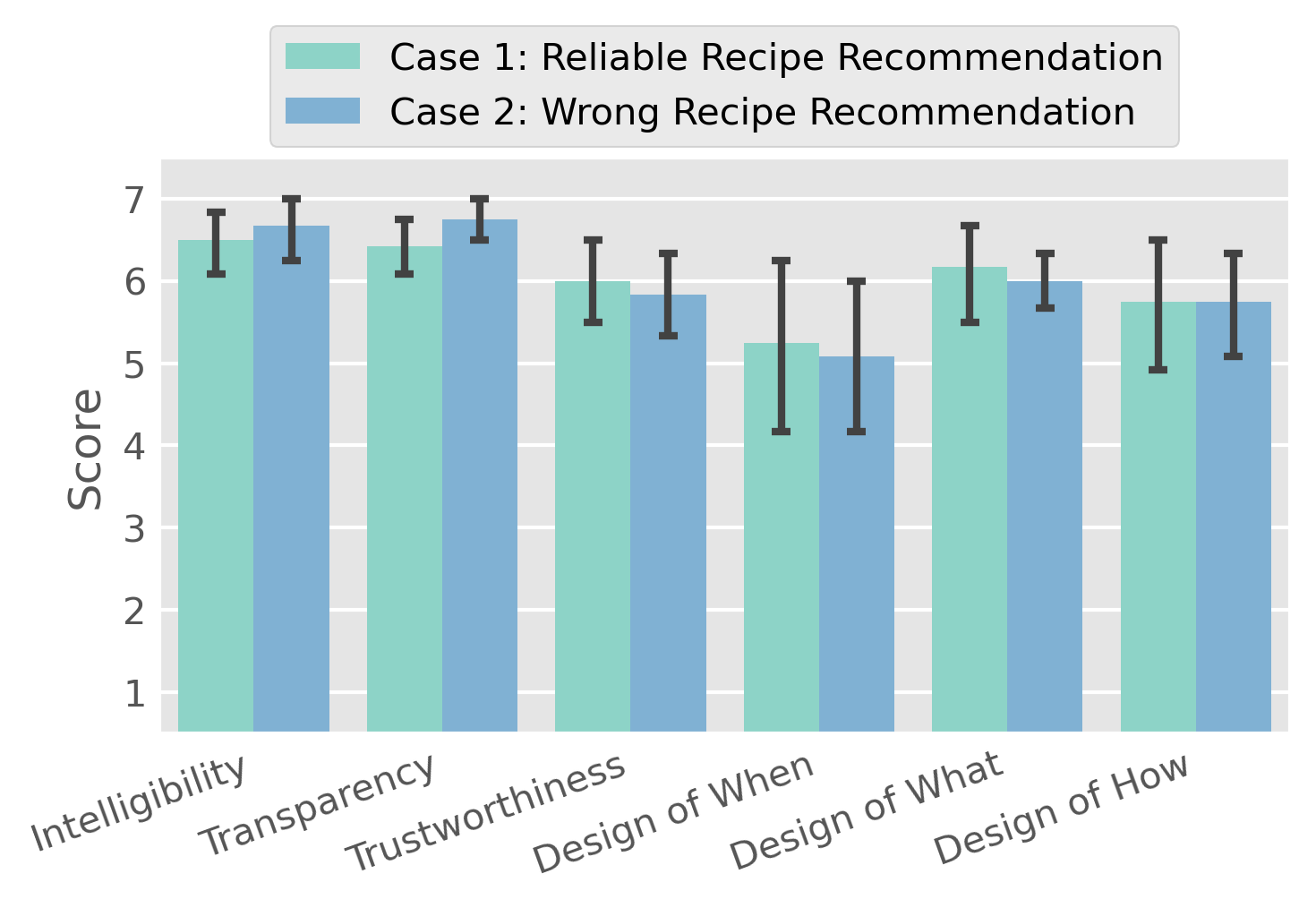}
    \caption{Evaluation Scores}
    \label{subfig:study_end_user_eval:scores}
    \end{subfigure}
    \vspace{-0.6cm}
    \caption{
    \review{End-user Evaluation of The AR System in Study 4. (a) Study Setup. (b) Evaluation Scores. Users had positive experience in both tasks. Note that tasks were evaluated separately and not meant to compare against each other.}
    }
    \label{fig:study_end_user_eval}
    \Description[End-user Evaluation of The AR System in Study 4.]{(a) The figure shows the study setup. A man was wearing a Hololens, looking at a shelf, with his hand reaching toward the shelf. There are a lot of vegetables and fruits on the shelf.
    (b) The figure shows a bar chart illustrating the result of our evaluation of the AR system derived from the XAIR framework in section 7.2. The Y-axis is the mean ratings from the end users on a 7-point Likert scale. The X-axis is the seven evaluation categories: intelligibility, transparency, trustworthiness, design of when, design of what, and design of how. In both task 1, reliable recipe recommendation, and task 2, recipe recommendation with error, our system obtained positive ratings from end users in terms of all categories. Note that among these categories, intelligibility, transparency, and trustworthiness obtained especially high ratings, with a mean of 6 and above. The figure demonstrates the strong usability of the XAIR framework in providing satisfying XAI experiences in AR, as described in section 7.4.1.
}
    \vspace{-0.4cm}
\end{figure}

\subsubsection{Participants and Apparatus}
Twelve participants (5 Female, 7 Male, Age 32 $\pm$ 3) volunteered to join the study. None of the them had participated in previous studies. The two cases had the same setup (except for the recognition error). \review{We prepared a number of food ingredients on a shelf (including steak, but no salmon) to simulate the opening-a-fridge moment, as shown in Fig.~\ref{subfig:study_end_user_eval:setup}.}

\subsubsection{Design and Procedure}
% In each task, we had the availability of explanation as the main factor. 
Since there is no existing XAI design for AR systems, we compared the design examples with a baseline condition that only presented recommendations without explanations.
Note that for Case 2's baseline condition, participants could still change the output by clicking a button that said ``Doesn't seem right? Click to see the next batch.'' to ensure a fair comparison \footnote{Another baseline could have been to compare against designers' old designs before using XAIR. However, we did not include this baseline since designers already explicitly preferred the new version that they created after using XAIR.}.

We used a within-subject design. Participants started with one case and completed both conditions. They took a break and completed a questionnaire to compare the two conditions. Then, they completed the two conditions in the second case and completed a similar questionnaire. The case order was counterbalanced. The study took about 30 minutes and ended with a brief interview.

The questionnaire contained six questions (1-7 Likert scale) comparing the two conditions. Three were from the XAI literature and measured the explanations' effect on the system's intelligibility, transparency, and trustworthiness. The other three questions asked about participants' preferences towards the design choices of \colorwhen{\textit{when}}, \colorwhat{\textit{what}}, and \colorhow{\textit{how}}\footnote{Since a factorial study design to compare all XAIR design options would involve a large number of conditions (\ie 2 options of \colorwhen{\textit{when}} $\times$ at least 2 options of \colorwhat{\textit{what}} $\times$ 2 options of \colorhow{\textit{how}}), asking participants to undergo several scenarios would be too costly. Order effects would also be hard to counterbalance. So the three questions about when/what/how described other design choices by showing examples and asked about participants' preferences. For instance, in Case 1's \colorwhen{\textit{when}} part, participants rated how much they agreed with the claim ``I prefer to have explanations triggered manually by me, compared to being triggered automatically.'', or vice-versa in Case 2.}.
The SUS was also administered to measure the usability of the system with explanations.

\subsubsection{Results}
Participants strongly preferred the condition with explanations in both cases, especially Case 2, \eg
\pquote{2}{Seeing the explanation automatically when the AR system makes mistakes is very helpful. It lets me know when I should adjust my expectation} and \pquote{9}{the mistake \text{[in Case 2]} is understandable... salmon and steak can have similar colors and shapes. But if I didn't see the explanation, I would be very confused.}
This sentiment was also reflected in participants' high rating of the system's intelligibility, transparency, and trustworthiness with the explanation (Fig.~\ref{subfig:study_end_user_eval:scores}).
Moreover, the AR system received high SUS scores: 86 $\pm$ 3 in Case 1, and 80 $\pm$ 3 in Case 2, both indicating excellent usability of the system.
Participants also liked the design of the system, which was supported by the positive ratings for the when/what/how questions (see Fig.~\ref{subfig:study_end_user_eval:scores}).
\review{
These results demonstrated that compared to the baseline, XAI design using XAIR can effectively improve the transparency and trustworthiness of AR systems for end-users.
}
\section{Discussion}
\label{sec:discussion}
XAIR defines the problem space structure of XAI design in AR and details the relationship that exists between the factors and the problem space. By highlighting the key factors that designers need to consider and providing a set of design guidelines for XAI in AR, XAIR not only serves as a reference for researchers, but also assists designers by helping them propose more effective XAI designs in AR scenarios.
The two evaluation studies in Sec.~\ref{sec:evaluation} illustrated that XAIR can inspire designers with more design opportunities and lead to transparent and trustworthy AR systems.
\review{In this section, we discuss how researchers and designers can apply XAIR, as well as potential future directions of the framework inspired by our studies. We also summarize the limitations of this work.}

\review{
\subsection{Applying XAIR to XAI Design for AR}
\label{sub:discussion:procedure}
Researchers and designers can make use of XAIR in their XAI design for AR scenarios by initially using their intuition to propose an initial set of designs.
Then, they can follow the framework to identify five key factors: \textit{User State}, \textit{Contextual Information}, \textit{System Goal}, \textit{User Goal}, and \textit{User Profile}. The example scenarios in Sec.~\ref{sec:applications} and Sec.~\ref{sec:evaluation} indicate how these factors can be specified.
Based on these factors, they would then work through the eight guidelines of \colorwhen{\textit{when}}, \colorwhat{\textit{what}}, and \colorhow{\textit{how}}, using Fig.~\ref{fig:overview_when}-Fig.~\ref{fig:overview_how} to inspect their initial design and make modifications if there is anything inappropriate or missing. Low-fidelity storyboards or prototypes of the designs can be tested via small-scale end-user evaluation studies. This would be an iterative process.
In the future, when sensing and AI technologies are more advanced, it is promising that the procedures of identifying factors and checking guidelines could be automated.
}

\subsection{Towards An Automatic Design Recommendation Toolkit}
\label{sub:discussion:toolkit}
In Study 3, more than one user mentioned the possibility of converting the framework into an automatic toolkit. 
For example, P3 was thinking aloud when using XAIR in the study, \textit{``If this framework is described as an algorithm, the five key factors can be viewed as the input of the algorithm... and the output is the design of the three questions.''}
There are a few decision-making steps in the current framework that involve human intelligence. For example, when designing the default explanations in \colorwhat{what - detail}, designers need to consider users' priority under a given context to determine which explanation content type to highlight. When picking the appropriate visual \colorhow{paradigm}, designers need to determine whether the explanation content is more appropriate in a textual or graphical format, as well as whether the content can be naturally embedded within the environment.
Assuming future intelligent models can assist with these decisions, XAIR could be transformed into a design recommendation tool that could enable designers and researchers to experiment with a set of \textit{User State}, \textit{Contexts}, \textit{System/User Goals}, and so on. 
This could achieve a more advanced version of XAIR, where XAIR are fully automated as an end-to-end model: determining the optimal XAI experience by inferring the five key factors in real time.
This is an appealing direction. However, although factors such as \textit{Context} and \textit{System Goal} are easier to predict with a system, the inference of \textit{User State/Goal} is still at an early research stage~\cite{arguel2017inside,duchowski_index_2018,huang2018predicting}. Moreover, extensive research is needed to validate the adequacy and comprehensiveness of the end-to-end algorithm.
This also introduces the challenge of nested explanations in XAIR (\ie explaining explanations)~\cite{mittelstadt2019explaining}, which calls for further study.

\subsection{The Customized Configuration of XAI Experiences in AR}
\label{sub:discussion:config}
The experts in Study 2 and the designers in Study 3 brought up the need for end-user to control XAI experiences in AR, \eg
\pquote{12, Study 2}{XAIR can provide a set of default design solutions, and users could further customize the system} and
\pquote{8, Study 3}{I personally agree with the guidelines, but I can also imagine some users may want different design options. So there should be some way that allows them to select when/what/how... For example, a user may want the interface to be in an explicit dialogue window all the time \text{[related to \colorhow{how}]}. We should support this.}
This need for control suggests that to achieve a personalized AR system, designers should provide users with methods to configure their system, so that they can set up specific design choices to customize their XAI experience.
Such personalization capabilities may also be used to support people with accessibility needs (also mentioned by P2 in Study 3), \eg visually impaired users can choose to always use the audio \colorhow{modality}.

% ``To this date, there is no agreement in prior work on whether to include all details of system logic in explanation interfaces. Moreover, there might not be a universal answer to this question, but it might rather depend on the product domain, user groups, and the context of use.''~\cite{eiband_bringing_2018}

\subsection{User-in-The-Loop and Co-Learning}
\label{sub:discussion:co_learning}
During the iterative expert workshops (Study 2, Sec.~\ref{sub:methodology:expert_workshops}), experts mentioned an interesting long-term co-learning process between the AR system and a user.
On the one hand, based on a user's reactions to AI outcomes and explanations, a system can learn from the data and adapt to the user. Ideally, as the AR system better understands the user, the AI models would be more accurate, thus reducing the need for mistake-related explanations (\eg cases where \textit{System Goal} as Error Management).
On the other hand, the user is also learning from the system. \pquote{4, Study 2}{Users' understanding of the system and AI literacy may change as they learn from explanations}. This may also affect the user's need for explanations. For example, the user may have less confusion (\textit{User Goal} as Resolving Surprise/Confusion) as they become more familiar with the system. Meanwhile, they may become more interested in exploring additional explanation types (\textit{User Goal} as Informativeness).
Such a long-term and co-learning process is an interesting research question worth more exploration.

\subsection{Limitations}
\label{sub:discussion:limitations}
There are a few limitations to this research.
First, although we highlighted promising technical paths within the framework in Sec.~\ref{sec:framework}, XAIR does not involve specific AR techniques. The real-time AR system in Study 4 implemented the ingredient recognition and recipe recommendation modules, but the detection of user state/goal was omitted.
\review{
Second, our studies might have some intrinsic biases. For example, Study 1 only involved AR recommendation cases. Since everyday AR HMDs are still not widely adopted in daily life, we grouped 500+ participants only based on AI experience instead of AR experience. The experts and designers of our studies were all employees of a technology company. Study 4 only evaluated two specific proposals from designers. Moreover, as there is no previous XAI design in AR, we were only able to compare our XAIR-based system against a baseline without explanation.
}
Third, other than when, what, and how, there could be more aspects in the problem space, \eg who and where to explain. Moreover, XAIR mainly focuses on non-expert end-users. Other potential users, such as developers or domain experts, were not included.
% There are other AR-specific information presentation frameworks that can be further incorporated~\cite{muller2016taxonomy,luo2022should}.
The scope of the five key factors may also not be comprehensive. For example, we do not consider user trust in AI, which is a part of \textit{User Profile} that may be dynamic along with user-system interaction.
These could limited the generalizability of our framework, but also suggests a few potential future work directions to expand and enhance XAIR.
\section{Conclusion}
\label{sec:conclusion}
In this paper, we propose XAIR, a framework to guide XAI design in AR.
Based on a literature review of multiple domains, we identified the problem space using three main questions, \ie when to explain, what to explain, and how to explain.
We combined the results from a large-scale survey with over 500 end-users (Study 1) and iterative workshops with 12 experts (Study 2) to develop XAIR and a set of eight design guidelines.
Using our framework, we walked through example XAI designs in two everyday AR scenarios.
To evaluate XAIR's utility, we conducted a study with 10 designers (Study 3).
The study revealed that designers found XAIR to be a helpful, comprehensive reference that could inspire new design thoughts and provide a backup of designer intuitions.
Moreover, to demonstrate the effectiveness of XAIR, we instantiated two design examples in a real-time AR system and conducted another user study with 12 end-users (Study 4). The results indicated excellent usability of the AR system.
XAIR can thus help future designers and researchers achieve effective XAI designs in AR and help them explore new design opportunities.

\bibliographystyle{ACM-Reference-Format}
% \bibliography{
% bib/Augmented_Reality,
% bib/Adaptive_UI,
% bib/Behavior_Intervention,
% bib/Human-AI_Interaction,
% bib/Interpretable_ML,
% bib/Modeling_Behavior-General,
% bib/Orson_Publication,
% bib/other
% }

\bibliography{bib/merged.bib}

\balance
\clearpage
\onecolumn

\appendix

\section{Appendix A: Earlier Versions of XAIR}
\label{sec:appendix:earlier_frameworks}

\begin{figure*}[!h]
    \centering
    \begin{subfigure}[bpth]{0.75\textwidth}
    \includegraphics[width=0.8\columnwidth]{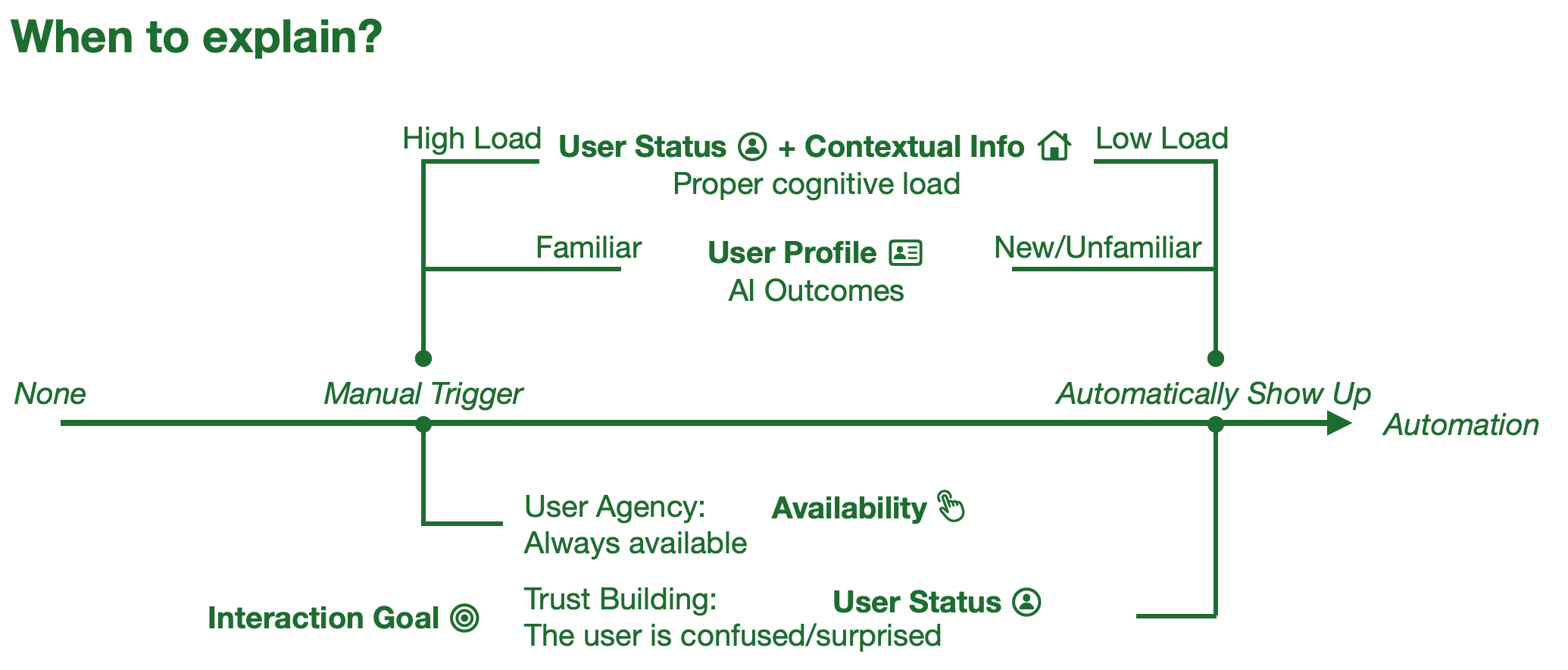}
    % \caption{Version 1 before 1st Iterative Expert Workshop}
    % \label{subfig:when_old:v1}
    \end{subfigure}
    \begin{subfigure}[bpth]{0.75\textwidth}
    \includegraphics[width=0.73\columnwidth]{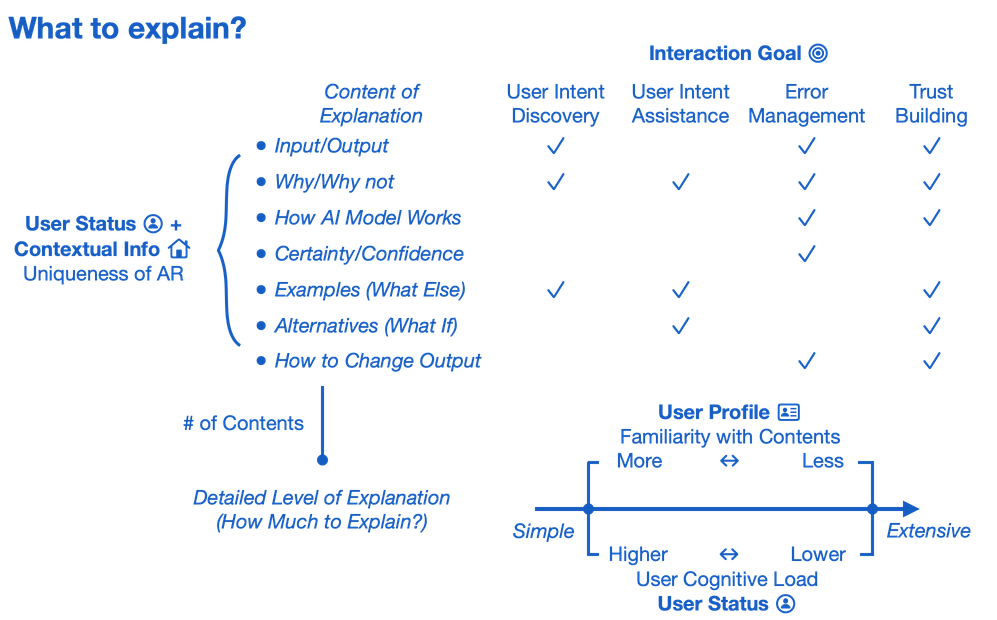}
    % \caption{Version 2 before 2nd Iterative Expert Workshop}
    % \label{subfig:what_old:v2}
    \end{subfigure}
    \begin{subfigure}[bpth]{0.75\textwidth}
    \centering
    \includegraphics[width=1\columnwidth]{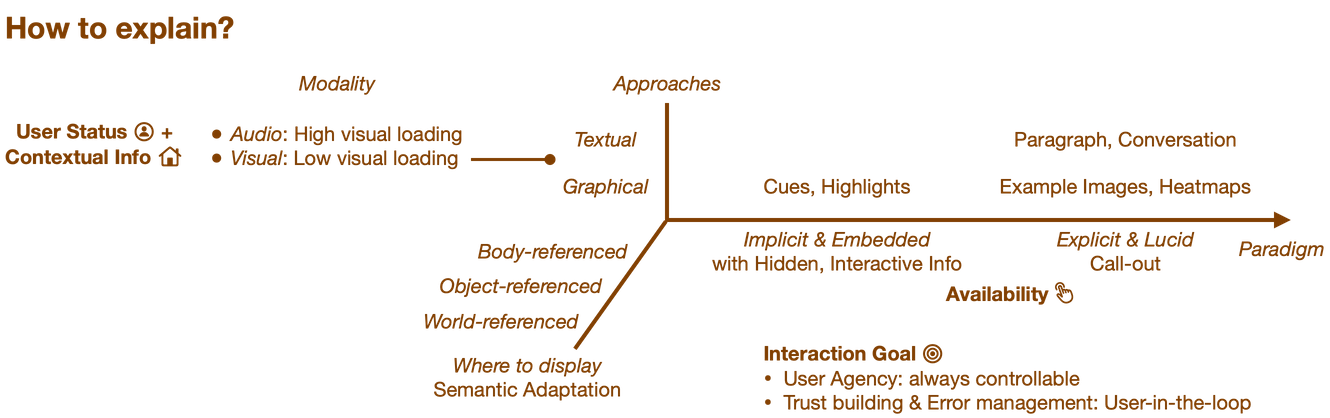}
    % \caption{Version 1 before 1st Iterative Expert Workshop}
    % \label{subfig:how_old:v1}
    \end{subfigure}
    \caption{Version 1 before The 1rd Iterative Expert Workshop (Study 2)}
    \label{fig:old_framework_v1}
\end{figure*}

\begin{figure*}[p!]
    \centering
    \begin{subfigure}[bpth]{0.87\textwidth}
    \includegraphics[width=0.8\columnwidth]{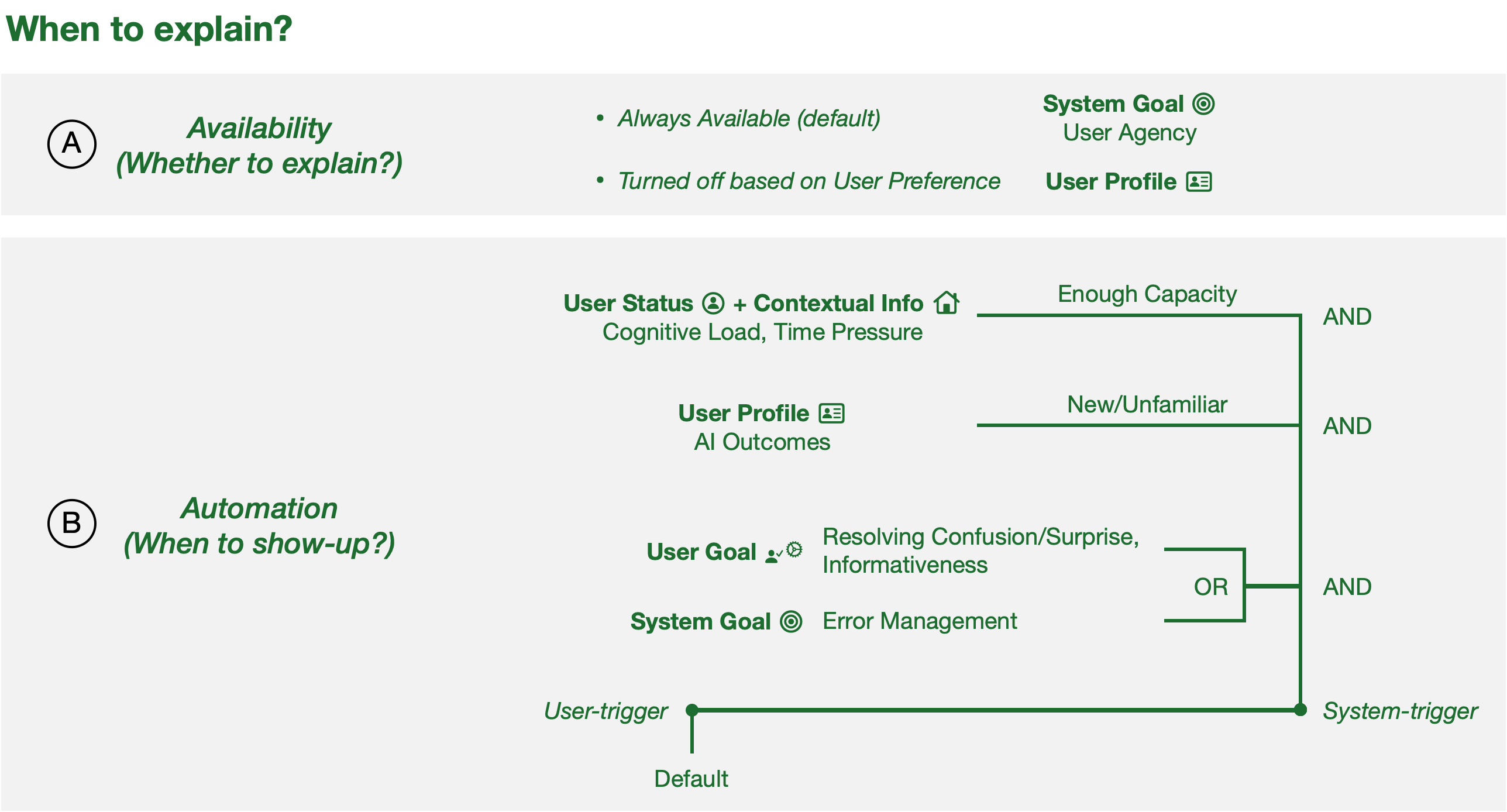}
    % \caption{Version 2 before 2nd Iterative Expert Workshop}
    % \label{subfig:when_old:v2}
    \end{subfigure}
    \begin{subfigure}[bpth]{0.87\textwidth}
    \centering
    \includegraphics[width=1\columnwidth]{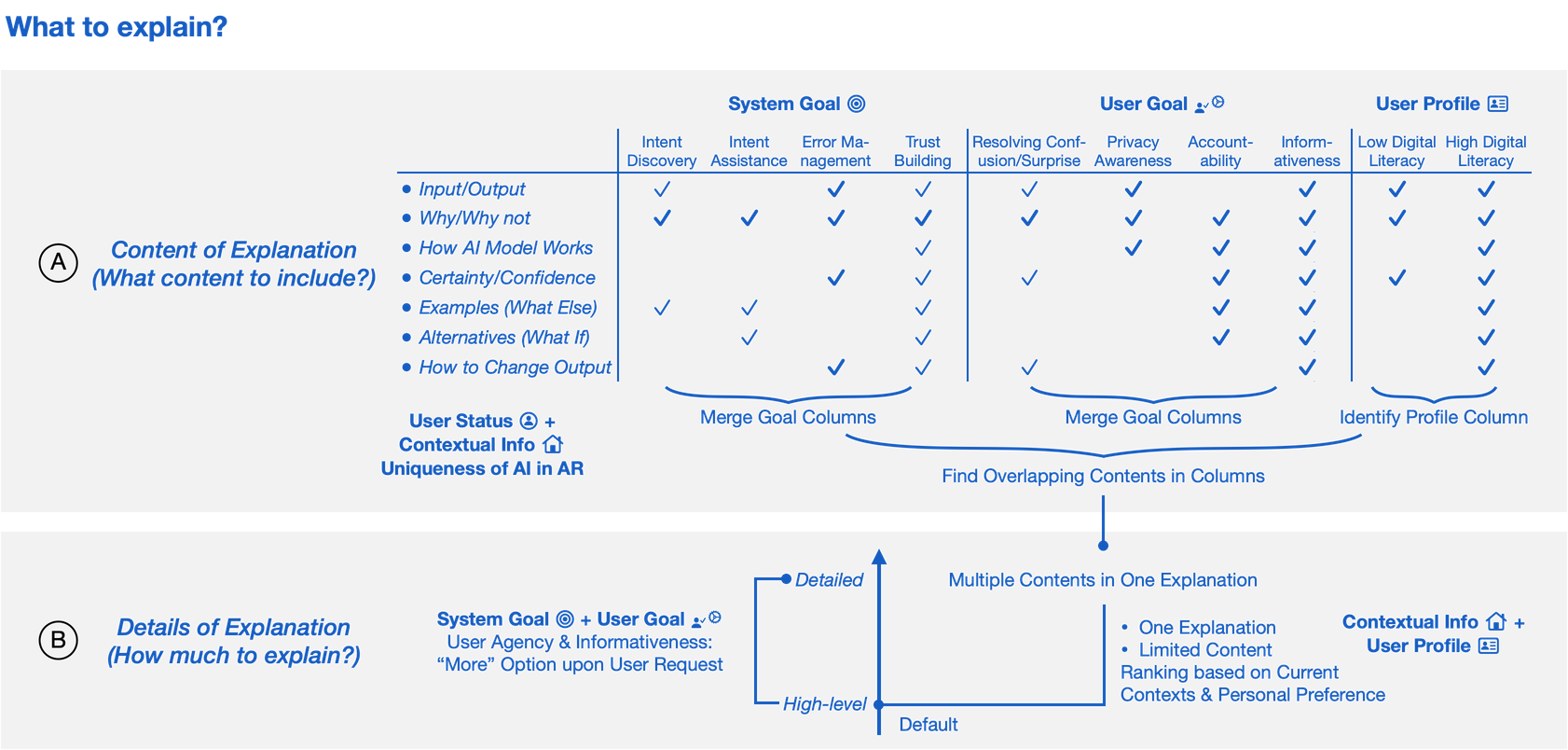}
    % \caption{Version 1 before 1st Iterative Expert Workshop}
    % \label{subfig:what_old:v1}
    \end{subfigure}
    \begin{subfigure}[bpth]{0.87\textwidth}
    \includegraphics[width=0.83\columnwidth]{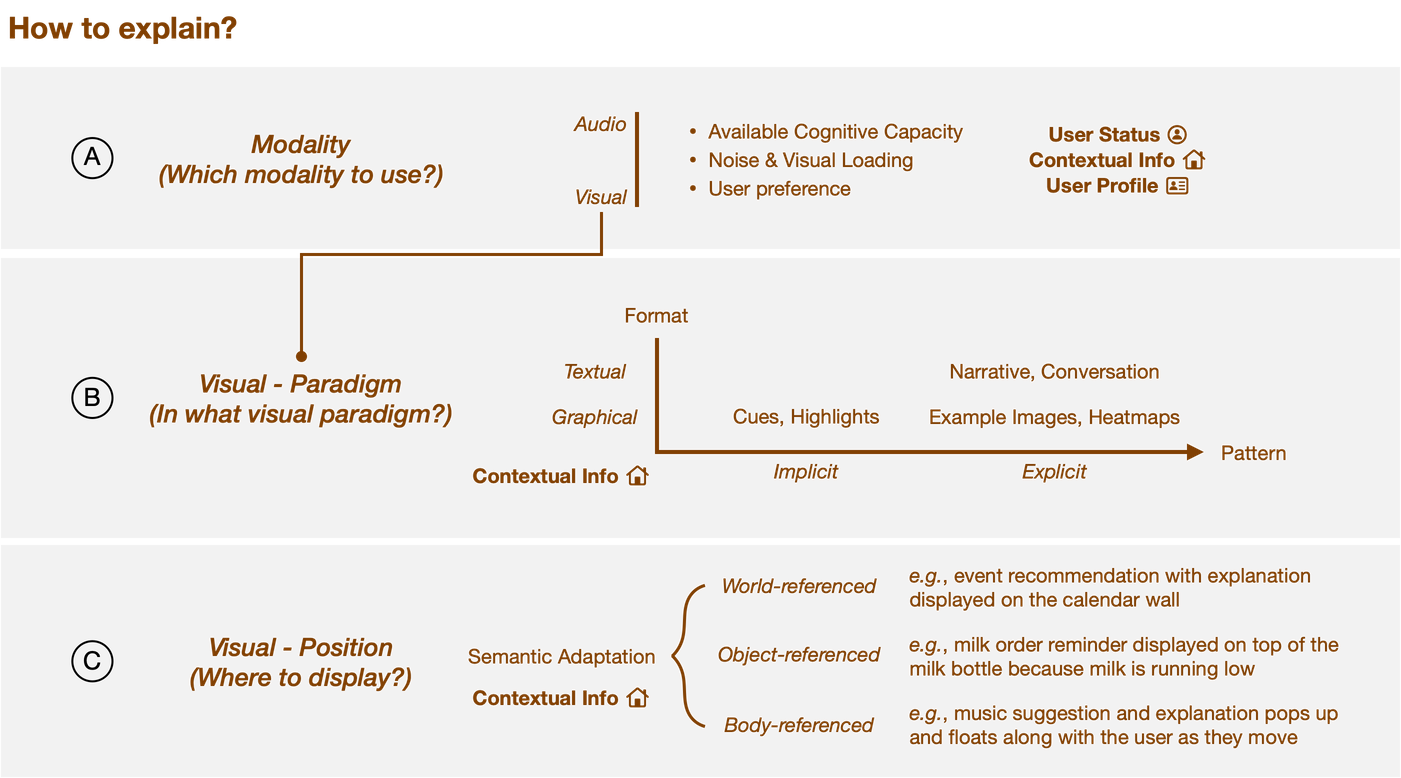}
    % \caption{Version 2 before 2nd Iterative Expert Workshop}
    % \label{subfig:how_old:v2}
    \end{subfigure}
    \caption{Version 2 before The 2rd Iterative Expert Workshop (Study 2). Main updates from Version 1: (\colorwhen{\textbf{\textit{When}}}) Add dimensions and update the connection between the key factors and the dimensions.
    (\colorwhat{\textbf{\textit{What}}}) Add \textit{User Goal} and \textit{User Profile} into the content type table.
    (\colorhow{\textbf{\textit{How}}}) Reorganize according to the dimensions and simply the structure.}
    \label{fig:old_framework_v2}
\end{figure*}

\begin{figure*}[p!]
    \centering
    \begin{subfigure}[bpth]{0.7\textwidth}
    \includegraphics[width=0.85\columnwidth]{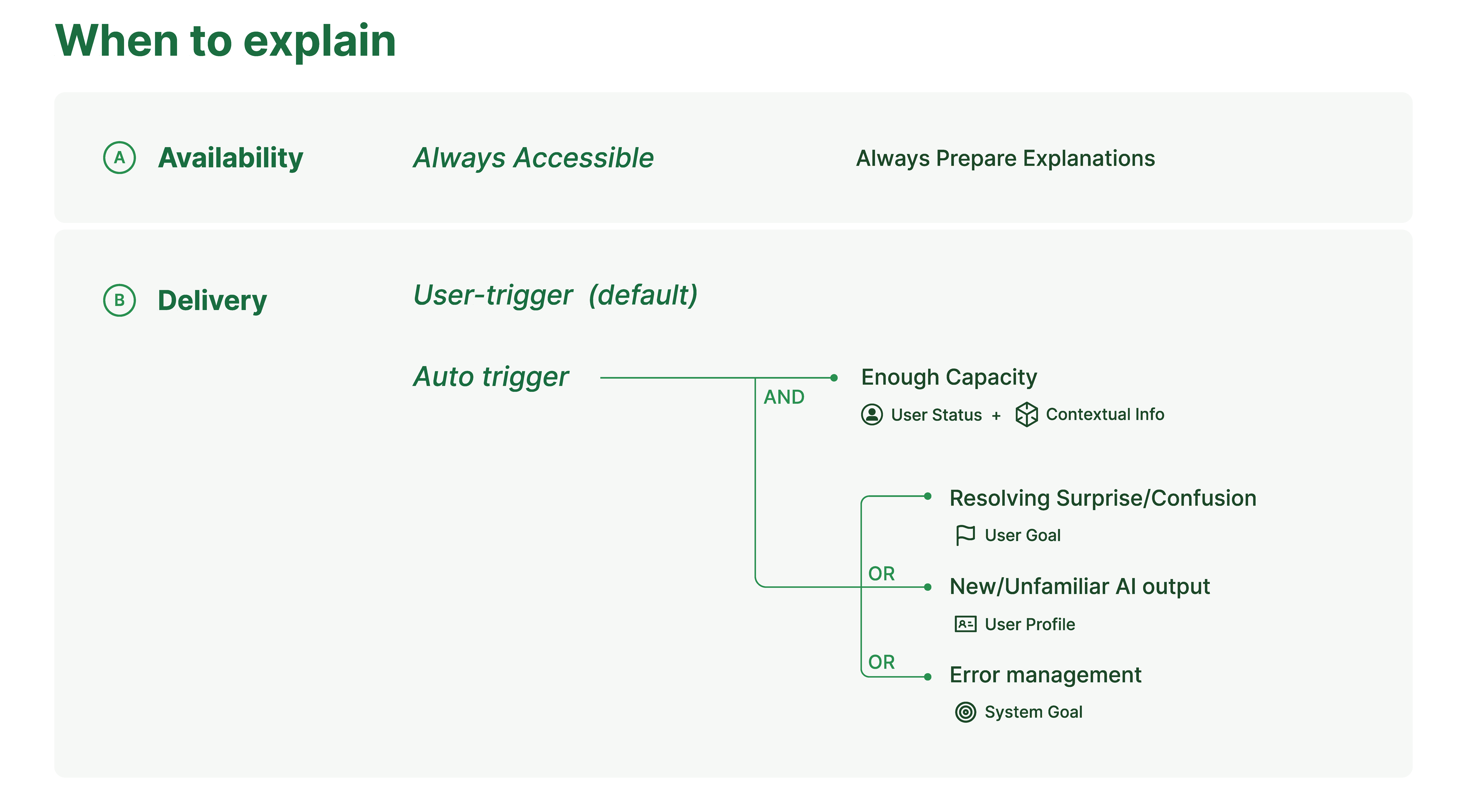}
    % \caption{Version 3 before 3rd Iterative Expert Workshop}
    % \label{subfig:when_old:v3}
    \end{subfigure}
    \begin{subfigure}[bpth]{0.7\textwidth}
    \centering
    \includegraphics[width=1\columnwidth]{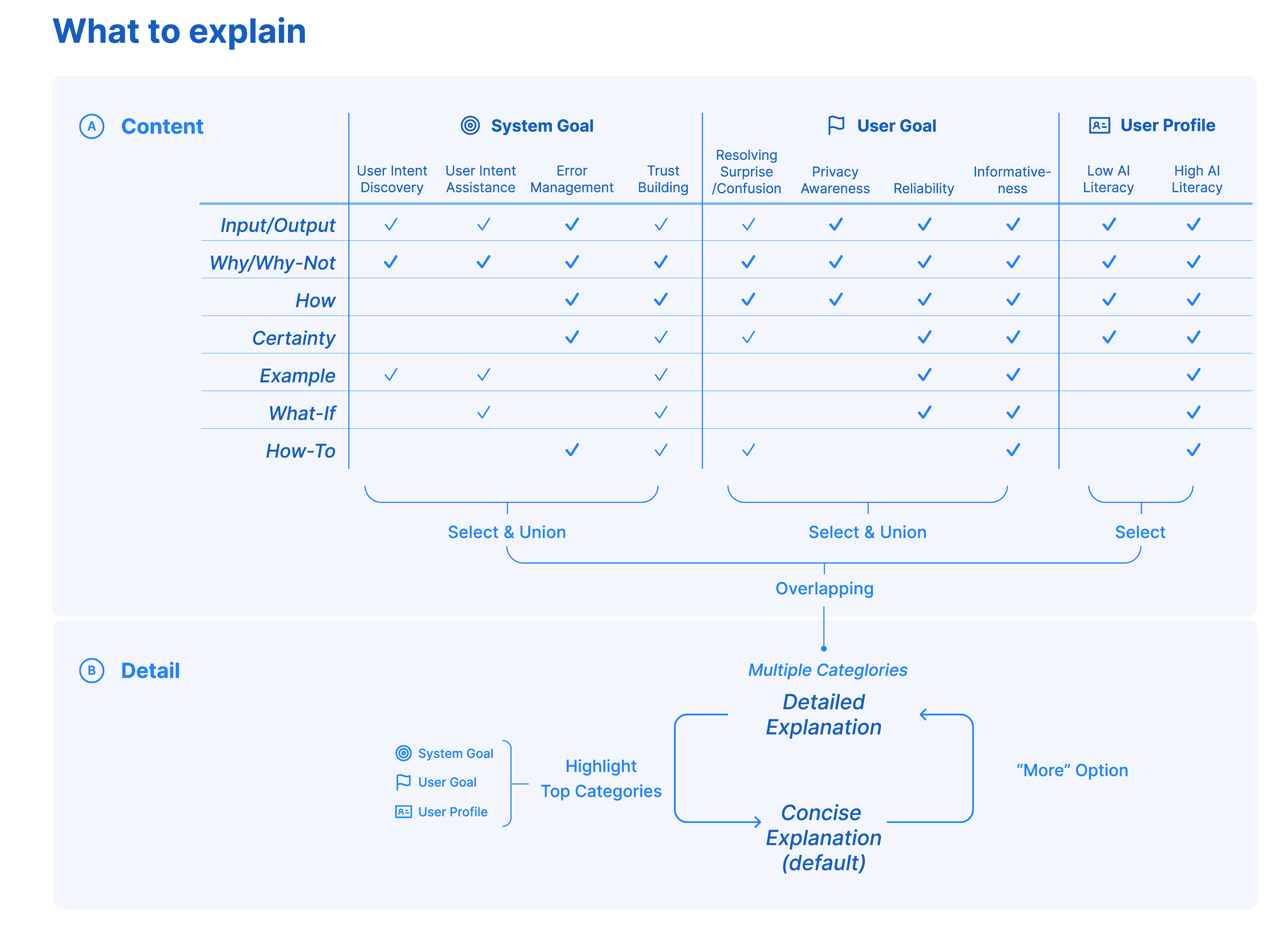}
    % \caption{Version 3 before 3rd Iterative Expert Workshop}
    % \label{subfig:what_old:v3}
    \end{subfigure}
    \begin{subfigure}[bpth]{0.7\textwidth}
    \includegraphics[width=0.85\columnwidth]{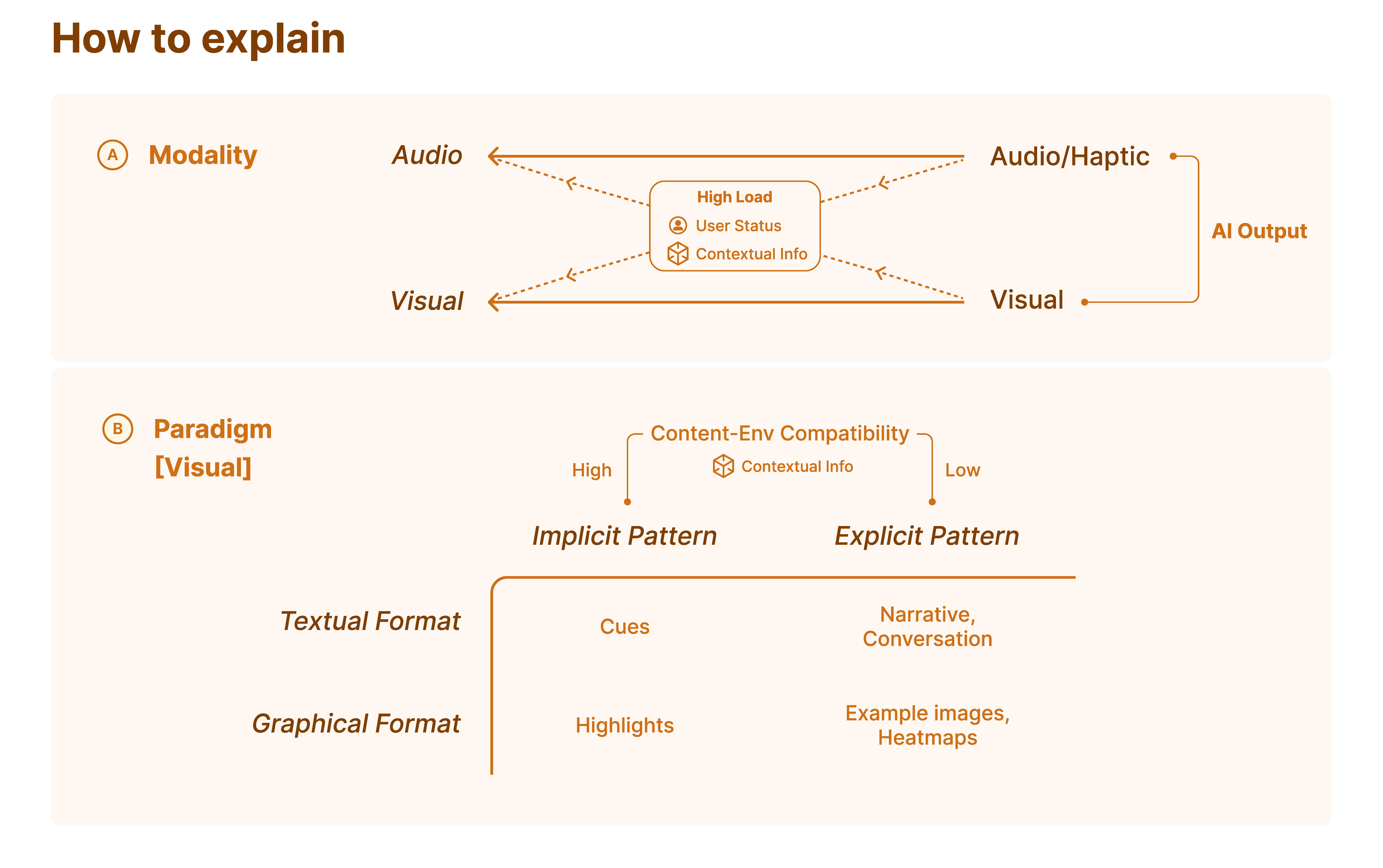}
    % \caption{Version 3 before 3rd Iterative Expert Workshop}
    % \label{subfig:how_old:v3}
    \end{subfigure}
    \caption{Version 3 before The 3rd Iterative Expert Workshop (Study 2). Main updates from Version 2: (\colorwhen{\textbf{\textit{When}}}) Update the connection between the key factors and the dimensions.
    (\colorwhat{\textbf{\textit{What}}}) Simplify the structure and provide default option as guidance.
    (\colorhow{\textbf{\textit{How}}}) Remove the ``location'' dimension and improve the visual design.}
    \label{fig:old_framework_v3}
\end{figure*}

We provide the initial versions of the framework that were used at the beginning of the three iterative workshops (from Fig.~\ref{fig:old_framework_v1} to Fig.~\ref{fig:old_framework_v3}). These examples show how XAIR improved throughout the workshops.

\newpage

\section{Appendix B: Details of Application Scenarios}
\label{sec:appendix:details_scenarios}

\subsection{Explanation Details for The Two Applications}
\label{sub:appendix:details_scenarios:more_details_of_application}

We present a more structured summary of the two scenarios in Sec.~\ref{sec:applications}, together with examples of all explanation content type (Tab.~\ref{tab:application_scenario_1}).

\renewcommand{\arraystretch}{1.4}
\begin{table}[hbtp]
\centering
\resizebox{1\textwidth}{!}{
\begin{tabular}{c|c|m{7cm}|m{7cm}}
\hline \hline
\multicolumn{2}{c|}{\makecell{\text{ }}} & \includegraphics[width=7cm]{figures/application_scenario1.png} & \includegraphics[width=7cm]{figures/application_scenario2.png}\\
\multicolumn{2}{c|}{\makecell{\textbf{Scenario Info}}} & \makecell{\textbf{Scenario 1: Route Suggestion}} & \makecell{\textbf{Scenario 2: Plant Fertilization Reminder}} \\
 \hline
\multicolumn{2}{c|}{\makecell{\textbf{Scenario}}} & Nancy (AI expert, high AI literacy) is jogging in the morning on a quiet trail. Since it is the cherry-blossom season and Nancy loves cherries, her AR glasses display a map beside her and recommend a detour. Nancy is surprised since this route is different from her regular one, but she is happy to explore it. She is also curious to know the reason this new route was recommended. & Sarah (general end-user, low AI literacy) was chatting with her neighbor about gardening. After she returned home and sat on the sofa, her AR glasses recommended instructions about plant fertilization by showing a care icon on the plant. Sarah is concerned about technology invading her privacy, and wants to know the reason behind the recommendation. \\ \hline
\multirow{3}{*}{\makecell[c]{\\\textbf{Platform-}\\\textbf{Agnostic}\\\textbf{Key Factors}}}  & \textbf{System Goal} & \textbf{User Intent Discovery (new route)} & \textbf{Trust Building (clarification)} \\
 & \textbf{User Goal} & \textbf{Resolve Surprise} & \textbf{Privacy Awareness} \\
 & \textbf{User Profile} & \makecell[l]{\textbf{User Preference}: Cherry blossom tree lover;\\\textbf{History}: Regular jogging in the morning;\\\textbf{AI Literacy}: Expert, high} & \makecell[l]{\textbf{User Preference}: Plant enthusiast;\\\textbf{History}: Did not care of the plant for a while;\\\textbf{AI Literacy}: General end-user, low} \\ \hdashline
\multirow{2}{*}{\makecell[c]{\\\textbf{AR-Specific}\\\textbf{Key Factors}}} & \textbf{Contextual Info} & \makecell[l]{\textbf{Location}: Outdoor; \textbf{Time}: Morning;\\\textbf{Environment}: Trails, streets} & \makecell[l]{\textbf{Location}: Home; \textbf{Time}: Afternoon;\\\textbf{Environment}: Living room furniture, the plant} \\
 & \textbf{User State} & \makecell[l]{\textbf{Activity}: Jogging;\\\textbf{Cognitive Load}: Low} & \makecell[l]{\textbf{Activity}: Sitting on the sofa;\\\textbf{Cognitive Load}: Low} \\ \hline
\multirow{7}{*}{\makecell[c]{\\\\\\\textbf{Explanation}\\ \textbf{Content Type}\\\textbf{Examples}}} & \textbf{Input/Output} & This route is recommended based on seasons, your routine and preferences. & The system checks the plant’s current status by visually scanning the plant. \\
 & \textbf{Why/Why-Not} & The route has cherry blossom trees that you can enjoy. The length of the route is appropriate and fits your morning schedule. & The plant has abnormal spots on the leaves, which indicates fungi or bacteria infection. \\
 & \textbf{How} & This algorithm finds and ranks possible routes based your location and other people who share similar preferences to you. & The system checks the plant's visual appearance, then searches online to find ways to cure it.\\
 & \textbf{Certainty} & Match rate between this route’s condition and your preference: 93\% & The chance of the plant having disease is high (94\%). \\
 & \textbf{Example} & These photos captured memories about jogging during cherry blossom season. & These are some images of other plants with similar symptoms. \\
 & \textbf{What-If} & The recommended route will be a shorter one if you jog in the evening. & N/A \\
 & \textbf{How-To} & Disable the “season option” to return to the old route. & N/A \\
\hline \hline
\end{tabular}
}
\caption{Details of The Two Application Scenarios in Sec.~\ref{sec:applications}. ``N/A'' indicates that this particular explanation type is not applicable for this case. The same below.}
\label{tab:application_scenario_1}
\end{table}
\renewcommand{\arraystretch}{1.0}

\newpage

\subsection{Additional Application Scenarios}
\label{sub:appendix:details_scenarios:more_scenarios}

We further applied XAIR to additional everyday AR scenarios to illustrate the practicability of XAIR.
The four scenarios cover extra indoor \& outdoor recommendations (Tab.~\ref{tab:more_scenario_1}), as well as AR-based intelligent instructions and automation aside from recommendations (Tab.~\ref{tab:more_scenario_2}).

\vspace{-0.1cm}
\renewcommand{\arraystretch}{1.6}
\begin{table}[hbtp]
\centering
\resizebox{0.86\textwidth}{!}{
\begin{tabular}{c|c|m{7cm}|m{7cm}}
\hline \hline
\multicolumn{2}{c|}{\makecell{\text{ }}} & \includegraphics[width=7cm]{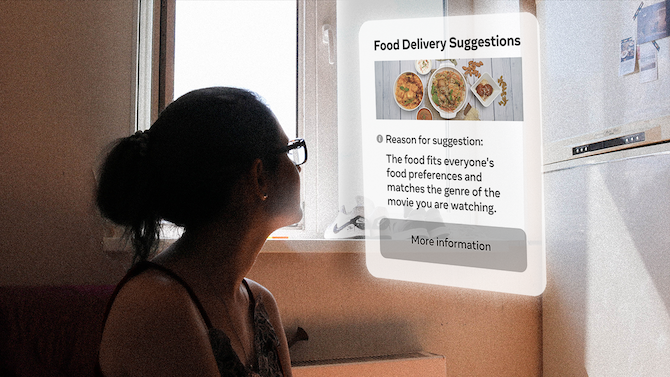} & \includegraphics[width=7cm]{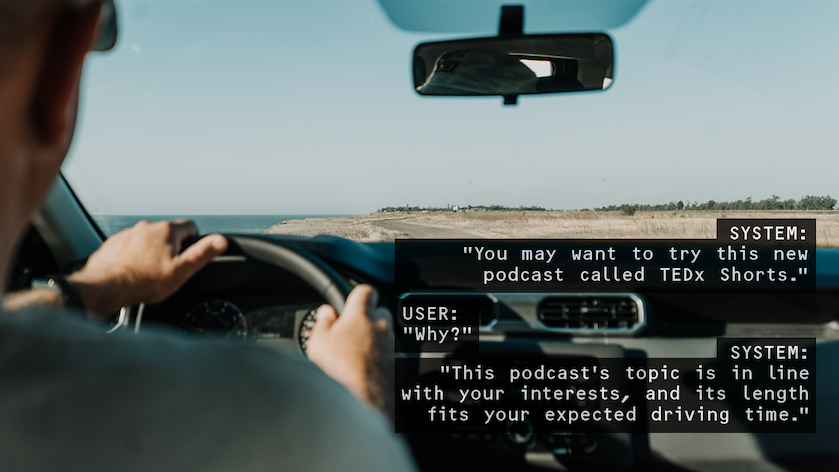}\\
\multicolumn{2}{c|}{\makecell{\textbf{Scenario Info}}} & \makecell{\textbf{Scenario 3: Food Rec for A Movie Night}} & \makecell{\textbf{Scenario 4: Podcast Rec while Driving}} \\
 \hline
\multicolumn{2}{c|}{\makecell{\textbf{Scenario}}} & Emma (general end-user, low AI literacy) has a few friends over for a small party. They decide to watch a Bollywood movie and now they are about to order food. The AR glasses recommends ordering from an Indian restaurant. Mary never heard of this restaurant before, but she loves this idea. She is also curious about the reason of this recommendation. & Jeff (general end-user, low AI literacy) is about to drive to work. The AR glasses recommends a new podcast ``TEDx Shorts'' that Jeff is unfamiliar with. However, the topic is interesting and Jeff wants to give it a try. Meanwhile, Jeff is curious to know the reason for this recommendation. \\ \hline
\multirow{3}{*}{\makecell[c]{\\\textbf{Platform-}\\\textbf{Agnostic}\\\textbf{Key Factors}}}  & \textbf{System Goal} & \textbf{User Intent Assistance} (find good food) & \textbf{User Intent Discovery} (new podcast) \\
 & \textbf{User Goal} & \textbf{Reliability, Informativeness} & \textbf{Informativeness} \\
 & \textbf{User Profile} & \makecell[l]{\textbf{User Preference}: Everyone's food preferences;\\\textbf{History}: Just decided to watch a Bollywood movie;\\\textbf{AI Literacy}: General end-user, low} & \makecell[l]{\textbf{User Preference}: Topic interests;\\\textbf{History}: Morning driving routine;\\\textbf{AI Literacy}: General end-user, low} \\ \hdashline
\multirow{2}{*}{\makecell[c]{\\\textbf{AR-Specific}\\\textbf{Key Factors}}} & \textbf{Contextual Info} & \makecell[l]{\textbf{Location}: Indoor; \textbf{Time}: Evening;\\\textbf{Environment}: Home with a group of friends} & \makecell[l]{\textbf{Location}: Outdoor; \textbf{Time}: Morning;\\\textbf{Environment}: Street conditions}\\
 & \textbf{User State} & \makecell[l]{\textbf{Activity}: Hanging out with friends;\\\textbf{Cognitive Load}: Low to Medium} & \makecell[l]{\textbf{Activity}: About to Start Driving to Work;\\\textbf{Cognitive Load}: High} \\ \hline
\multirow{7}{*}{\makecell[c]{\\\\\\\textbf{Explanation}\\ \textbf{Content Type}\\\textbf{Examples}}} & \textbf{Input/Output} & This restaurant is recommended based on everyone's food preference and movie genre. & The recommendation takes your playlist history and driving routine into account. \\
 & \textbf{Why/Why-Not} & The food fits everyone's food preferences and matches the genre of the movie you are watching. & This podcast's topic is in line with your interest, and its length fits your expected driving time. \\
 & \textbf{How} & The algorithm filters the restaurants by food preferences and then finds the best match between the food and the related activity. & The algorithm detects that it’s morning and you are driving to work, then recommends the new podcast whose topic may be of interest to you. \\
 & \textbf{Certainty} & Match score between the restaurant and the food preference and the movie: 90\% & The podcast was liked by 85\% of people with similar interests as you. \\
 & \textbf{Example} & Last time, everyone enjoyed Chinese food while watching a Chinese movie.  & ``The Daily'' and ``Fresh Air'' are other appropriate  examples when you drove to work \\
 & \textbf{What-If} & If movie genre is disabled, other cuisines would be recommended. & If the commute is longer, there are other episodes that may be of interest to you. \\
 & \textbf{How-To} & N/A & To listen to previous podcasts, you can set history as the main recommendation factor.\\
\hline
\multirow{2}{*}{\makecell[c]{\\\\\textbf{XAI Designs}\\\textbf{in AR}: \colorwhen{\textbf{\textit{When}}}}} & \colorwhen{Availability} (\gone) & Always available & Always available\\
 & \colorwhen{Delivery} (\gtwo) & Auto-trigger as both conditions is met (enough capacity and the user is not familiar with the recommendation). & Manual-trigger (high cognitive load during driving). \\ \hdashline
\multirow{3}{*}{\makecell[c]{\\\\\textbf{XAI Designs}\\\textbf{in AR}: \colorwhat{\textbf{\textit{What}}}}} & \colorwhat{Content} (\gfour) & Input/Output \& Why/Why-Not & Input/Output \& Why/Why-Not\\
 & \colorwhat{Detail - Concise} (\gfive) & The Why part of the explanation examples. & The Why part of the explanation examples.\\
 & \colorwhat{Detail - Detailed} (\gsix) & A list of the two explanation content types, plus images of the movie and food to support the Why part. & A list of the two explanation types.\\ \hdashline
\multirow{3}{*}{\makecell[c]{\\\\\textbf{XAI Designs}\\\textbf{in AR}: \colorhow{\textbf{\textit{How}}}}} & \colorhow{Modality} (\gseven) & Visual modality. & \makecell[l]{Audio modality}\\
 & \colorhow{Paradigm - Format} (\geight) & Textual format, plus graphical format in the detailed explanations. & N/A\\
 & \colorhow{Paradigm - Pattern} (\gnine) & Explicit pattern, presenting texts in the same window as the recommendations. & N/A\\
\hline \hline
\end{tabular}
}
\caption{Additional Application Examples of XAIR on Two Recommendation Scenarios.}
\label{tab:more_scenario_1}
\end{table}
\renewcommand{\arraystretch}{1.0}

\renewcommand{\arraystretch}{1.6}
\begin{table}[hbtp]
\centering
\resizebox{0.86\textwidth}{!}{
\begin{tabular}{c|c|m{7cm}|m{7cm}}
\hline \hline
\multicolumn{2}{c|}{\makecell{\text{ }}} & \includegraphics[width=7cm]{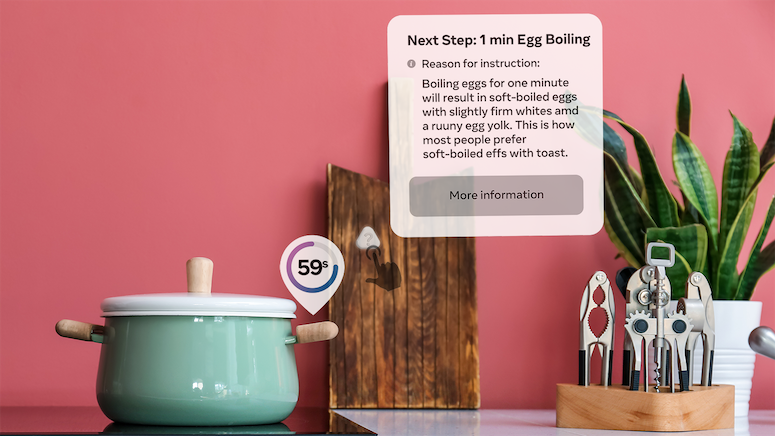} & \includegraphics[width=7cm]{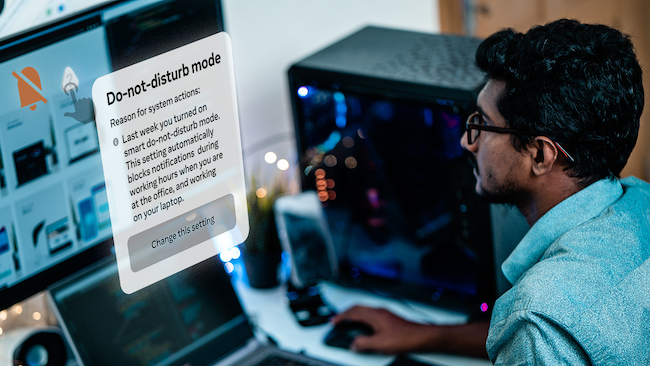}\\
\multicolumn{2}{c|}{\makecell{\textbf{Scenario Info}}} & \makecell{\textbf{Scenario 5: Cooking Instructions}} & \makecell{\textbf{Scenario 6: Automatic Do-Not-Disturb Mode}} \\
 \hline
\multicolumn{2}{c|}{\makecell{\textbf{Scenario}}} & Lisa (general end-user, low AI literacy) has recently been learning how to cook. She wants to try out a new recipe for today's lunch. She picks ``Poached Egg on Avacado Toast'' and starts to follow the instructions. After she takes the eggs out of the fridge, the AR glasses prompts to boil the egg for 1 min. Lisa is curious about the time recommendation and wants to understand what the prompt is based on. & Jeff (general end-user, low AI literacy) is about to drive to work. The AR glasses recommends a new podcast ``TEDx Shorts'' that Jeff is unfamiliar with. However, the topic is interesting and Jeff wants to give it a try. Meanwhile, Jeff is curious to know the reason for this recommendation. \\ \hline
\multirow{3}{*}{\makecell[c]{\\\textbf{Platform-}\\\textbf{Agnostic}\\\textbf{Key Factors}}}  & \textbf{System Goal} & \textbf{User Intent Assistance} (learn the recipe) & \textbf{User Intent Discovery} (new podcast) \\
 & \textbf{User Goal} & \textbf{Reliability} & \textbf{Informativeness} \\
 & \textbf{User Profile} & \makecell[l]{\textbf{User Preference}: The purpose of learning how to cook;\\\textbf{History}: Following this instruction for the first time;\\\textbf{AI Literacy}: General end-user, low} & \makecell[l]{\textbf{User Preference}: Topic interests;\\\textbf{History}: Morning driving routine;\\\textbf{AI Literacy}: General end-user, low} \\ \hdashline
\multirow{2}{*}{\makecell[c]{\\\textbf{AR-Specific}\\\textbf{Key Factors}}} & \textbf{Contextual Info} & \makecell[l]{\textbf{Location}: Kitchen; \textbf{Time}: Noon;\\\textbf{Environment}: Cookwares and ingredients } & \makecell[l]{\textbf{Location}: Outdoor; \textbf{Time}: Morning;\\\textbf{Environment}: Street conditions}\\
 & \textbf{User State} & \makecell[l]{\textbf{Activity}: Cooking;\\\textbf{Cognitive Load}: High} & \makecell[l]{\textbf{Activity}: About to Start Driving to Work;\\\textbf{Cognitive Load}: High} \\ \hline
\multirow{7}{*}{\makecell[c]{\\\\\\\textbf{Explanation}\\ \textbf{Content Type}\\\textbf{Examples}}} & \textbf{Input/Output} & The guidance is based on the instruction and your current stage. & Last week you turned on smart do-not-disturb mode. The mode is based on your location, time, and your ongoing activity. \\
 & \textbf{Why/Why-Not} & Boiling eggs for one minute will result in soft-boiled eggs with slightly firm whites and a runny egg yolk. This is how people prefer soft-boiled eggs with toast. & This setting automatically blocks notifications when you are at the office during the working hour and working on the laptop. \\
 & \textbf{How} & The algorithm detects your activity and recognizes which stage you are in, then it provides the guidance for the next step. & The system detects your current context and activity, and checks whether they meet your authored settings. If so, the Do-Not-Disturb mode will be turned on. \\
 & \textbf{Certainty} & The recognition of activity has a high certainty (88\%). & The recognition of the time, location and current activity has a high certainty of 92\%. \\
 & \textbf{Example} & N/A  & N/A \\
 & \textbf{What-If} & Other possible ways of cooking eggs, such as scrambled eggs, if you want to explore other recipe instructions. & When you are not in the office, or it is out of working hours, or you are not working in front of the laptop, the setting will not be turned on. \\
 & \textbf{How-To} & N/A & You can update any of the three conditions to change the moment the setting it’s activated.\\
\hline
\multirow{2}{*}{\makecell[c]{\\\\\textbf{XAI Designs}\\\textbf{in AR}: \colorwhen{\textbf{\textit{When}}}}} & \colorwhen{Availability} (\gone) & Always available & Always available\\
 & \colorwhen{Delivery} (\gtwo) & Manual-trigger (high cognitive load during cooking). & Manual-trigger (limited capacity in office). \\ \hdashline
\multirow{3}{*}{\makecell[c]{\\\\\textbf{XAI Designs}\\\textbf{in AR}: \colorwhat{\textbf{\textit{What}}}}} & \colorwhat{Content} (\gfour) & Input/Output \& Why/Why-Not & Input/Output, Why/Why Not, How to, Confidence, How\\
 & \colorwhat{Detail - Concise} (\gfive) & The Why part of the explanation examples. & A summary of Input, Why, and How-To as the user gets confused and wants to change the output.\\
 & \colorwhat{Detail - Detailed} (\gsix) & A list of the two explanation types, plus images of the soft-boiled eggs to support the Why part. & A list of the five types.\\ \hdashline
\multirow{3}{*}{\makecell[c]{\\\\\textbf{XAI Designs}\\\textbf{in AR}: \colorhow{\textbf{\textit{How}}}}} & \colorhow{Modality} (\gseven) & Visual modality. & \makecell[l]{Visual modality}\\
 & \colorhow{Paradigm - Format} (\geight) & Textual format, plus graphical format in the detailed explanations. & Textual format\\
 & \colorhow{Paradigm - Pattern} (\gnine) & Explicit pattern, presenting texts in the window besides the timer. & Explicit pattern, presenting texts in front of the user.\\
\hline \hline
\end{tabular}
}
\caption{Additional Application Examples of XAIR on Intelligent Instructions and Automation.}
\label{tab:more_scenario_2}
\end{table}
\renewcommand{\arraystretch}{1.0}

\newpage

\subsection{Explanation Details of the Scenarios in Study 3 \& 4}
\label{sub:appendix:details_scenarios:more_details_of_study}

Tab.~\ref{tab:case_explanation_details} shows the explanation examples presented to designers in Study 3.

\renewcommand{\arraystretch}{2}
\begin{table}[hbtp]
\centering
\resizebox{1\textwidth}{!}{
\begin{tabular}{c|c|m{9cm}|m{9cm}}
\hline \hline
\multicolumn{2}{c|}{\makecell{\text{ }}} & \includegraphics[width=9cm]{figures/design_case1_example1.png} & \includegraphics[width=9cm]{figures/design_case2_example1.png}\\
\multicolumn{2}{c|}{\makecell{\textbf{Scenario Info}}} & \makecell{\textbf{Case 1: Reliable Recipe Recommendation}} & \makecell{\textbf{Case 2: Wrong Recipe Recommendation}} \\
 \hline
% \multirow{2}{*}{\makecell[c]{\\\textbf{AR-Specific}\\\textbf{Key Factors}}} & \textbf{User Status} & \makecell[l]{\textbf{Activity}: Jogging;\\\textbf{Cognitive Load}: Low} & \makecell[l]{\textbf{Activity}: Sitting on the sofa;\\\textbf{Cognitive Load}: Low} \\
%  & \textbf{Contextual Info} & \makecell[l]{\textbf{Location}: Outdoor; \textbf{Time}: Morning;\\\textbf{Environment}: Trails, streets} & \makecell[l]{\textbf{Location}: Home; \textbf{Time}: Afternoon;\\\textbf{Environment}: Living room furniture, the plant} \\ \hdashline
% \multirow{3}{*}{\makecell[c]{\\\textbf{Platform-}\\\textbf{Agnostic}\\\textbf{Key Factors}}}  & \textbf{System Goal} & \textbf{User Intent Discovery (new route)} & \textbf{Trust Building (clarification)} \\
%  & \textbf{User Goal} & \textbf{Resolve Surprise} & \textbf{Privacy Awareness} \\
%  & \textbf{User Profile} & \makecell[l]{\textbf{User Preference}: Cherry blossom tree lover;\\\textbf{History}: Regular jogging in the morning;\\\textbf{AI Literacy}: Expert, high} & \makecell[l]{\textbf{User Preference}: Plant enthusiast;\\\textbf{History}: Did not care of the plant for a while;\\\textbf{AI Literacy}: General end-user, low} \\ \hline
\multirow{7}{*}{\makecell[c]{\\\\\\\textbf{Explanation}\\\textbf{Content Type}\\\textbf{Examples}}} & \textbf{Input/Output} & This recipe comes from the items detected in the fridge: egg and shrimp, and take your diet into account. & This recipe is based on friends' food preferences and the detected ingredients in your fridge: salmon and carrot. \\
 & \textbf{Why/Why-Not} & This recipe fits your diet and food preference. It is recommended based on the rich amount of protein: 32g. & This recipe matches your friends' preference. It is recommended based on the popularity: 3201 people liked it. \\
 & \textbf{How} & The algorithm recognizes ingredients in the fridge, finds and ranks recipes based on the available ingredients and your diet preference. & The algorithm first recognizes ingredients in the fridge, finds and ranks recipes based on the available ingredients and food preference.\\
 & \textbf{Certainty} & Match rate between the recipe and the food preference \& ingredients : 82\%. & The recognition of salmon is uncertain (confidence 71\%). It is not sure whether salmon or steak (recognition confidence 29\%). \\
 & \textbf{Example} & N/A & N/A \\
 & \textbf{What-If} & More recipes if you want to try other cuisines. & Different recipes if your friends want to try other cuisines. \\
 & \textbf{How-To} & Disable the diet option to see previous recipes before you went on the high-protein diet. & Select the right ingredients to change the recommendations: salmon or steak [clickable buttons]. \\
\hline \hline
\end{tabular}
}
\caption{Details of The Two Cases in Sec.~\ref{sec:evaluation}. Examples are ``N/A'' as they are already multiple examples in the recommendations.}
\label{tab:case_explanation_details}
\end{table}
\renewcommand{\arraystretch}{1.0}

\end{document}